\documentclass[10pt,a4paper]{article}
\usepackage{times}
\usepackage{mathrsfs}
\usepackage{latexsym,amsopn,amsmath,amssymb}
\usepackage{amsfonts,amsbsy,amscd,stmaryrd}
\usepackage{theorem}
\usepackage{geometry}
\newcommand{\mbc}{\mathbb{C}}

\newcommand{\mbn}{\mathbb{N}}
\newcommand{\mbr}{\mathbb{R}}

\newcommand{\mbz}{\mathbb{Z}}
\newcommand{\ca}{\mathcal{A}}

\newcommand{\cc}{\mathcal{C}}

\newcommand{\ce}{\mathcal{E}}
\newcommand{\cf}{\mathcal{F}}

\newcommand{\ch}{\mathcal{H}}

\newcommand{\ck}{\mathcal{K}}
\newcommand{\cl}{\mathcal{L}}
\newcommand{\cm}{\mathcal{M}}
\newcommand{\cn}{\mathcal{N}}
\newcommand{\co}{\mathcal{O}}
\newcommand{\cp}{\mathcal{P}}
\newcommand{\cR}{\mathcal{R}}
\newcommand{\cs}{\mathcal{S}}

\def\rond{\mathscr}
\newcommand{\ra}{\rond{A}}

\newcommand{\rc}{\rond{C}}
\newcommand{\rd}{\rond{D}}

\newcommand{\rf}{\rond{F}}

\newcommand{\rh}{\rond{H}}

\newcommand{\rj}{\rond{J}}
\newcommand{\rk}{\rond{K}}
\newcommand{\rl}{\rond{L}}
\newcommand{\rM}{\rond{M}}
\newcommand{\rn}{\rond{N}}

\newcommand{\rp}{\rond{P}}

\newcommand{\rt}{\rond{T}}

\newcommand{\rw}{\rond{W}}
\newcommand{\rx}{\rond{X}}

\def\rmc{\text{c}}
\newcommand{\dd}{\mbox{\rm d}}
\def\rmd{\text{d}}
\def\rme{\text{e}}
\def\rmi{\text{i}}

\def\g#1{\mathfrak #1}
\def\gs{\mathfrak{S}}

\newcommand{\Co}{{C_0}}
\newcommand{\Cc}{C_{\rmc}}
\newcommand{\gfin}{\Gamma_{\text{fin}}}
\def\veps{\varepsilon}

\newcommand{\ind}{1}    
\newcommand{\proof}{\noindent{\bf Proof: }}

\def\rag{\rangle}
\def\lag{\langle}
\def\se{\sigma_{\text{ess}}}
\def\sd{\sigma_{\text{d}}}

\def\rarrow{\rightarrow}
\def\rarrowup{\rightharpoonup}

\def\wtilde{\widetilde}
\def\wwtilde#1{\widetilde{#1\,}}

\def\Id{\mbox{\rm Id}}

\def\slim{\mbox{\rm s-}\!\lim}
\def\wlim{\mbox{\rm w-}\!\lim}
\def\nin{\notin}

\def\llb{\llbracket}
\def\rrb{\rrbracket}

\def\nin{\notin}

\def\ccup{\textstyle\bigcup}

\def\qed{\hfill \vrule width 8pt height 9pt depth-1pt \medskip}

\def\build#1_#2^#3{\mathrel{\mathop{\kern 0pt#1}\limits_{#2}^{#3}}}
\newcounter{PAR}[section]

\def\@begintheorem#1#2{\it \trivlist
\item[\hskip \labelsep{\bf #1\ #2}]}
\def\@endtheorem{\endtrivlist}
\newtheorem{theorem}{\rm \bf Theorem}[section]
\newtheorem{lemma}[theorem]{Lemma}
\newtheorem{proposition}[theorem]{Proposition}
\newtheorem{corollary}[theorem]{Corollary}

\newtheorem{definition}[theorem]{Definition}
\newtheorem{example}[theorem]{Example}
\newtheorem{remark}[theorem]{Remark}

\long\def\symbolfootnote[#1]#2{\begingroup%
\def\thefootnote{\fnsymbol{footnote}}\footnote[#1]{#2}\endgroup} 
{
\theorembodyfont{\rmfamily} \theoremstyle{plain}

}
\begin{document}
\title{On the Spectral Analysis of Quantum Field Hamiltonians}
\author{ Vladimir Georgescu\thanks{CNRS (UMR 8088) and
    Department of Mathematics, University of Cergy-Pontoise,
    2 avenue Adolphe Chauvin, 95302 Cergy-Pontoise Cedex, France.
    E-mail address: \texttt{vlad@math.cnrs.fr}}}
\maketitle 
\begin{abstract}
\noindent
We define $C^*$-algebras on a Fock space such that the Hamiltonians
of quantum field models with positive mass are affiliated to
them. We describe the quotient of such algebras with respect to the
ideal of compact operators and deduce consequences in the spectral
theory of these Hamiltonians: we compute their essential spectrum
and give a systematic procedure for proving the Mourre estimate.
\end{abstract}
\tableofcontents

\section{Introduction}
\protect\setcounter{equation}{0}

This paper is motivated and related to the work on the spectral and
scattering theory of quantum field models initiated in \cite{HS,Ge1}
and further developed in \cite{DG1,DG2,DJ}. Our purpose is to show
that abstract $C^*$-algebra techniques allow one to obtain in this
context quite general results in a rather simple and systematic way
which avoids ad-hoc and intricate constructions.  We use ideas
introduced in \cite{BG1,BG2} in the context of the $N$-body problem
and in a more general setting in \cite{GI1}. The main point of this
approach is that understanding the quotient of a $C^*$-algebra with
respect to the ideal of compact operators\symbolfootnote[2]{\ More
general ideals also play a r\^ole, cf.\ \cite{BG1,BG2,ABG}.}  gives
a lot of information relevant to the spectral analysis of the
operators affiliated to the algebra. 
In \cite{GI1,GI2} the relevant $C^*$-algebras are  generated by a
set of ``elementary'' Hamiltonians specific to a certain physical
situation. The ``real'' Hamiltonians are then the self-adjoint
operators affiliated to the algebra. We adopt here the same strategy.

In order to avoid any misunderstanding we emphasize that the topics
considered in this paper are quite far from the theory of
relativistic quantum fields. As in the references quoted above (and
in the Reference section) our results are relevant only for quantum
field models with a spatial cutoff and living in a Fock space
(hopefully this last restriction will be removed in the near
future). On the other hand, our approach clearly covers many
physically interesting models of the many-body theory, our focus
being on the study of systems with an infinite number of degrees of
freedom and without particle number conservation.

Our results on the spectral analysis of quantum field Hamiltonians
(QFH) are consequences of the theorem stated
below\symbolfootnote[3]{\ In this introduction we shall freely use
notions, notations and terminology which are defined in precise
terms in the body of the paper,  see especially Sections
\ref{s:fock}, \ref{s:ffock} and \ref{s:saa}. 
}. 
Let $\ch$ be a
complex Hilbert space and let $\Gamma(\ch)$ be the symmetric or
antisymmetric Fock space over $\ch$. The field operators $\phi(u)$
and the Segal operators $\Gamma(A)$ are defined as usual. If
$U=(u_1,\dots,u_n)$ belongs to the Cartesian power $\ch^n$ we set
$\phi(U)=\phi(u_1)\dots\phi(u_n)$; in the case $n=0$ this is
interpreted as $\phi(\emptyset)=1_{\Gamma(\ch)}$.  If $\|A\|<1$ then
$\phi(U)\Gamma(A)$ is a well defined bounded operator.  Let
$\rk(\ch)$ be the space of all compact operators on $\Gamma(\ch)$.

\begin{theorem}\label{th:max}
Let $\co$ be an abelian $C^*$-algebra on $\ch$ such that its strong
closure does not contain finite rank projections. Let
$\rf(\co)\subset B(\Gamma(\ch))$ be the $C^*$-algebra generated by
the operators $\phi(U)\Gamma(A)$ with $U$ as above and $A\in\co$
with $\|A\|<1$.  Then there is a unique morphism
$\cp:\rf(\co)\rarrow\co\otimes\rf(\co)$ such that
$\cp\left[\phi(U)\Gamma(A)\right]=
A\otimes\left[\phi(U)\Gamma(A)\right]$ for all $U,A$.  We have
$\ker\cp=\rk(\ch)$, which defines a canonical embedding
\begin{equation}\label{eq:cmoi}
\rf(\co)/\rk(\ch)\hookrightarrow \co\otimes\rf(\co).
\end{equation} 
\end{theorem}
This statement has the advantage that it is simple and covers both
the bosonic and fermionic cases. Alternative, technically more
convenient, versions of Theorem \ref{th:max} are Theorem
\ref{th:cmo} (see also Lemma \ref{lm:cmd}) and Theorem
\ref{th:fcmo}.  Instead of working separately with the Bose and
Fermi case one may consider a supersymmetric (or $\mbz_2$-graded)
Hilbert space $\ch$ as in \cite{De} which gives a unified approach
to the subject. Since this requires more preliminary developments,
and since one gets the same result by taking a tensor product of the
bosonic and fermionic Fock space, we did not present this version.

In spite of the simplicity of its statement, Theorem \ref{th:max}
has important consequence in the spectral theory of QFH: it
immediately gives a description of the essential spectrum of these
Hamiltonians and also gives a systematic and simple way of proving
the Mourre estimate for them with conjugate operators of the form
$A=\rmd\Gamma(\g{a})$. Such an estimate allows one to prove absence
of singular continuous spectrum and is an important step in the
proof of asymptotic completeness, cf.\ \cite{DG1,DG2,Am2,Am3}.

The first difficulty one meets in the algebraic approach we use is
the isolation of the correct ``algebra of energy observables'', in
the terminology of \cite{GI1,GI2}. In fact, if the algebra we start
with is too large, then its quotient with respect to the compacts
will probably be too complicated to be useful. On the other hand, we
cannot choose it too small because then physically relevant
Hamiltonians will not be affiliated to it. Since we have chosen the
algebras $\rf(\co)$ in such a way that general classes of QFH are
self-adjoint operators affiliated to them, it seems to as quite
remarkable that the description of the quotient given in
\eqref{eq:cmoi} is so simple. 

One can also give a priori
justifications of the choice of $\rf(\co)$, we describe two of them
below.
First, the algebra $\rf(\co)$ can be obtained by a procedure
completely analogous to that used in \cite{GI1} in the setting of
quantum systems with a finite number of degrees of freedom. We
interpret $\ch$ as the one particle Hilbert space and $\co$ as the
$C^*$-algebra generated by the one particle kinetic energy
Hamiltonians\symbolfootnote[2]{\ We assume here that $\co$ acts
non-degenerately on $\ch$, the only situation of physical interest.
The model one should always have in mind is $\ch=L^2(X)$ with $X$ a
locally compact abelian group, the configuration space of the
particle, and $\co=\Co(X^*)$, the algebra of continuous, convergent
to zero at infinity, functions of the momentum operator. }.  We take
as algebra of kinetic energy observables of the field $\Gamma(\co) =
C^*(\Gamma(A)\!\mid\! A\in\co,\|A\|<1)$, because this is the
$C^*$-algebra generated by the operators of the form $\dd\Gamma(h)$
with $h$ a self-adjoint operator operator affiliated to $\co$ with
$\inf h=m>0$ (in this paper we restrict ourselves to the case of
particles with strictly positive mass). Now we have to decide what
kind of interactions we take into account.  It is characteristic to
quantum fields that the interaction term is some kind of generalized
polynomial in the field operators.  In the fermionic case we define
the ``algebra of elementary interactions'' $\rf(\ch)$ as the
$C^*$-algebra generated by polynomials in the field operators. Since
in the bosonic case the field operators are not bounded, we define
$\rf(\ch)$ in this case as the $C^*$-algebra generated by operators
of the form $\int_E\rme^{\rmi\phi(u)}f(u)d_Eu$, where $E$ is a
finite dimensional vector subspace of $\ch$, $d_Eu$ is the measure
associated to the Euclidean structure we have on $E$, and $f$ is an
integrable function on $E$. Finally, the algebra of energy
observables of the field should be the norm closed linear space of
operators on $\Gamma(\ch)$ generated by the products $FS$ with
$F\in\rf(\ch)$ and $S\in\Gamma(\co)$. It is easy to see that this is
exactly $\rf(\co)$.

A second characterization of the algebra $\rf(\co)$ is physically
more satisfactory.  Let us call \emph{elementary quantum field
Hamiltonian of type $\co$} a self-adjoint operator of the form
$H=\dd\Gamma(h)+V$, where $h$ is a self-adjoint operator on $\ch$
affiliated to $\co$ such that $h\geq m$ for some real $m>0$ and
$V\in\rf(\ch)$ is a symmetric operator. This seems to be the
smallest class of self-adjoint operators which may naturally be
thought as QFH. But $\rf(\co)$ is just the $C^*$-algebra generated
by these QFH (Proposition \ref{pr:woo}).

On the other hand, the condition which characterizes $\cp$ in
Theorem \ref{th:max} can be stated in the following equivalent form:
$\cp(H)=h\otimes\ind_{\Gamma(\ch)}+\ind_\ch\otimes H$ for each
elementary QFH (Proposition \ref{pr:carp} and Lemma \ref{lm:cmd}).
But this relation has a simple physical interpretation: it says that
by taking the quotient with respect to the compacts one gets the
Hamiltonian of the system consisting of a free particle with kinetic
energy $h$ and of the initial field (the interaction between them
being cutoff). So one particle has been pull out from the field
without modifying the Hamiltonian of the field, which is possible
because it consists of a (potentially) infinite number of particles.

As we said above, the embedding \eqref{eq:cmoi} has interesting
consequences in the spectral analysis of the self-adjoint operators
affiliated to $\rf(\co)$. Thus it is important to show that
physically realistic QFH belong to this class and this is not at all
obvious because the elementary QFH which generate the algebra are
just toy models, they only look like real QFH. In Section
\ref{s:saa} we give several general criteria for an operator to be
affiliated to $\rf(\co)$ which show that the class of affiliated
Hamiltonians is large. As an application, we point out in Section
\ref{s:lgr} an abstract class of operators affiliated to $\rf(\co)$
which covers the Hamiltonian of the $P(\varphi)_2$ model with a
spatial cutoff.  In Section \ref{s:cs}, where we show how to treat
coupled systems in our framework, we prove that massive Pauli-Fierz
Hamiltonians are affiliated to $\rf(\co)\otimes K(\rl)$ ($\rl$ is
the Hilbert space of the confined system) and deduce the location of
their essential spectrum and the Mourre estimate under conditions on
the form factor weaker than usual (see assumption (PF)
\pageref{p:pf}).

We shall describe now in colloquial terms the kind of results we get
concerning the spectral properties of the operators affiliated to
$\rf(\co)$ (precise statements and details are in Sections
\ref{s:saa}, \ref{s:me} and \ref{s:cs}). In what concerns the
essential spectrum, the following is an immediate consequence of
\eqref{eq:cmoi}: if $H$ is a self-adjoint operator affiliated to
$\rf(\co)$ then
\begin{equation}\label{eq:essi}
\se(H)=\sigma(\cp(H)).
\end{equation}
Here $\widetilde H\equiv\cp(H)$ is a self-adjoint
operator\symbolfootnote[2]{\ Or, in very singular situations that do
not concern us here, a slightly more general object, since its
domain could be not dense. }  on $\ch\otimes\Gamma(\ch)$ affiliated
to $\co\otimes\rf(\co)$. If $\rx$ is the spectrum of the abelian
algebra $\co$ then $\co\otimes\rf(\co)\cong \Co(\rx;\rf(\co))$ and
$\wtilde H$ is identified with a continuous family $\{\wtilde
H(x)\}_{x\in\rx}$ of self-adjoint operators on $\Gamma(\ch)$
affiliated to $\rf(\co)$. Then \eqref{eq:essi} can be written as:
\begin{equation}\label{eq:ox*i}
\se(H)=\textstyle\bigcup_{x\in\rx}\sigma(\wtilde H(x)).
\end{equation}
The Hamiltonians of the quantum field models usually considered in
the literature are, however, much more specific than just
affiliated to $\rf(\co)$: they are bounded from below and have the
property that there is a self-adjoint operator $h$ affiliated to
$\co$ with $h\geq m>0$ such that
$\cp(H)=h\otimes\ind_{\Gamma(\ch)}+\ind_\ch\otimes H$.  We call such
QFH \emph{standard} (Definition \ref{df:dqfh}). The simplest
standard QFH are the elementary ones, but the class is much larger,
for example the $P(\varphi)_2$ and Pauli-Fierz models as well as the
fermionic models considered in \cite{Am2,Am3} belong to this class.
Now for a standard $H$ we clearly have:
\begin{equation}\label{eq:ses}
\se(H)=\sigma(h)+\sigma(H).
\end{equation}
This formula covers the models treated in \cite{DG1,DG2,DJ,Am2,Am3}
(in historical order). The version \eqref{eq:fnes} for systems with
a particle number cutoff covers the spin-boson model \cite{HS,Ge1}.

We then study the Mourre estimate for \emph{standard} QFH (in this
case the result is quite explicit, but more general situations may
be treated, see Remark \ref{re:disp}). As in
\cite{BFS,BFSS,DG1,DG2,Sk} we consider only conjugate operators of
the form $A=\dd\Gamma(\g{a})$ where $\g{a}$ is a self-adjoint
operator on $\ch$. We assume $\rme^{-\rmi t\g{a}}\co\rme^{\rmi
t\g{a}}=\co$ for all real $t$ and that $t\mapsto\rme^{-\rmi
t\g{a}}S\rme^{\rmi t\g{a}}$ is norm continuous if $S\in\co$ (these
conditions are easy to check in applications). Moreover, $H$ and $h$
must satisfy usual regularity conditions with respect to $A$ and
$\g{a}$ respectively (see Theorem \ref{th:mou}). Finally, the
commutator $[h,\rmi\g{a}]$ must satisfy a weak local positivity
condition (this is assumption $\rho^{\g{a}}_h\geq0$ in Theorem
\ref{th:mou}), namely:
\begin{equation}\label{eq:mip}
\text{for each real } \lambda \text{ and } \delta>0
\text{ there  is } \varepsilon>0 \text{ such that }
E_h(\lambda,\varepsilon)[h,\rmi\g{a}]E_h(\lambda,\varepsilon)\geq
-\delta E_h(\lambda,\varepsilon )
\end{equation}
where $E_h(\lambda,\varepsilon)$ is the spectral projection of $h$
associated to the interval
$[\lambda-\varepsilon,\lambda+\varepsilon]$. In fact, in
applications one chooses $\g{a}$ such that $[h,\rmi\g{a}]\geq0$, so
this condition is trivial to check. 

Let us define the \emph{threshold set $\tau_{\g{a}}(h)$ of $h$ with
respect to $\g{a}$} as the complement in $\mbr$ of the set of
$\lambda$ where the Mourre estimate holds, i.e.\ $\lambda$ is such
that:
\begin{equation}\label{eq:mips}
\text{there are } \varepsilon,\delta>0  
\text{ and a compact operator }  K
\text{ such that }
E_h(\lambda,\varepsilon)[h,\rmi\g{a}]E_h(\lambda,\varepsilon)\geq
\delta E_h(\lambda,\varepsilon ) +K.
\end{equation}
Denote $\tau^n_{\g{a}}(h)=\tau_{\g{a}}(h)+\dots+\tau_{\g{a}}(h)$
($n$ terms) and let
\begin{equation}\label{eq:thsi}
\tau_A(H)=
\big[\ccup_{n=1}^\infty \tau^n_{\g{a}}(h)\big]+\sigma_{\text{p}}(H),
\end{equation}
Then Theorem \ref{th:mou} says that \emph{$\tau_A(H)$ is the
threshold set of $H$ with respect to $A$}. So at each point outside
$\tau_A(H)$ the operator $A$ is conjugate to $H$ in the sense of
Mourre (i.e.\ an estimate similar to \eqref{eq:mips} holds).

The relation \eqref{eq:thsi} is quite intuitive physically
speaking. It says that an energy $\lambda$ is an $A$-threshold for
$H$ if and only if one can write it as a sum
$\lambda=\lambda_1+\dots+\lambda_n+\mu$ where the $\lambda_k$ are
$\g{a}$-threshold energies of the free particle of kinetic energy
$h$ and $\mu$ is the energy of a bound state of the field. So at
energy $\lambda$ one can extract $n$ free particles from the field
such that each one has an $\g{a}$-threshold energy and such that the
field remains in a bound state.

We wish to make some historical comments concerning the methods we
use. First, the fact that quotients of $C^*$-algebras with respect
to the ideal of compact operators play an important r\^ole is an old
and quite natural idea in the context of the theory of
pseudo-differential operators; the references \cite{Co,Ty} seem
particularly relevant for us. Second, the first use of $C^*$-algebra
methods in the spectral analysis of physically interesting models
appears, as far as we know, in the work of J.\ Bellissard
\cite{Be1,Be2} on solid state physics (see \cite{Be3} for more
recent results and references). But the $C^*$-algebras and the
$C^*$-algebra techniques used by Bellissard and his collaborators
are very different from ours, e.g.\ $K$-theory plays an important
r\^ole in their works but are probably irrelevant here (it would be
nice if somebody would show the contrary).  The usefulness of
techniques like computation of quotients of $C^*$-algebras in the
spectral theory of many body systems and quantum field models seems
to have been first noticed in \cite{BG1,BG2}.  Note that some of the
results described here were announced in \cite{Ge,GI1,GI2}.

The paper is organized as follows.  In Section \ref{s:fock} we
summarize the most important notations and results from the theory
of symmetric Fock spaces following \cite{BR,BSZ,Gu} and also the
more recent \cite{DG1,DG2}. We prefer to define the scalar product
\eqref{eq:sc} on a Fock space as in \cite{Ni} and the definition
\eqref{eq:an} of the annihilation and creation operators in terms of
the field operators is slightly unusual, which explains some
differences in the numerical factors. Similar conventions are
adopted in the antisymmetric case presented in Section \ref{s:ffock}
where we use \cite{PR} as main reference.  

In Section \ref{s:wo} we define the algebras $\rf(\co)$ and present
some of their properties and alternative characterizations.  In
Section \ref{s:wa} we prove the main theorem for the algebra
$\ra(\ch)\equiv\rf(\mbc\ind_\ch)$, which is an important technical
step but also has an intrinsic mathematical interest because we show
that the quotient $\ra(\ch)/\rk(\ch)$ is canonically isomorphic
to $\ra(\ch)$. We also give there some consequences of this fact in
the spectral analysis of the elements of $\ra(\ch)$. In Section
\ref{s:cmo} we prove our main technical result, Theorem
\ref{th:cmo}.  We consider only the bosonic case until Section
\ref{s:ffock} where we describe briefly the corresponding results
in the fermionic case (which is nicer but easier).

Sections \ref{s:saa}-\ref{s:cut} are devoted to applications in the
spectral analysis of quantum field models of Theorem
\ref{th:max}. In Section \ref{s:saa} we give criteria for
affiliation to $\rf(\co)$ and a general formula for the essential
spectrum of the operators affiliated to this algebra (Theorem
\ref{th:ess} and relation \eqref{eq:ox*}). We also introduce there
the important class of standard QFH and describe their essential
spectrum. The main result of Section \ref{s:me} is Theorem
\ref{th:mou} which gives the Mourre estimate for such Hamiltonians.
In Section \ref{s:lgr} we show that a general class of QFH,
including the $P(\varphi)_2$ model, are standard in the sense
defined before, hence all these results apply to them. In Section
\ref{s:cs} we sketch a method of analyzing several fields with
couplings between them and external systems and consider in detail
the massive Pauli-Fierz model. Note that the Pauli-Fierz Hamiltonian
is also standard. In the last section we treat models with a
particle number cutoff, which have some interesting features.  We do
not treat explicitly the fermionic case because it is easy to see
that models like those considered in \cite{Am2,Am3} are standard in
our sense so their spectral properties (essential spectrum and
Mourre estimate) follow from the general theorems of Sections
\ref{s:saa} and \ref{s:me}.

\bigskip

\noindent{\bf Acknowledgments:} I am grateful to Christian
G\'erard and George Scandals for very helpful discussions, cf.\ the 
Remarks \ref{re:skan} and \ref{re:cpro} and the comment after
Theorem \ref{th:byte}.

\section{Bosonic Fock space}\label{s:fock}
\protect\setcounter{equation}{0}

\noindent{\bf 1.}  Our notations are rather standard but we recall
here some of them to avoid any ambiguity. If $\ce,\cf$ are vector
spaces then $L(\ce,\cf)$ is the space of linear maps $\ce\rarrow\cf$
and we abbreviate $L(\ce)=L(\ce,\ce)$.  If $\ce,\cf$ are Banach
spaces then $B(\ce,\cf)$ and $K(\ce,\cf)$ are the subspaces of
$L(\ce,\cf)$ consisting of continuous or compact maps respectively
and we set $B(\ce)=B(\ce,\ce)$, $K(\ce)=K(\ce,\ce)$.  When needed
for the clarity of the argument we denote by $\ind_\ce$ the identity
operator on a Banach space $\ce$ or the identity element of an
algebra $\ce$. The domain of an operator $T$ is denoted $D(T)$.  The
Hilbert spaces are complex Hilbert spaces unless the contrary is
explicitly mentioned and the scalar product is linear in the second
variable. If a symbol like $T^{(*)}$ appears in a relation, this
means that the relation holds both for $T$ and $T^*$.  We denote by
$C^*(T\mid T\in\rt, P_1,P_2,\dots)$ the $C^*$-algebra generated by a
family $\rt$ of operators $T$ which have the properties $P_1,P_2$,
etc.  The $C^*$-algebra generated by a self-adjoint operator $H$ is
$\Co(H)=\{f(H)\mid f\in\Co(\mbr)\}$. More generally, the
$C^*$-algebra generated by a family of self-adjoint operators is the
smallest $C^*$-algebra which contains the re solvents of these
operators. A \emph{morphism} between two $C^*$-algebras is a
$*$-morphism.  $\Co(X)$ is the space of continuous complex valued
functions on the locally compact space $X$ that converge to zero at
infinity and $\Cc(X)$ that of continuous functions with compact
support.

We need a version of the polarization formula. Let $X,Y$
be vector spaces, $Q:X\times\dots\times X\rarrow Y$ an $n$-linear
symmetric map, and let us set $q(x)=Q(x,\dots,x)$. Denote $|a|$ the
cardinal of a set $a$. Then:
\begin{equation}\label{eq:pz}
(-1)^nn! Q(x_1,\dots,x_n)= \textstyle\sum_{a\subset\{1,\dots,n\}}
(-1)^{|a|} q\left(\sum_{i\in a}x_i\right).
\end{equation}

\noindent{\bf 2.} Let $\ch$ be a complex Hilbert space with scalar
product $\lag\cdot|\cdot\rag$ and let $U(\ch)$ be the group of
unitary operators on $\ch$. A \emph{(regular) representation of the
CCR over $\ch$}, or a \emph{Weyl system over $\ch$}, is a couple
$(\rh,W)$ consisting of a Hilbert space $\rh$ and a map
$W:\ch\rarrow U(\rh)$ which satisfies
\begin{equation}\label{eq:ws}
W(u+v)=\rme^{\rmi\Im\lag u|v\rag}W(u)W(v) \hspace{2mm} 
\textrm{ for all } u,v\in\ch
\end{equation}
and such that the restriction of $W$ to each finite dimensional
subspace is strongly continuous. Then
\begin{equation}\label{eq:wr}
W(0)=1,\hspace{2mm} W(u)^*=W(-u),\hspace{2mm}
W(u)W(v)=\rme^{-2\rmi\Im\lag u|v\rag}W(v)W(u).
\end{equation}
We denote $\rw(\ch)$ the $C^*$-algebra generated by the operators
$W(u)$ and we call it \emph{Weyl algebra over $\ch$}: \label{p:w}
\begin{equation}\label{eq:wah}
\rw(\ch)=C^*(W(u)\mid u\in\ch).
\end{equation}
The $C^*$-algebras $\rw(\ch)$ associated to two Weyl systems are
canonically isomorphic, see \cite{BR} for a proof. This also gives
canonical embeddings $\rw(\ck)\subset\rw(\ch)$ for closed
subspace $\ck$ of $\ch$.

The \emph{field operator} associated to the one particle state
$u\in\ch$ is defined as the unique self-adjoint operator
$\phi(u)$ on $\rh$ such that $W(tu)=\rme^{\rmi t\phi(u)}$ for
all real $t$. We have for all $u,v\in\ch$:
\begin{equation}\label{eq:ph}
W(u)\phi(v)W(u)^*=\phi(v)-2\Im\lag u|v \rag
\hspace{2mm} \text{ and } \hspace{2mm}
[\phi(u),\phi(v)]=2\rmi\Im\lag u|v \rag.
\end{equation}
The space $\rh^\infty$ of vectors $f\in\rh$ such that $u\mapsto
W(u)f$ is a $C^\infty$ map on each finite dimensional subspace of
$\ch$ is a dense subspace of $\rh$ stable under all the operators
$W(u)$ and $\phi(u)$. Moreover, $\rh^\infty$ is a core for each
$\phi(u)$ (by Nelson Lemma) and the second relation in \eqref{eq:ph}
holds in operator sense on $\rh^\infty$.  The map
$u\mapsto\phi(u)\in L(\rh^\infty)$ is clearly \emph{not} linear but
only $\mbr-$linear, as it follows from \eqref{eq:ws} after replacing
$u,v$ by $tu,tv$ with $t$ real and then taking derivatives at $t=0$.

The \emph{annihilation} and \emph{creation} operators associated to
the one particle state $u$ are defined by
\begin{equation}\label{eq:an}
a(u) = (\phi(u)+\rmi\phi(\rmi u))/2, \hspace{2mm}
a^*(u)= (\phi(u)-\rmi\phi(\rmi u))/2
\end{equation}
on $\rh^\infty$ and then extended by taking closures. On
$\rh^\infty$ we have $\phi(u) = a(u)+a^*(u)$.  The map $u\mapsto
a^*(u)\in L(\rh^\infty)$ is linear, $u\mapsto a(u)\in L(\rh^\infty)$
is antilinear, and:
\begin{equation}\label{eq:ac}
[a(u),a^*(v)]=\lag u|v \rag, \hspace{2mm}
 [a(u),a(v)]=0, \hspace{2mm}
[a^*(u),a^*(v)]=0 \hspace{2mm} \text{ on } \rh^\infty.
\end{equation}
On the other hand, from \eqref{eq:ph} we also get:
\begin{equation}\label{eq:wa}
W(u)a^{(*)}(v)W(u)^*=a^{(*)}(v)-\lag v|\rmi u \rag^{(*)}, 
\hspace{2mm}
[a^{(*)}(v),W(u)]=\lag v|\rmi u\rag^{(*)}W(u).
\end{equation}
 
Some of our later constructions will depend only on the existence of
a \emph{particle number operator for the Weyl system $W$}, which is
a self-adjoint operator $N$ on $\rh$ such that
\begin{equation}\label{eq:no}
\rme^{\rmi tN}W(u)\rme^{-\rmi tN}=W(\rme^{\rmi t}u) \hspace{2mm}
\textrm{ for all } t\in\mbr \textrm{ and } u\in\ch. 
\end{equation}
Such an operator is clearly not uniquely defined and it is easy to
prove that if it exists then $N$ can be chosen such that its
spectrum be either $\mbn=\{0,1,2,\dots\}$ or $\mbz$, see \cite{Ch1}.
In \cite{Ch2} it is shown that we are in the first situation if and
only if $W$ is a direct sum of Fock representations (cf.\
below). Since
$$
W(\rme^{\rmi t}u)=W(u\cos t+\rmi\sin t)=
\rme^{\frac{\rmi}{2}\|u\|^2\sin2t}W(u\cos t)W(\rmi u\sin t)
$$
by taking derivatives in \eqref{eq:no} at $t=0$ we get
(this is easy to justify in the Fock representation):
\begin{equation}\label{eq:wn}
W(u)NW(u)^*=N-\phi(\rmi u)+\|u\|^2, \hspace{2mm}
[N,W(u)]=W(u)(\phi(\rmi u)+\|u\|^2).
\end{equation}
Replacing $u$ by $tu$ in the last equation and then taking the
derivatives at $t=0$ we get
\begin{equation}\label{eq:nf}
[N,\rmi\phi(u)]=\phi(\rmi u), \hspace{2mm}
(N+1)a(u)=a(u)N, \hspace{2mm} (N-1)a^*(u)=a^*(u)N.
\end{equation}
 
A \emph{vacuum state} for the Weyl system $W$ is a vector
$\Omega\in\rh$ with $\|\Omega\|=1$, $\Omega\in D(\phi(u))$ for
all $u\in\ch$, and such that the map $u\mapsto\phi(u)\Omega$ is
linear. It is easy to prove that a vacuum state belongs to
$\rh^\infty$ and that a vector $\Omega$ of norm one is a vacuum
state if and only if $\Omega\in\cap_uD(a(u))$ and $a(u)\Omega=0$ for
all $u$, see for example  \cite[Proposition 4.1]{DG2}.

A \emph{Fock representation} \label{p:fock} of the CCR over $\ch$ is
a triple $(\rh,W,\Omega)$ consisting of a Weyl system $(\rh,W)$ over
$\ch$ and a vacuum state $\Omega$ which is cyclic for $W$.  It is
easy to show that two Fock representations are canonically
isomorphic, more precisely if $(\rh',W',\Omega')$ is a second Fock
representation then there is a unique bijective isometry
$J:\rh\rarrow\rh'$ such that $J\Omega=\Omega'$ and $JW(u)=W'(u)J$
for all $u\in\ch$. For this reason one may say {\it the} Fock
representation and speak about ``realizations'' of this
representation. The realizations are constructed such as to
diagonalize various sets of operators. If $\ch$ is infinite
dimensional then there are irreducible representations of the CCR
which are not Fock.

The Fock space realization that we describe below is motivated by
the following observations.  Let $\rh^0=\mbc\Omega$ and for each
integer $n\geq1$ let $\rh^n$ be the closed linear subspace of $\rh$
generated by the vectors of the form $a^*(u_1)\dots a^*(u_n)\Omega$
with $u_k\in\ch$. From \eqref{eq:ac} and since $\Omega$ is cyclic we
get $\rh=\oplus_{n=0}^\infty\rh^n$ (Hilbert direct sum) and
$\|a^*(u)^n\Omega\|=\sqrt{n!}\|u\|^{n}$. Let us denote $\gs(n)$ the
set of permutations of $\{1,\dots,n\}$. Then, since the operators
$a^*(u)$ are pairwise commuting, we have:
\begin{equation}\label{eq:scp}
\lag a^*(u_1)\dots a^*(u_n)\Omega | a^*(v_1)\dots a^*(v_n)\Omega
\rag 
=\textstyle\sum_{\sigma\in\gs(n)}\lag u_1 | v_{\sigma(1)} \rag
\dots \lag u_n | v_{\sigma(n)} \rag
\end{equation}

\noindent{\bf 3.} Let $\ch^{\vee}_{\text{alg}}$ be the symmetric
\label{p:sa} algebra\symbolfootnote[2]{\ This is a complex abelian
unital algebra in which $\ch$ is linearly embedded and which is
uniquely determined (modulo canonical isomorphisms) by the following
universal property: if $\xi:\ch\rarrow\ca$ is a linear map with
values in a unital algebra $\ca$ such that
$\xi(u)\xi(v)=\xi(v)\xi(u)$ for all $u,v$ then there is a unique
extension of $\xi$ to a morphism of unital algebras
$\ch^{\vee}_{\text{alg}}\rarrow\ca$ (see \cite{Bo} for example).
Concerning the notation $uv$ we use for the product we note that in
concrete situations, when some other product $uv$ is already
defined, this notation could be ambiguous. Then we replace it by
$u\vee v$ and denote by $u^{\vee n}$ the powers of $u$. } over the
vector space $\ch$. We denote by $uv$ the product of two elements
$u,v$ of $\ch^{\vee}_{\text{alg}}$ and by $u^n$ the $n$-th power of
an element $u\in\ch^{\vee}_{\text{alg}}$. The unit element is
denoted either $1$ or $\Omega$.  Let $\ch^{\vee n}_{\text{alg}}$ be
the linear subspace spanned by the powers $u^n$ with $u\in\ch$. Note
that $\ch^{\vee 0}_{\text{alg}}=\mbc \Omega$.  Then
$\ch^{\vee}_{\text{alg}}=\sum_{n\in\mbn}\ch^{\vee n}_{\text{alg}}$
(direct sum of linear spaces) and for $f\in\ch^{\vee
n}_{\text{alg}}$ and $g\in\ch^{\vee m}_{\text{alg}}$ we have
$fg\in\ch^{\vee (n+m)}_{\text{alg}}$.  We set $\ch^{\vee
n}_{\text{alg}}=\{0\}$ for $n<0$, so $\ch^{\vee}_{\text{alg}}$
becomes a $\mbz$-graded algebra.

We shall equip $\ch^{\vee}_{\text{alg}}$ with the unique
scalar product such that 
$\ch^{\vee n}_{\text{alg}}\perp\ch^{\vee m}_{\text{alg}}$ 
if $n\neq m$ and 
\begin{equation}\label{eq:sc}
\lag u_1\dots u_n| v_1\dots v_n \rag
=\textstyle\sum_{\sigma\in\gs(n)}\lag u_1 | v_{\sigma(1)} \rag
\dots \lag u_n | v_{\sigma(n)} \rag
\end{equation}
From the polarization formula \eqref{eq:pz} we see that this scalar
product is uniquely determined by the condition $\lag u^n | v^m
\rag=n!\lag u | v \rag^n\delta_{nm}$ for all $u,v\in\ch$ and
$n,m\geq0$ (see also the characterization given on page
\pageref{p:csc}). Then it is easy to prove that:
\begin{equation}\label{eq:prod}
\|uv\|\leq\binom{n+m}{n}^{1/2}\|u\|\,\|v\| \hspace{2mm}
\text{ if } u\in\ch^{\vee n}_{\text{alg}} \text{ and } 
v\in\ch^{\vee m}_{\text{alg}}.
\end{equation}

We define the \emph{Fock space $\Gamma(\ch)\equiv\ch^\vee$ over
$\ch$} as the completion of $\ch^{\vee}_{\text{alg}}$ for the scalar
product defined by \eqref{eq:sc}.  Let $\ch^{\vee n}$ be the closure
of $\ch^{\vee n}_{\text{alg}}$ in $\Gamma(\ch)$.  Then we can write
$\Gamma(\ch)=\bigoplus_{n}\ch^{\vee n}$, a Hilbert space direct
sum. We shall also use the notations
$\Gamma_n(\ch)=\sum_{k=0}^n\ch^{\vee n}$ and
$\Gamma_{\text{fin}}(\ch)=\bigcup_n\Gamma_n(\ch)$.  Note that
$\ch^{\vee 0}\equiv\mbc\Omega$. The vector $\Omega$ is the
\emph{vacuum state} and the orthogonal projection on it is
$\omega=|\Omega\rag\lag\Omega|$.

Using \eqref{eq:prod} we can extend by continuity the multiplication
and get a structure of unital abelian algebra on
$\Gamma_{\text{fin}}(\ch)$ such that $\ch^{\vee n}\ch^{\vee
m}\subset\ch^{\vee(n+m)}$. Then \eqref{eq:prod} remains valid for
all $u\in\ch^{\vee n}$ and $v\in\ch^{\vee m}$.  We keep the notation
$uv$ for the product of two elements $u$ and $v$ of $\gfin(\ch)$.

We denote by $\ind^n$ and $\ind_n$ the orthogonal projections of
$\Gamma(\ch)$ onto the subspaces $\ch^{\vee n}$ and $\Gamma_n(\ch)$
respectively.  Thus $\ind_n=\ind^0+\dots+\ind^n$ and
$\ind^0=\omega$.  The \emph{number operator} is defined by
$N=\sum_nn\ind^n$.

For each $u\in\ch$ the \emph{creation operator} $a^*(u)$ is the
closure of the operator of multiplication by $u$ on $\Gamma(\ch)$
and the \emph{annihilation operator} $a(u)$ is its the adjoint
of. Then $\Gamma_{\text{fin}}(\ch)$ is included in the domains of
$a^*(u)$ and $a(u)$, is left invariant by both operators, and the
operator $a(u)$ is a derivation of the algebra
$\Gamma_{\text{fin}}(\ch)$.  The \emph{field operator}
$\phi(u)=a(u)+a^*(u)$ is essentially self-adjoint on
$\Gamma_{\text{fin}}(\ch)$ and the following elementary estimate
\begin{equation}\label{eq:fest}
\|\phi(u)^pv\|\leq\|2u\|^p \|\sqrt{(N+1)\dots(N+p)}v\|
\end{equation}
valid for all $u\in\ch$, $v\in\Gamma(\ch)$, and $p\geq1$ integer.
Then $W(u)=\rme^{\rmi\phi(u)}$ defines a Weyl system over $\ch$.

\noindent{\bf 4.}  If $A_i\in B(\ch)$ for $i=1,\dots,n$ are given
then there is a unique operator $A_1\vee\dots\vee A_n\in B(\ch^{\vee
n})$ such that $(A_1\vee\dots\vee A_n)u^n=(A_1u)\dots(A_nu)$ for all
$u\in \ch$. We extend it to $\Gamma(\ch)$ by identifying
$A_1\vee\dots\vee A_n\equiv A_1\vee\dots\vee A_n\ind^n$.  
By convention $A^{\vee 0}=\omega$.

If $A\in B(\ch)$ then there is a unique unital endomorphism
$\Gamma(A)$ of the algebra $\Gamma_{\text{fin}}(\ch)$ such that
$\Gamma(A)u=Au$ for all $u\in\ch$ and such that the restriction of
$\Gamma(A)$ to each $\Gamma_n(\ch)$ be continuous.  One has
$\Gamma(A)u^n=(Au)^n$ if $u\in\ch$ and $\Gamma(A)=\oplus_{n\geq0}
A^{\vee n}$ in an obvious sense.  The operator $\Gamma(A)$ is
bounded on $\Gamma(\ch)$ if and only if $\|A\|\leq1$ (we keep the
notation $\Gamma(A)$ for its closure). Then $\|\Gamma(A)\|=1$,
$\Gamma(AB)=\Gamma(A)\Gamma(B)$, $\Gamma(1)=1$ and
$\Gamma(0)=\omega$. Note that $z^N=\Gamma(z)$ for $z\in\mbc$.

Moreover, there is a unique derivation $\rmd\Gamma(A)$ of the
algebra $\Gamma_{\text{fin}}(\ch)$ such that $\rmd\Gamma(A)u=Au$ for
all $u\in\ch$.  Thus we have $\rmd\Gamma(A)u^n=n(Au)u^{n-1}$ if
$n\geq1$ and $\rmd\Gamma(A)\Omega=0$.  This operator is closable and
we denote its closure by the same symbol. If $A$ is self-adjoint
then $\Gamma(\rme^{\rmi A})=\rme^{\rmi \rmd\Gamma(A)}$.

The definition of $\rmd\Gamma(A)$ is extended as usual to operators
$A$ which are infinitesimal generators of contractive
$C_0$-semigroups $\{\rme^{tA}\}$ on $\ch$: the operator
$\rmd\Gamma(A)$ is defined by the rule 
$\Gamma(\rme^{tA})=\rme^{t\rmd\Gamma(A)}$.

The following identities hold on
$\Gamma_{\text{fin}}(\ch)$ for all $A\in B(\ch)$ and $u\in\ch$:
\begin{equation}\label{eq:ga}
\Gamma(A)a^*(u)=a^*(Au)\Gamma(A), \hspace{2mm}
\Gamma(A)a(A^*u)=a(u)\Gamma(A).
\end{equation}
If $A^*A=1$ we also get $\Gamma(A)a(u)=a(Au)\Gamma(A)$ by replacing
$u$ by $Au$ in the second identity, hence
\begin{equation}\label{eq:gf}
\Gamma(A)\phi(u)=\phi(Au)\Gamma(A) \text{ and }
\Gamma(A)W(u)=W(Au)\Gamma(A) \hspace{2mm} \text{ if } A^*A=1.
\end{equation}
More generally, if $A^*:\ch\rarrow\ch$ is a surjective map then
there is an operator $A^\dagger\in B(\ch)$ such that
$A^*A^\dagger=1$ and then, if we denote
$\phi_A(u)=a(A^\dag)+a^*(Au)$ we get:
\begin{equation}\label{eq:ga*}
\Gamma(A)a(u)=a(A^\dagger u)\Gamma(A) 
\hspace{2mm} \text{ and } \hspace{2mm} 
\Gamma(A)\phi(u)=\phi_A(u)\Gamma(A).
\end{equation}
Observe that if $A\in B(\ch)$ is invertible then
$A^\dagger=(A^*)^{-1}$.

\noindent{\bf 5.}  Let $\ck\subset\ch$ be a linear subspace. Then we
have a canonical embedding
$\ck^{\vee}_{\text{alg}}\subset\ch^{\vee}_{\text{alg}}$ obtained by
identifying $\ck^{\vee}_{\text{alg}}$ with the unital subalgebra of
$\ch^{\vee}_{\text{alg}}$ generated by $\ck$. If $\cl\subset\ch$ is
another linear subspace then 
$\ck^\vee_{\text{alg}}$ and $\cl^\vee_{\text{alg}}$ are subalgebras
of the abelian algebra $\ch^\vee_{\text{alg}}$ so we have a natural
unital morphism
$\ck^\vee_{\text{alg}}\otimes\cl^\vee_{\text{alg}}\rarrow
\ch^\vee_{\text{alg}}$ (algebraic tensor product) which is injective
if and only if $\ck\cap\cl=0$ and surjective if and only if
$\ck+\cl=\ch$. Thus $(\ck\oplus\cl)^\vee_{\text{alg}}=
\ck^\vee_{\text{alg}}\otimes\cl^\vee_{\text{alg}}$.

Let $\ck\subset\ch$ be a closed subspace. Then the embedding
$\ck^\vee_{\text{alg}}\subset\ch^\vee_{\text{alg}}$ obviously
extends to an isometric embedding
$\Gamma(\ck)\subset\Gamma(\ch)$. Moreover, the canonical algebraic
identification $\ch^\vee_{\text{alg}}=
\ck^\vee_{\text{alg}}\otimes\ck^{\perp\vee}_{\text{alg}}$ extends to
a Hilbert space identification
$\Gamma(\ch)=\Gamma(\ck)\otimes\Gamma(\ck^\perp)$. Indeed, the
scalar product \eqref{eq:sc} has been chosen such that the
identification map be isometric (the norm of a tensor product of
Hilbert spaces being defined in the standard way).  In fact
\label{p:csc} \eqref{eq:sc} is the unique scalar product on
$\ch^\vee_{\text{alg}}$ such that $\|\Omega\|=1$, a vector $u\in\ch$
has the same norm in $\ch$ and in $\ch^\vee_{\text{alg}}$, and for
each closed subspace $\ck\subset\ch$:
$$
\lag uv \mid u'v'\rag=\lag u \mid u'\rag\lag v \mid v'\rag=
\lag u\otimes v \mid u'\otimes v'\rag  \hspace{2mm}
\text{ for all } 
u\in\ck^\vee_{\text{alg}},v\in(\ck^\perp)^\vee_{\text{alg}}. 
$$
In order to avoid ambiguities we indicate, when necessary, by a
subindex the Hilbert space on which the various objects depend, for
example $W_\ch,N_\ch$ and so on. We also use abbreviations like
$N_\ck'=N_{\ck^\perp},\Omega'_\ck=\Omega_{\ck^\perp}$, etc.  Then,
relatively to the factorization
$\Gamma(\ch)=\Gamma(\ck)\otimes\Gamma(\ck^\perp)$, we have for
$u\in\ck$: 
\begin{equation}\label{eq:kh}
W_\ch(u)=W_\ck(u)\otimes1, \hspace{2mm}
\phi_\ch(u)=\phi_\ck(u)\otimes1, \hspace{2mm}
a^{(*)}_\ch(u)=a^{(*)}_\ck(u)\otimes1.
\end{equation}
Note also the relations $\Omega_\ch=\Omega_\ck\otimes\Omega'_{\ck}$
and $\omega_\ch=\omega_\ck\otimes\omega'_{\ck}$.
If $A=B\oplus C$ in $\ch=\ck\oplus\ck^\perp$ then:
\begin{equation}\label{eq:abc}
\Gamma(A)=\Gamma(B)\otimes\Gamma(C), \hspace{2mm}
\rmd\Gamma(A)=\rmd\Gamma(B)\otimes1+1\otimes\rmd\Gamma(C).
\end{equation}
In particular $z^{N_\ch}=z^{N_\ck}\otimes z^{N_{\ck}'}$ for
$|z|\leq1$ and $N_\ch=N_\ck\otimes1+1\otimes N_{\ck}'$.

After the identification
$\Gamma(\ch)=\Gamma(\ck)\otimes\Gamma(\ck^\perp)$ the embedding
$\Gamma(\ck)\subset\Gamma(\ch)$ is nothing else but
$\Gamma(\ck)\equiv\Gamma(\ch)\otimes\Omega'_\ck$.  Then
extending an operator $T$ defined on the subspace $\Gamma(\ck)$ by
zero on the orthogonal subspace of $\Gamma(\ch)$ amounts to
identifying $T\equiv T\otimes\omega'_\ck$. This is coherent with the
first relation in \eqref{eq:abc}:
$\Gamma(B\oplus0)=\Gamma(B)\otimes\Gamma(0)=
\Gamma(B)\otimes\omega'_\ck$.

Let $\rk(\ch)=K(\Gamma(\ch))$ be the $C^*$-algebra of compact
operators on $\Gamma(\ch)$. Clearly:
\begin{equation}\label{eq:facc}
\rk(\ch)=\rk(\ck)\otimes\rk(\ck^\perp) 
\end{equation}
As explained above, we have a natural identification of $\rk(\ck)$
with a $C^*$-subalgebra $\rk_\ck(\ch)$ of $\rk(\ch)$, a compact
operator on $\Gamma(\ch)$ being identified with its extension by
zero on $\Gamma(\ck)^\perp$:
\begin{equation}\label{eq:emb}
\rk_\ck(\ch)\equiv\rk(\ck)\otimes\omega'_\ck
\subset\rk(\ch). 
\end{equation}

\begin{lemma}\label{lm:cind}
$\{\rk_E(\ch)\}$, where $E$ runs over the set of finite
dimensional subspaces of $\ch$, is an increasing family of
$C^*$-algebras and the closure of its union is $\rk(\ch)$.
\end{lemma}
\proof It suffices to note that the spaces $\Gamma(E)$, with
$E\subset\ch$ finite dimensional, form an increasing family of
closed subspaces of $\Gamma(\ch)$ whose union is dense in
$\Gamma(\ch)$.
\qed

\section{The algebras $\rf(\co)$} \label{s:wo}
\protect\setcounter{equation}{0}

We fix a complex Hilbert space $\ch$ and to each $C^*$-algebra $\co$
of operators on it we associate a $C^*$-algebra of operators on the
bosonic Fock space $\Gamma(\ch)$ according to the following rule:
\begin{equation}\label{eq:kef}
\Gamma(\co) = C^*(\Gamma(A)\mid A\in\co, \|A\|<1).
\end{equation}
Since $\Gamma(A)\Gamma(B)=\Gamma(AB)$ and $\Gamma(A)^*=\Gamma(A^*)$
this is in fact the norm closed linear space generated by the
operators $\Gamma(A)$ with  $A\in\co$ and $\|A\|<1$. We shall prove
in a moment that
\begin{equation}\label{eq:late}
\Gamma(\co)=\text{ closure of the linear space generated by the } 
\Gamma(A) \text{ with } A\in\co \text{ and } 0\leq A\leq\|A\|<1.
\end{equation}

\begin{proposition}\label{pr:gopr}
The map $\co\mapsto\Gamma(\co)$ is increasing and we have:
\begin{equation}\label{eq:gno}
\Gamma(\{0\})=\mbc\omega, \hspace{3mm} 
\Gamma(\mbc\ind_\ch)=\Co(N)=\{\theta(N)\mid\theta\in\Co(\mbn)\}.
\end{equation}
If $\ch=\ch_1\oplus\ch_2$ and $\co=\co_1\oplus\co_2$ for some
$C^*$-subalgebras $\co_i\subset B(\ch_i)$, then
\begin{equation}\label{eq:gds}
\Gamma(\co)=\Gamma(\co_1)\otimes\Gamma(\co_2)
\end{equation}
where the tensor product is defined by the identification
$\Gamma(\ch)=\Gamma(\ch_1)\otimes\Gamma(\ch_2)$.
\end{proposition}
\proof The first assertion is obvious and the first relation in
\eqref{eq:gno} follows from $\Gamma(0)=\omega$.  Since the closed
subspace generated by the functions $\lambda\mapsto\lambda^n$ with
$0<\lambda<1$ is dense in $\Co(\mbn)$ we see that the second
relation in \eqref{eq:gno} is true. To prove \eqref{eq:gds} we use
\eqref{eq:abc} and the fact that for $A=A_1\oplus A_2$ we have
$\|A\|=\sup(\|A_1\|,\|A_2\|)$ so that $\|A\|<1$ if and only if
$\|A_1\|<1$ and $\|A_2\|<1$.  \qed

We shall give a more explicit description of $\Gamma(\co)$ for an
arbitrary $\co$ below.  Observe first that the linear subspace of
$B(\ch^{\vee n})$ generated by the operators of the form
$A_1\vee\dots\vee A_n$ with $A_i\in\co$ is a $*$-algebra. Indeed,
this follows from 
$(A_1\vee\dots\vee A_n)^*=A_1^*\vee\dots\vee A_n^*$ and
\begin{equation}\label{eq:opr}
n!(A_1\vee\dots\vee A_n)(B_1\vee\dots\vee B_n)=\textstyle
\sum_{\sigma\in\gs(n)}
(A_1B_{\sigma(1)})\vee\dots\vee(A_nB_{\sigma(n)}).
\end{equation}
which is obvious if $A_1=\dots=A_n$ and $B_1=\dots=B_n$ and the
general case follows by applying twice the polarization formula
\eqref{eq:pz}. Thus the norm closed linear space generated by the
operators $A_1\vee\dots\vee A_n$ with $A_i\in\co$ is a $C^*$-algebra
that we shall denote $\co^{\vee n}$. We make the convention
$\co^{\vee 0}=\mbc\ind^0=\mbc\omega$.

\begin{proposition}\label{pr:kef}
$\co^{\vee n}$ is the norm closed linear space of operators on
$\ch^{\vee n}$ generated by the operators $A^{\vee n}$ with
$A\in\co$ and $A\geq0$. Moreover, we have \eqref{eq:late} and:
\begin{equation}\label{eq:kef*}
\Gamma(\co)=\textstyle\bigoplus_n\co^{\vee n}\equiv
\{\textstyle\sum_n A_n\ind^n\mid
A_n\in\co^{\vee n},\ \|A_n\|\rarrow0\}.
\end{equation}
\end{proposition}
\proof Let $\cl$ be the linear space of operators on $\ch^{\vee n}$
generated by the operators $A^{\vee n}$ with $A\in\co$ and $A\geq0$.
From the polarization formula \eqref{eq:pz} we first deduce that the
operators $A_1\vee\dots\vee A_n$ with $A_i\in\co$ and $A_i\geq0$
belong to $\cl$ and then, by $n$-linearity, that the same assertion
holds without the condition $A_i\geq0$.  This proves the first
assertion of the proposition.

Let $\rl$ be the norm closed linear space generated by the operators
$\Gamma(A)$ such that $A\in\co$ and $0\leq A\leq a$ for some
$a<1$. Let $A\geq0$ with $\|A\|<1$. For $0\leq t\leq1$ we then have
$\Gamma(tA)=\sum t^nA^{\vee n}$, so the map
$t\mapsto\Gamma(tA)\in\rl$ is of class $C^\infty$ and its derivative
of order $n$ at $t=0$ is equal to $n!A^{\vee n}$. Clearly then we
get $A^{\vee n}\in\rl$ for all $A\in\co$, $A\geq0$. From what we
proved before we get $\co^{\vee n}\subset\rl$. Then if
$A\in\co,\|A\|<1$ we have $\Gamma(A)\ind_n\in\rl$ and
$\|\Gamma(A)-\Gamma(A)\ind_n\|\leq\|A\|^{n+1}\rarrow0$, so
$\Gamma(A)\in\rl$. This clearly proves $\rl=\Gamma(\co)$, i.e.\
\eqref{eq:late}. The inclusion $\subset$ in \eqref{eq:kef*} is
obvious and the inverse inclusion follows from the preceding
arguments.  
\qed

We are mainly interested in $C^*$-algebras of operators on
$\Gamma(\ch)$ of the following form:
\begin{equation}\label{eq:wo}
\rf(\co) = C^*(W(u)\Gamma(A)\mid u\in\ch,A\in\co,\|A\|<1).
\end{equation}
Observe that $\Gamma(\co)\subset\rf(\co)$.

\begin{proposition}\label{pr:elpr}
{\rm(1)} If $\co_1\subset\co_2$ are $C^*$-subalgebras of $B(\ch)$
then $\rf(\co_1)\subset\rf(\co_2)$.\\ 
{\rm(2)} We have $\rf(\{0\})=\rk(\ch)$, in particular
$\rk(\ch)\subset\rf(\co)$ for all $\co$.\\
{\rm(3)} If $\ch=\ch_1\oplus\ch_2$ and $\co=\co_1\oplus\co_2$ for
some $C^*$-subalgebras $\co_i\subset B(\ch_i)$, then
\begin{equation}\label{eq:ds}
\rf(\co)=\rf(\co_1)\otimes\rf(\co_2)
\end{equation}
where the tensor product is defined by the identification
$\Gamma(\ch)=\Gamma(\ch_1)\otimes\Gamma(\ch_2)$.
\end{proposition}
\proof The first assertion is obvious and an easy proof of (2)
involves coherent vectors \cite{Gu}. Indeed:
$$
W(u)\Omega=\rme^{-\|u\|^2/2}\rme^{\rmi u}\equiv
\rme^{-\|u\|^2/2}\textstyle\sum_{n}\frac{\rmi^n}{n!} u^n
$$ and the linear span of these vectors is dense in $\Gamma(\ch)$.
Thus the norm closed linear subspace of $B(\Gamma(\ch))$ generated
by the operators $W(u)\omega=|W(u)\Omega\rag\lag\Omega|$ is equal to
the space of rank one operators of the form $|u\rag\lag\Omega|$ with
$u\in\Gamma(\ch)$. But the $C^*$-algebra generated by these
operators is exactly $\rk(\ch)$. Finally, to prove (3) we argue as
in the proof of Proposition \ref{pr:gopr} by using \eqref{eq:kh} and
\eqref{eq:abc} in order to get
$W(u)\Gamma(A)=[W(u_1)\Gamma(A_1)]\otimes [W(u_2)\Gamma(A_2)]$ if
$u=u_1\oplus u_2$ and $A=A_1\oplus A_2$.  \qed

If $\co\subset B(\ch)$ is a $C^*$-subalgebra then let $\ch_\co$ be
the closed linear space generated by the vectors $Au$ with
$A\in\co,u\in\ch$.  One says that $\co$ is \emph{non-degenerate} (or
acts non-degenerately on $\ch$) if $\ch_\co=\ch$. Denote $\co_0$ the
algebra $\co$ when viewed as a $C^*$-algebra of operators on
$\ch_\co$. Thus $\co_0$ acts non-degenerately on $\ch_\co$ and we
have $\co|\ch_\co^\perp=\{0\}$, hence by (2) and (3) of Proposition
\ref{pr:elpr}:
\begin{equation}\label{eq:ndeg}
\rf(\co)=\rf(\co_0)\otimes\rk(\ch_\co^\perp) \hspace{2mm}
\text{ relatively to }
\Gamma(\ch)=\Gamma(\ch_\co)\otimes\Gamma(\ch_\co^\perp).
\end{equation}
In some of our results we shall assume that $\co$ is non-degenerate
but one may use \eqref{eq:ndeg} to extend them to possibly
degenerate algebras. We shall not do it explicitly in order to
simplify the arguments and also because this is of no interest in
the applications we have in mind.  In fact, we interpret $\co$ as
the $C^*$-algebra generated by the allowed one particle Hamiltonians
of the field, in particular there should be self-adjoint operators
$h$ on $\ch$ affiliated to $\co$. But this implies that $\co$ is
non-degenerate (see Section \ref{s:saa}).

\begin{proposition}\label{pr:wot}
If $\co$ is non-degenerate then $\rf(\co)$ is the norm closed linear
subspace generated by the operators of the form $W(u)\Gamma(A)$ with
$u\in\ch$ and $A\in\co$ such that $A\geq0$ and $\|A\|<1$.
\end{proposition}
\proof Let $\rM$ be the norm closed linear subspace generated by the
operators of the form $W(u)\Gamma(A)$ with $A$ as in the statement
of the lemma. Clearly $\rM\subset\rf(\co)$ and \eqref{eq:late}
implies that $\rM$ contains a set which generates $\rf(\co)$ as a
$C^*$-algebra, so it suffices to show that $\rM$ is a $*$-algebra.
Proposition \ref{pr:kef} shows that $W(u)\Gamma(A)\ind^n\equiv
W(u)A^{\vee n}\in\rM$ if $u\in\ch$ and $A\in\co$. By computing
derivatives with respect to $t_1,\dots,t_p$ of
$W(t_1u_1+\dots+t_pu_p)$ and by using the estimate
\begin{equation}\label{eq:fest1}
\|\phi(u)^p\ind_n\|\leq \sqrt{p!}\|2\sqrt{n+1}u\|^p
\end{equation}
which is a consequence of \eqref{eq:fest} we get
$\phi(u_1)\dots\phi(u_p)\Gamma(A)\ind^n\in\rM$ for all
$u_1,\dots,u_p\in\ch$. And this is equivalent to
$a^*(u)^pa(v)^q\ind^n\Gamma(A)\in\rM$ for all $u,v,p,q,n$.

Now let $A,B\in\co$ be positive and $\varepsilon>0$ real. Then
\eqref{eq:ga} and $\ind^na^*(u)^pa(v)^q=a^*(u)^pa(v)^q\ind^{n-p+q}$
imply:
$$
\ind^{n}\Gamma(A+\varepsilon B)a^*(u)^pa((A+\varepsilon B)v)^q=
a^*((A+\varepsilon B)u)^pa(v)^q\ind^{n-p+q}\Gamma(A)\in\rM.
$$

Thus $\ind^{n}\Gamma(A+\varepsilon B)a^*(u)^pa(w)^q\in\rM$ for each
$w$ in the closure of the range of an operator of the form
$A+\varepsilon B$ (because the preceding expression is norm
continuous as function of $w$). Now let $J_\nu$ be an approximate
unit for $\co$ \cite[pages 77-78]{Mu}, let $\cR_\nu$ be the closure
of the range of $A+\varepsilon J_\nu$, and
$\cn_\nu=\ker(A+\varepsilon J_\nu)$, so that $\cR=\cn_\nu^\perp$.
We have $v\in\cn_\nu$ if and only if $\lag v|Av\rag=\lag v|J_\nu
v\rag=0$ hence $\cn_\mu\subset\cn_\nu$ if $\mu\geq\nu$. Moreover,
$\cn_\nu$, and hence $\cR_\nu$, is independent of $\varepsilon$. And
we have $\ind^{n}\Gamma(A+\varepsilon J_\nu)a^*(u)^pa(w)^q\in\rM$
for each $w\in\cR_\nu$ by what we proved before. If we make here
$\varepsilon\rarrow0$ then we get norm convergence and so
$\ind^{n}\Gamma(A)a^*(u)^pa(w)^q\in\rM$ for $w\in\cR_\nu$.  On the
other hand $\cap_\nu\cn_\nu=\{0\}$ because $\co$ is non-degenerate
and so $\lim_\nu J_\nu v=v$ for all $v\in\ch$. It follows that
$\{\cR_\nu\}$ is an increasing family of closed subspaces of $\ch$
whose union is dense in $\ch$. Thus we have
$\ind^{n}\Gamma(A)a^*(u)^pa(w)^q\in\rM$ for $w$ in the union and
then by norm continuity for all $w\in\ch$.  Clearly then we get
$\ind^{n}\Gamma(A)\phi(u)^p\in\rM$ for all $A\in\co$ with $A\geq0$
and $u\in\ch$.  From \eqref{eq:fest1} we see that
$\ind^nW(u)=\sum_p\ind^n(\rmi\phi(u))^p/p!$ the series being
convergent in norm. Hence $\ind^{n}\Gamma(A)W(u)\in\rM$ for all
$u\in\ch$ and positive $A\in\co$. By arguments already used in the
proof of Proposition \ref{pr:kef} we obtain $\ind^nA^{\vee
n}W(u)\in\rM$ for arbitrary $A\in\co$. This clearly implies
$\Gamma(A)W(u)\in\rM$ if $A\in\co$ and $\|A\|<1$.

To summarize, $\rM$ is equal to the norm closed linear subspace
generated by the operators $W(u)\Gamma(A)$ with $A\in\co$,
$\|A\|<1$, and we have proved that $\Gamma(A)W(u)\in\rM$ under the
same conditions. Thus $\rM$ is stable under taking adjoints. For a
product $W(u)\Gamma(A)W(v)\Gamma(B)$ we write $\Gamma(A)W(v)$ as
limit of linear combinations of operators $W(w)\Gamma(C)$ with
$C\in\co$, $\|C\|<1$, and use \eqref{eq:ws} and
$\Gamma(C)\Gamma(B)=\Gamma(CB)$. This gives
$W(u)\Gamma(A)W(v)\Gamma(B)\in\rM$, hence $\rM$ is a $C^*$-algebra.
\qed

\begin{remark}\label{re:wo}{\rm
The arguments of the preceding proof show that if $\co$ is
non-degenerate then $\rf(\co)$ is the norm closed linear span of the
operators $\phi(u)^n\Gamma(A)$ with $u\in\ch,n\in\mbn$ and $A\in\co$
with $\|A\|<1$.  }\end{remark}

\begin{remark}\label{re:wop}{\rm
Proposition \ref{pr:wot} is not valid if $\co$ is
degenerate. Indeed, with the notations of \eqref{eq:ndeg} and if
$u=u_0+u_1$ with $u_0\in\ch_\co, u_1\in\ch_\co^\perp$, then for
$A\in\co$ with $\|A\|<1$ we have
$$
W(u)\Gamma(A)=[W(u_0)\Gamma(A_0)]\otimes[W(u_1)\Gamma(0)]=
[W(u_0)\Gamma(A_0)]\otimes|W(u_1)\Omega\rag\lag\Omega|
$$
and the operators $|W(u_1)\Omega\rag\lag\Omega|$ do not generate
linearly $\rk(\ch_\co^\perp)$.
}\end{remark}

\begin{lemma}\label{lm:eles} Assume $A,B\in B(\ch)$ and 
$\|A\|\leq c,\|B\|\leq c$  with $c<1$. If we set
$\widetilde{c}=\sup_{k\geq1} kc^{k-1}$ then
$$
\|\Gamma(A)-\Gamma(B)\|\leq\wtilde{c}\|A-B\|.
$$ 
For $u,v\in\ch$ and $n\in\mbn$ we have:
$$
\|(W(u)-W(v))\ind_n\|\leq |\Im\lag u|v\rag|+2\sqrt{n+1}\|u-v\|.
$$
If $\|A\|<1$ the map $u\mapsto W(u)\Gamma(A)$ is norm
continuous on $\ch$ and $\|\phi(u)^p\Gamma(A)\|<\infty$
for all $p$.
\end{lemma}
\proof To prove the first part it suffices to show that $\|A^{\vee
k}-B^{\vee k}\|\leq kc^{k-1}\|A-B\|$ if $k\geq1$.  But this follows
from $A^{\vee k}-B^{\vee k}= \sum_{j=0}^{k-1}B^{\vee j}\vee(A-B)\vee
A^{\vee(k-1-j)}$.  For the proof of the second estimate we note that
\eqref{eq:ws} implies $ \|(W(u)-W(v))\ind_n\|\leq |\rme^{\rmi\Im\lag
v|u\rag}-1|+ \|(W(u-v)\ind_n-\ind_n\| $ and then we use
$$
\|(W(u)\ind_n-\ind_n\|=\|\int_0^1W(tu)\rmi\phi(u)\ind_n dt\|\leq
\|\phi(u)\ind_n\|\leq2\sqrt{n+1}\|u\|.
$$
Next observe that
$W(u)\Gamma(A)=W(u)\ind_n\Gamma(A)+W(u)\Gamma(A)\ind_n^\perp$ and
$\|W(u)\Gamma(A)\ind_n^\perp\|\leq\|A\|^{n+1}$. Finally, the
estimate 
\begin{equation}\label{eq:fest2}
\|\phi(u)^p\lambda^N\| \leq
\|2u\|^p\|\sqrt{(N+1)\dots(N+p)}\lambda^N\|\leq
\sqrt{p!}\|2u\|^p\|(N+1)^{\frac{p}{2}}\lambda^N\|
\end{equation}
is a straightforward consequence of \eqref{eq:fest}, and this proves
the last assertion of the lemma.
\qed

We define now an analog in the present setting of the graded Weyl
algebra which has been introduced and studied for finite dimensional
symplectic spaces $\ch$ in \cite{BG2,GI3}. The following
construction makes sense for an arbitrary Weyl system $(\rh,W)$. A
finite dimensional real vector subspace $E$ of $\ch$ inherits an
Euclidean structure so it is equipped with a canonical translation
invariant measure $d_Eu$ and the corresponding $L^1(E)$ space
is well defined.  Since the map $u\mapsto W(u)$ is strongly
continuous on $E$, we can define 
$W(f)=\int_E W(u)f(u)d_Eu\in B(\rh)$ if $f\in L^1(E)$. Let: 
\begin{equation}\label{eq:we}
\rf(E,\ch)= \textrm{ norm closure of }  
\{W(f) \mid f\in L^1(E)\}.
\end{equation}
From \eqref{eq:ws} one may deduce that $\rf_E(\ch)$
is a $C^*$-algebra and that we have (the proof given in \cite{BG2}
for finite dimensional $\ch$ extends without any modification to our
context):
\begin{eqnarray*}
&\text{(i) }& 
\rf(E,\ch)\cdot\rf(F,\ch)\subset
\rf(E+F,\ch),\\
&\text{(ii)}& \text{ if } \cl \text{ is a finite family of finite
dimensional real subspaces of } \ch \text{ then } 
\textstyle\sum_{E\in\cl}\rf(E,\ch)\\
&\phantom{\text{(ii)}}& 
\text{ is a norm closed subspace and the sum is a direct of linear
spaces}. 
\end{eqnarray*}
We define the \emph{graded Weyl algebra}
$\rf(\ch)\equiv\rw_{\text{gr}}(\ch)$
\label{p:wa} as the norm closure of 
$\sum_{E}\rf(E,\ch)$, where $E$ runs over the set of all finite
dimensional \emph{complex} subspaces of $\ch$. Then
$\rf(\ch)$ is equipped with a graded $C^*$-algebra
structure in the sense of \cite[Definition 3.1]{DG}. $\rf(\ch)$ is
unital because $\rf(\{0\},\ch)=\mbc$.

In the Fock representation we have a quite explicit description of
the algebras $\rf(E,\ch)$. This follows, as explained
in \cite{BG2}, from the fact that a complex finite dimensional
subspace of $\ch$ is symplectic:
\begin{equation}\label{eq:wec}
\rf(E,\ch)=
\rk(E)\otimes1 \text{ relatively to the tensor
factorization }\Gamma(\ch)=\Gamma(E)\otimes\Gamma(E^\perp).
\end{equation}

Finally, we define $\rw_{\text{max}}(\ch)$, the largest
$C^*$-algebra of operators which can be naturally associated to the
Weyl system in the Fock representation. In particular,
$\rw_{\text{max}}(\ch)$ contains $\rw(\ch)$ and
$\rf(\ch)$. If $f$ is a bounded Borel regular measure on
$\ch$ (for the norm topology) and $v\in\gfin(\ch)$ then the integral
$W(f)v=\int_\ch W(u)v df(u)$ is well defined because, by Lemma
\ref{lm:eles}, the map $u\mapsto W(u)v$ is bounded and continuous on 
$\ch$. Clearly $\|W(f)v\|\leq\|v\| \|f\|$ where $\|f\|$ is the
variation of $f$, so $v\mapsto W(f)v$ extends to a bounded operator
$W(f)$  on $\Gamma(\ch)$. It is easy to show that the set of
operators $W(f)$ is a $*$-algebra and we define
$\rw_{\text{max}}(\ch)$ as its norm closure. Clearly

If $\rM,\rn$ are $C^*$-subalgebras of a given $C^*$-algebra we
denote by $\rM\cdot\rn$ the linear subspace consisting of the
operators of the form $S_1T_1+\dots+S_nT_n$ with
$S_i\in\rM,T_i\in\rn$ and $n\geq1$, and by $\llb\rM\cdot\rn\rrb$ the
norm closure of this linear subspace.

\begin{proposition}\label{pr:wo}
If $\co$ is non-degenerate then
\begin{equation}\label{eq:wol}
\mbox{\rm  
$\rf(\co)=\llb\rf(\ch)\cdot\Gamma(\co)\rrb=
\llb\rw(\ch)\cdot\Gamma(\co)\rrb=
\llb\rw_{\text{max}}(\ch)\cdot\Gamma(\co)\rrb.$}
\end{equation}
\end{proposition}
\proof We first observe that
$W(u)\Gamma(A)\in\llb\rf(\ch)\cdot\Gamma(\co)\rrb$ if
$u\in\ch$ and $\|A\|<1$. Indeed, since $W(tu)\Gamma(A)$ is a norm
continuous function of $t$ (see Lemma \ref{lm:eles}), the sequence
$\int_\mbr W(tu)f_k(t)dt\Gamma(A)$ converges in norm to
$W(u)\Gamma(A)$ if $f_k$ is a sequence in $L^1(\mbr)$ which
converges to the Dirac measure at $t=1$.  Thus
$\rf(\co)\subset\llb\rf(\ch)\cdot\Gamma(\co)\rrb$ by
Proposition \ref{pr:wot}. The converse inclusion follows from the
norm continuity of the map $u\mapsto W(u)\Gamma(A)$ (use again Lemma
\ref{lm:eles}). For the same reason we have
$W(f)\Gamma(A)\in\llb\rw(\ch)\cdot\Gamma(\co)\rrb$ for an arbitrary
bounded Borel regular measure on $\ch$.  \qed

Proposition \ref{pr:woo} will justify the physical interpretation of
the algebra $\rf(\co)$ as $C^*$-algebra of energy observables of the
field with one particle kinetic energy affiliated to $\co$. Recall
that QFH is an abbreviation for ``quantum field Hamiltonian''.

\begin{definition}\label{df:qfh}
We shall call \emph{elementary quantum field Hamiltonian of type
$\co$} a self-adjoint operator of the form $H=\dd\Gamma(h)+V$ where:
(i) $h$ is a self-adjoint operator on $\ch$ with $h\geq m$ for some
real $m>0$ and $h^{-1}\in\co$; (ii) $V$ a symmetric operator such
that $V=W(f)$ with $f\in L^1(E)$ for some finite dimensional linear
space $E\subset\ch$.
\end{definition}

For a self-adjoint operator $h$ such that $h\geq m>0$ the relations
$h^{-1}\in\co$ and $\rme^{-h}\in\co$ are equivalent and imply
$\theta(h)\in\co$ for all $\theta\in\Co(\mbr)$.  If an elementary
QFH of type $\co$ exists then $\co$ contains a positive injective
operator, e.g.\ $A=h^{-1}$, and this clearly implies that $\co$ is
non-degenerate.  Reciprocally:

\begin{proposition}\label{pr:woo}
If $\co$ contains a positive injective operator then $\rf(\co)$ is
the $C^*$-algebra generated by the elementary QFH of type $\co$. In
particular: $\rf(\co)=C^*(\rme^{-H}\mid H \text{ is an elementary
QFH })$.
\end{proposition}
\proof 
Let $H_s=\rmd\Gamma(h)+sV\equiv H_0+sV$ where $h,V$ are as in
Definition \ref{df:qfh} and $s$ is a real number. If $z$ is far
enough from the spectrum of $H_0$ then we have a norm convergent
expansion for $R_s=(z-H)^{-1}$:
\begin{equation}\label{eq:res}
R_s=R_0\left(1-VR_0\right)^{-1}=
\textstyle\sum_{n\geq0} s^nR_0\left(VR_0\right)^n.
\end{equation}
We have $\rme^{-tH_0}=\Gamma(\rme^{-th})\in\Gamma(\co)$ if $t>0$
because $\rme^{-th}\in\co$ and has norm $<1$, so
$R_0\in\Gamma(\co)$.  From Proposition \ref{pr:wo} we then get
$R_s\in\rf(\co)$, hence the $C^*$-algebra $\rc$ generated by the
elementary QFH is contained in $\rf(\co)$.

We now prove the converse inclusion. Let $h$ and $H_s$ be as above,
so that $R_s\in\rc$ for all $s$. By taking the first order
derivative at $s=0$ in \eqref{eq:res} we get $R_0VR_0\in\rc$. By
definition we have $\theta(H_0)\in\rc$ for any $\theta\in\Co(\mbr)$,
hence we also have $\theta(H_0)R_0VR_0\in\rc$. By choosing $\theta$
conveniently in $\Cc(\mbr)$ and then by an approximation argument we
get $\eta(H_0)VR_0\in\rc$ for all $\eta\in\Co(\mbr)$.

Let $\eta_n$ be a sequence of continuous functions with
$0\leq\eta_n\leq1$, $\eta_n(x)=1$ if $|x|\leq n$, and $\eta_n(x)=0$
if $|x|\geq n+1$. Our next purpose is to prove that
$\eta_n(H_0)VR_0\rarrow VR_0$ in norm. The operator $(N+1)R_0$ is
bounded, hence it is easy to see that it suffices to show that
$\|\ind_n^\perp V(N+1)^{-1}\|\rarrow0$ as $n\rarrow\infty$. We have
$V=W(f)=\int_EW(u)f(u)d\lambda_E(u)$ for some subspace $E$ of finite
dimension and $f\in L^1(E)$ and it is clear that for the proof of
this assertion it suffices to assume that $f$ has compact
support. We have
$$
\|\ind_n^\perp W(u)(N+1)^{-1}\|\leq
(n+1)^{-1}+\|\ind_n^\perp [W(u),(N+1)^{-1}]\|.
$$
On the other hand $[N,W(u)]=W(u)(\phi(\rmi u)+\|u\|^2)$
hence by using \eqref{eq:fest1} we get:
\begin{eqnarray*}
\|\ind_n^\perp [W(u),(N+1)^{-1}]\| &=&
\|\ind_n^\perp(N+1)^{-1}W(u)(\phi(\rmi u)+\|u\|^2)(N+1)^{-1}\|\\
&\leq&
(n+1)^{-1}\|(\phi(\rmi u)+\|u\|^2)(N+1)^{-1}\|
\leq
(n+1)^{-1}(2\|u\|+\|u\|^2).
\end{eqnarray*}
Thus we have 
$$\|\ind_n^\perp W(u)(N+1)^{-1}\|\leq(1+\|u\|)^2(n+1)^{-1}$$ from
which we get $\|\ind_n^\perp V(N+1)^{-1}\|\rarrow0$.  This finishes
the proof of $\lim\eta_n(H_0)VR_0\rarrow VR_0$ in norm
which in turn implies $VR_0\in\rc$.

Thus we have $VR_0\in\rc$ and then
$V\rme^{-H_0}=VR_0\cdot(z-H_0)\rme^{-H_0}\in\rc$. Since
$\rme^{-H_0}=\Gamma(\rme^{-h})$ we obtain $V\Gamma(A)\in\rc$ for any
operator $A$ of the form $A=\rme^{-h}$ with $h$ a self-adjoint
operator on $\ch$ such that $h\geq m>0$ and $\rme^{-h}\in\co$. In
other terms, we have $V\Gamma(A)\in\rc$ for any operator $A\in\co$
such that $A$ is positive and injective and such that
$\|A\|<1$. Indeed, it suffices then to choose $h=-\log A$. Now let
$A\in\co$ be positive and $\|A\|<1$. By assumption, $\co$ contains a
positive injective operator $S$. If $\varepsilon>0$ is small enough
then $A_\varepsilon=A+\varepsilon S$ belongs to $\co$, is positive
and injective, and $\|A_\varepsilon\|\leq c<1$ uniformly in
$\varepsilon$. Then $V\Gamma(A_\varepsilon)\in\rc$ and from Lemma
\ref{lm:eles} we get $V\Gamma(A)\in\rc$. Finally, \eqref{eq:late}
shows that $VT\in\rc$ for all $T\in\Gamma(\co)$.  From Proposition
\ref{pr:wo} we obtain $\rf(\co)\subset\rc$.  
\qed

\section{$\ra(\ch)$ and its canonical endomorphism}
\label{s:wa}
\protect\setcounter{equation}{0}

We set $\ra(\ch)=\rf(\mbc\ind_\ch)$.  From Proposition \ref{pr:wo}
we get:
\begin{equation}\label{eq:char}
\ra(\ch)=\llb\rf(\ch)\cdot\Co(N)\rrb=
\llb\rw(\ch)\cdot\Co(N)\rrb=
\llb\rw_{\text{max}}(\ch)\cdot\Co(N)\rrb.
\end{equation}
Alternative descriptions of $\ra(\ch)$ are consequences of the
results form Section \ref{s:wo}.  For example, $\ra(\ch)$ is the
norm closed subspace generated by each of the following classes of
operators: {\rm(i)} $\phi(u)^n\theta(N)$ with $u\in\ch,n\in\mbn$
and $\theta\in\Cc(\mbr)$; {\rm(ii)} $a^*(u)^pa(v)^q\ind^n$ with
$u,v\in\ch$ and $p,q,n\geq0$.

\begin{proposition}\label{pr:comp}
$\rk(\ch)\subset\ra(\ch)$ and $\rk(\ch)=\ra(\ch)$ if and only if
$\ch$ is finite dimensional.
\end{proposition}
\proof The first assertion is clear by Proposition
\ref{pr:elpr}. $\ch$ is finite dimensional if and only if
$\ind^1\in\rk(\ch)$ and then $\Co(N)\subset\rk(\ch)$. Since
$\ind^1\in\rf$, the second assertion of the proposition follows.
\qed

If $E$ is a finite dimensional
(complex) subspace of $\ch$ let us define
\begin{equation}\label{eq:shar}
\ra_E(\ch) =\llb\rw(E)\cdot\Co(N)\rrb=
\llb\rf(E,\ch)\cdot\Co(N)\rrb.
\end{equation}
The equality follows from the arguments of the proof of Proposition
\ref{pr:wo}.  Note that $\ra_{\{0\}}(\ch)=\Co(N)$.  With the
notation $N_E'=N_{E^\perp}$ introduced in Section \ref{s:fock},
we have:
\begin{proposition}\label{pr:en}
$\ra_E(\ch)=\rk(E)\otimes\Co(N_E')$ relatively to
$\Gamma(\ch)=\Gamma(E)\otimes\Gamma(E^\perp)$. In other terms:
\begin{equation}\label{eq:enf}
\ra_E(\ch)=\textstyle\bigoplus_n\rk(E)\otimes\ind^n_{E^\perp}=
\textstyle\{\sum_nK_n\otimes\ind^n_{E^\perp}\mid K_n\in\rk(E),
\|K_n\|\rarrow0\}
\end{equation}
where $\ind^n_{E^\perp}$ is the projection onto the $n$ particle
subspace of $\Gamma(E^\perp)$, in particular
$\ind^0_{E^\perp}=\omega'_E$.  If $\ch$ is infinite dimensional:
\begin{equation}\label{eq:eno}
\ra_E(\ch)\cap\rk(\ch)=\rk(E)\otimes\omega'_E\equiv\rk_E(\ch).
\end{equation}
\end{proposition}
\proof By an argument used before $\ra_E$ is the closed linear space
generated by the operators $T\lambda^N$ with
$T\in\rw(E|\ch)$ and $0<\lambda<1$. By \eqref{eq:wec} this
is the same as the closed linear space generated by
$(K\lambda^{N_E})\otimes\lambda^{N_E'}$ with $K$ compact on
$\Gamma(E)$. Replacing $K$ by $K\theta(N_E)\lambda^{-N_E}$ with
$\theta$ with compact support and then making $\theta\rarrow1$ we
see that $\ra_E$ is generated by the operators
$K\otimes\lambda^{N_E'}$, which proves the assertion of the
proposition.  \qed

We now prove that $\ra(\ch)$ is the inductive limit of the family of
$C^*$-algebras $\{\ra_E(\ch)\}$.

\begin{proposition}\label{pr:ef}
If $E\subset F$ are finite dimensional subspaces of $\ch$ then
$\ra_E(\ch)\subset\ra_F(\ch)$. And we have
\begin{equation}\label{eq:ind}
\ra(\ch)=\overline{\textstyle\bigcup_E\ra_E(\ch)}.
\end{equation}
\end{proposition}
\proof
We begin with a general remark. Let $\ck$ be a closed subspace of
$\ch$. If $z$ is a complex number such that $|z|<1$ then
$z^N=z^{N_\ck}\otimes z^{N'_\ck}\in\Co(N_\ck)\otimes\Co(N'_\ck)$.
This clearly implies:
\begin{equation}\label{eq:facn}
\Co(N)\subset\Co(N_\ck)\otimes\Co(N'_\ck)
\end{equation}
Now let us set $G=F\ominus E$. From $\ch=E\oplus G\oplus F^\perp$ we
get $\Gamma(\ch)=\Gamma(E)\otimes\Gamma(G)\otimes\Gamma(F^\perp)$,
hence:
\begin{eqnarray*}
\ra_E(\ch) &=& \rk(E)\otimes\Co(N_E')\subset
\rk(E)\otimes\Co(N_G)\otimes\Co(N_F')\\
&\subset& \rk(E)\otimes\rk(G)\otimes\Co(N_F')=
\rk(F)\otimes\Co(N_F')=\ra_F(\ch).
\end{eqnarray*}
We have used \eqref{eq:facn}, the fact that $\Co(N_G)\subset\rk(G)$
since $G$ is finite dimensional, and \eqref{eq:facc}.
\qed

If $\cp$ is an endomorphism of $\ra(\ch)$, then the following
conditions are equivalent:
\begin{itemize}
\item[{\rm(i)}]
$\cp\left(W(u)\lambda^N\right)=\lambda W(u)\lambda^N$ for each
$u\in\ch$ and $0<\lambda<1$;
\item[{\rm(ii)}]
$\cp\left(W(u)\theta(N)\right)=
W(u)\theta(N+1)$ for each $u\in\ch$ and $\theta\in\Co(\mbn)$.
\end{itemize}

Indeed,  since $\theta_\lambda(n)=\lambda^n$ defines a function in
$\Co(\mbn)$, we see that (ii)\,$\Rightarrow$\,(i). To prove the
converse, it suffices to note that the closed subspace generated by
the functions $\theta_\lambda,0<\lambda<1$ is dense in $\Co(\mbn)$.

If a morphism $\cp:\ra(\ch)\rarrow\ra(\ch)$ satisfying the
conditions (i) or (ii) above exists then \emph{it is unique and
surjective} by  \eqref{eq:char}. We shall call it \emph{the
canonical endomorphism of $\rf$}. If $\ch$ is finite dimensional
then $\ra(\ch)=\rk(\ch)$ has no nontrivial ideals, so the canonical
endomorphism cannot exist.

\begin{theorem}\label{th:mini}
If $\ch$ is infinite dimensional then the canonical endomorphism of
$\ra(\ch)$ exists and its kernel is $\rk(\ch)$. Hence we have a
canonical identification
\begin{equation}\label{eq:cani}
\ra(\ch)/\rk(\ch)\cong\ra(\ch).
\end{equation}
\end{theorem}
\proof Let $\tau$ be the endomorphism of $\Co(\mbn)$ defined by
$(\tau\theta)(m)=\theta(m+1)$. If $\ck\neq\{0\}$ then $\Co(N_\ck)$
is isomorphic with $\Co(\mbn)$ hence we get a realization of $\tau$
as endomorphism of $\Co(N_\ck)$.  For each finite dimensional
subspace $E$ let $\cp_E=1\otimes\tau$, which is an endomorphism of
$\ra_E=\rk(E)\otimes\Co(N'_E)$. We have
$\ker\cp_E=\rk(E)\otimes\ker\tau$ because tensor product with
$\rk(E)$ preserves exact sequences \cite[Theorem 6.5.2]{Mu}. Since
$\tau\theta(N'_E)=\theta(N'_E+1)$ we have $\ker\tau=\mbc\omega'_E$,
so $\ker\cp_E=\rk(E)\otimes\omega'_E=\ra_E\cap\rk(\ch)$ because of
\eqref{eq:eno}.

Let $F$ be a second finite dimensional subspace such that $E\subset
F$. Then we have $\ra_E\subset\ra_F$ and we shall prove that $\cp_E$
is the restriction of $\cp_F$ to $\ra_E$. From \eqref{eq:shar} and
arguments used before we see that $\ra_E$ is the norm closed linear
space generated by the operators $T=W(u)\lambda^N$ with $u\in E$ and
$0<\lambda<1$, hence it suffices to show that $\cp_E$ and $\cp_F$
are equal on such elements. We have
$T=\left(W(u)\lambda^{N_E}\right)\otimes\lambda^{N'_E}$ relatively
to the tensor factorization
$\Gamma(\ch)=\Gamma(E)\otimes\Gamma(E^\perp)$ hence
$$
\cp_E(T)=\left(W(u)\lambda^{N_E}\right)\otimes\lambda^{N'_E+1}=
W(u)\lambda^{N+1}.
$$ An identical computation gives $\cp_F(T)= W(u)\lambda^{N+1}$,
which proves our assertion.

Now from Proposition \ref{pr:ef} it follows that there is a unique
endomorphism $\cp$ of $\ra$ such that $\cp|\ra_E=\cp_E$. It is clear
that $\cp$ is the canonical endomorphism of $\rf$. From Lemma
\ref{lm:cind} it follows that $\cp(K)=0$ if $K$ is a compact
operator. Reciprocally, assume that $\cp(K)=0$ and let
$\varepsilon>0$. From \eqref{eq:ind} it follows that there is $E$
and $K_E\in\ra_E$ such that $\|K-K_E\|<\varepsilon$. Thus
$\|\cp_E(K_E)\|<\varepsilon$.  The kernel of $\cp_E$ is
$\rk_E=\ra_E\cap\rk(\ch)$ and $\cp_E$ induces an isometric map from
the quotient $\ra_E/\rk_E$ onto $\ra_E$. From the definition of the
quotient norm it follows that there is $L\in\rk_E$ such that
$\|K_E-L\|<2\varepsilon$. This implies $\|K-L\|<3\varepsilon$ and
since $L$ is a compact operator and $\varepsilon $ is arbitrary, we
see that $K$ is compact.  \qed

\begin{remark}\label{re:skan}{\rm
The following explicit expression of $\cp$ has been noticed by
George Skandalis:
\begin{equation}\label{eq:anc}
\cp(T)u=\slim_{e\rarrowup0}a(e)Ta^*(e)u
\hspace{3mm} \text{ for all } T\in\ra(\ch) \text{ and }
u\in\gfin(\ch).
\end{equation}
This is similar to relation (2.2) in \cite{BG2}.  The notation
$e\rarrowup0$ means that $\|e\|=1$ and that $e$ converges to zero in
the weak\symbolfootnote[2]{\ 
More precisely, the limit is taken along the filter consisting of
the intersections of the neighborhoods of zero in the weak topology
with the unit sphere of $\ch$. One may also replace it by the finer
filter consisting of the subsets of the unit sphere which are
orthogonal to finite dimensional subspaces of $\ch$, the proof is
then simpler. }  
topology. \eqref{eq:anc} follows easily from \eqref{eq:wa},
\eqref{eq:nf} and
$\slim_{e\rarrowup0}a(e)a^*(e)\ind_n=\ind_n$.
}
\end{remark}

We give an application of Theorem \ref{th:mini} in spectral
theory. Let $\ch$ be infinite dimensional.

\begin{lemma}\label{lm:minj}
If $T\in\ra(\ch)$ then $\lim_{k\rarrow\infty}\|\cp^k(T)\|=0$. 
Moreover: $1^n\in\ra(\ch)$ and $\cp^k(1^n)=1^{n-k}$.
\end{lemma}
\proof From the characterizations of $\ra(\ch)$ given in
\eqref{eq:char} we see that it suffices to consider $T$ of the form
$T=W(u)\theta(N)$ with $\theta\in\Cc(\mbn)$. Then
$\cp^k(T)=W(u)\theta(N+k)=0$ for $k$ large.  \qed

\begin{proposition}\label{pr:aes}
The spectrum of an element $\ra(\ch)$ is countable. If
$T\in\ra(\ch)$ then its essential spectrum is equal to the spectrum
of $\cp(T)$.
\end{proposition}
\proof Let $\se(T)$ be the essential spectrum of an operator $T$ and
$\sd(T)$ its discrete spectrum, so that $\sigma(T)$ is equal to the
\emph{disjoint} union $\sd(T)\sqcup\se(T)$ and $\sd(T)$ does not
have accumulation points outside $\se(T)$. If $T\in\ra(\ch)$ then
$\se(T)=\sigma(\cp(T))$ hence we get by induction:
$$
\sigma(T)=\sd(T)\sqcup\sigma(\cp(T))=
\left[\sqcup_{k=0}^n\sd(\cp^k(T))\right]\sqcup\sigma(\cp^{n+1}(T))
$$
which proves the assertion of the proposition.
\qed

\begin{remark}\label{re:cpro}{\rm
The following comments on the algebra $\ra(\ch)$ play no role in
this paper but are of some intrinsic interest.  The advantage in
using the graded Weyl algebra $\rf(\ch)$ instead of
other Weyl algebras which can be found in the literature is that $N$
implements a norm continuous action of the unit circle on
it. Indeed, \eqref{eq:no} gives for $z\in\Sigma=\{z\in\mbc|\
|z|=1\}$ and $u\in\ch$:
\begin{equation}\label{eq:z}
z^NW(u)\bar{z}^N=W(zu)
\end{equation}
If $E$ is a (complex) finite dimensional subspace of $\ch$ then $E$
is stable under multiplication by $z$ and for $f\in L^1(E)$ we have
$$
z^NW(f)\bar{z}^N=\int_E W(zu)f(u) d_Eu=
\int_E W(u)f(\bar{z}u) d_Eu\equiv W(f_z).
$$
Since $\|W(f_z)-W(f)\|\leq\|f_z-f\|_{L^1}\rarrow0$ as $z\rarrow1$
we see that $z\mapsto z^NW(f)\bar{z}^N$ is norm continuous.

Thus $\alpha_z(T)=z^NT\bar{z}^N$ induces a norm continuous action of
$\Sigma$ on $\rf(\ch)$ which is compatible with the
grading (i.e.\ each $\rf(E,\ch)$ is stable under the
action). In particular, the crossed product $C^*$-algebra
$\rf(\ch)\rtimes\Sigma$ is well defined. The algebra
$\ra(\ch)$ is a quotient of this crossed product: there is a unique
morphism $\rf(\ch)\rtimes\Sigma\rarrow\ra(\ch)$ such
that the image of $T\otimes\eta$ be $T\wtilde\eta(N)\equiv
T\int_\Sigma z^N\eta(z)dz$ for all $T\in\rf(\ch),\eta\in
L^1(\Sigma)$, see \cite[Theorem 2.9]{GI2}. This morphism is
surjective but not injective.

The similarly defined morphism
$\rf(E,\ch)\rtimes\Sigma\rarrow\ra_E(\ch)$  can be used in
order to give a more conceptual proof of the existence of the
morphism $\cp_E$ constructed at the beginning of the proof of
Theorem \ref{th:mini}. I am indebted to G.\ Skandalis for a
comment which clarified this point to me.

}\end{remark}

\section{Canonical morphism of  $\rf(\co)$}\label{s:cmo}
\protect\setcounter{equation}{0}

We now extend the results of Section \ref{s:wa} to a larger class of
$C^*$-algebras $\co$ of operators on $\ch$.

\begin{definition}\label{df:cmor}
If a morphism $\cp:\rf(\co)\rarrow\co\otimes\rf(\co)$ with the
property
\begin{equation}\label{eq:cmor}
\cp(W(u)\Gamma(A))=A\otimes[W(u)\Gamma(A)] \hspace{2mm} \text{ if }
u\in\ch \text{  and } A\in\co \text{ with } \|A\|<1
\end{equation}
exists, then it is uniquely determined and we call it the
\emph{canonical morphism} of $\rf(\co)$.
\end{definition}

\begin{example}\label{ex:cmor}{\rm
Assume that $\cp$ exists and recall that
$\Gamma(\co)\subset\rf(\co)$. Then
$\cp(\Gamma(A))=A\otimes\Gamma(A)$ if $A\in\co$ and $\|A\|<1$.
Replacing $A$ by $tA$ and taking derivatives at $t=0$ we obtain
$\cp(A^{\vee 0})\equiv\cp(\omega)=0$ and $\cp(A^{\vee n})=A\otimes
A^{\vee(n-1)}$ if $n\geq1$ (recall that $A^{\vee0}=\omega$).  From
the polarization formula we then get
\begin{equation}\label{eq:exa}
n\cp(A_1\vee\dots\vee A_n)=\textstyle\sum_k A_k\otimes
[A_1\vee\dots\vee A_{k-1}\vee A_{k+1}\vee\dots A_n].
\end{equation}
for all $A_1,\dots,A_n\in\co$.
}\end{example}

\begin{remark}\label{re:cmor}{\rm
If needed we denote $\cp_\co$ the morphism from Definition
\ref{df:cmor}.  Observe that if $\co_1\subset\co_2$ and if the
canonical morphism $\cp_{\co_2}$ exists, then $\cp_{\co_1}$ exists
too and we have $\cp_{\co_1}=\cp_{\co_2}|\rf(\co_1)$.  }\end{remark}

\begin{theorem}\label{th:cmo}
If $\co$ is an abelian $C^*$-algebra on $\ch$ and its strong
closure does not contain finite rank operators then the canonical
morphism $\cp$ exists and $\ker\cp=\rk(\ch)$. This gives
a canonical embedding
\begin{equation}\label{eq:cmo}
\rf(\co)/\rk(\ch)\hookrightarrow \co\otimes\rf(\co).
\end{equation} 
\end{theorem}

\begin{remark}\label{re:}{\rm
Observe that $\ch$ cannot be finite dimensional. In the rest of this
remark we assume $\co$ non-degenerate and denote $\co'$ and $\co''$
its commutant and bicommutant. Note that
\begin{equation}\label{eq:use}
\rk(\ch)\subset\rf(\co)\subset\rf(\co'').
\end{equation}
The strong closure of $\co$ is $\co''$, thus in Theorem \ref{th:cmo}
we have to assume that $\co''$ does not contain finite rank
operators. Clearly this is equivalent to $\co''\cap
K(\ch)=\{0\}$. Observe that if there is a sequence of unitary
operators $U_n\in\co'$ such that $\wlim_{n\rarrow\infty}U_n=0$ then
this assumption is satisfied. On the other hand, if $\ch$ is
separable then $\co''\cap K(\ch)=\{0\}$ if and only if there is a
self-adjoint operator $S\in\co'$ with purely absolutely continuous
spectrum; and if this is the case then $\rme^{\rmi tS}\in\co'$ and
$\wlim_{|t|\rarrow\infty}\rme^{\rmi tS}=0$.
}\end{remark}

\begin{lemma}\label{lm:ofin}
Let $\co$ be an abelian finite dimensional $C^*$-algebra on $\ch$
with $\ind_\ch\in\co$.  Let $P_1,\dots,P_n$ be the minimal
projections of $\co$ and $\ch_k=P_k\ch$. Then $\ch=\oplus_k\ch_k$
and we have
\begin{equation}\label{eq:ofin}
\rf(\co)=\otimes_k\ra(\ch_k) \hspace{2mm}\text{ relatively to }
\Gamma(\ch)=\otimes_k\Gamma(\ch_k).
\end{equation}
\end{lemma}
\proof Recall that we have $P_k\neq0$, $P_iP_j=0$ if $i\neq j$ and
$P_1+\dots+P_n=\ind_\ch$. Moreover, each element of $\co$ is a
linear combination of these projections. Thus we can write $\co$ as
a direct sum of $C^*$-algebras $\co\equiv\oplus_k\mbc P_k$ and then
we may use (2) of Proposition \ref{pr:elpr}. More explicitly, if
$A\in\co$ then $A=\sum_k z_kP_k$ and we have
$\|A\|=\sup_k|z_k|$. Assume $\|A\|<1$ and let $u\equiv\sum_ku_k$,
then we get from \eqref{eq:kh} and \eqref{eq:abc}
$$
W(u)\Gamma(A)=\otimes_k[W(u_k)\Gamma(z_kP_k)]
\equiv\otimes_k[W(u_k)\Gamma(z_k)]
$$
where we have identified $P_k=\ind_{\ch_k}$. Then \eqref{eq:ofin}
follows easily from this relation.
\qed

\begin{lemma}\label{lm:ofin*}
Theorem \ref{th:cmo} is true if $\co$ is finite dimensional and
$\ind_\ch\in\co$. 
\end{lemma}
\proof We keep the notations of Lemma \ref{lm:ofin} and observe that
each $\ch_k$ is infinite dimensional because $\co$ does not contain
finite dimensional projections. By Theorem \ref{th:mini} the
canonical endomorphism $\cp_k$ of $\ra(\ch_k)$ exists. We shall now
use Proposition \ref{pr:int}: define $\cp_k'$ as in that theorem and
note that $\rj_k=\rk(\ch_k)$ and
$\widetilde{\ra(\ch_k)}=\ra(\ch_k)$.  Proposition \ref{pr:en}
implies that each $\ra(\ch_k)$ is nuclear.  Taking into account
Lemma \ref{lm:ofin} and Proposition \ref{pr:int} we get a morphism
$$
\cp\equiv\textstyle\bigoplus_{k=1}^n\cp'_k:\rf(\co)\rarrow
\bigoplus_{k=1}^n
\ra(\ch_1)\otimes\dots\otimes\ra(\ch_k)
\otimes\dots\otimes\ra(\ch_n)\equiv \mbc^n\otimes\rf(\co)
\cong\co\otimes\rf(\co)
$$
whose kernel is $\rk(\ch_1)\otimes\dots\otimes\rk(\ch_n)=\rk(\ch)$.
Then, with the notations of the proof of Lemma \ref{lm:ofin}:
\begin{eqnarray*}
\cp(W(u)\Gamma(A)) &=& 
\textstyle\bigoplus_{k=1}^n\cp'_k
\left[\otimes_k[W(u_i)\Gamma(z_i)]\right]\\
&=& \textstyle\bigoplus_{k=1}^n
[W(u_1)\Gamma(z_1)]\otimes\dots\otimes[z_kW(u_k)\Gamma(z_k)]
\otimes\dots\otimes[W(u_n)\Gamma(z_n)]\\
&=&(z_1P_1+\dots+z_nP_n)\otimes(W(u)\Gamma(A))=
A\otimes(W(u)\Gamma(A)). 
\end{eqnarray*}
Thus $\cp$ is the canonical morphism of $\rf(\co)$.
\qed

\noindent{\bf Prof of Theorem \ref{th:cmo}:} If the theorem has been
proved for non-degenerate $\co$ then the general case is a
consequence of the factorization \eqref{eq:ndeg} and of Proposition
\ref{pr:int} with $n=2$, $\rc_1=\rf(\co_0)$,
$\rc_2=\rj_2=\rk(\ch_\co^\perp)$. Thus we may assume that $\co$ is
non-degenerate. Then, due to Remark \ref{re:cmor}, it suffices to
assume that $\co$ is a Von Neumann algebra, i.e.\ $\co=\co''$. Let
$\rl$ be the set of all finite dimensional $*$-subalgebras of $\co$
which contain $\ind_\ch$.  Then $\rl$ is a lattice for the order
relation given by inclusion.  Indeed, $\rl$ is stable under
(arbitrary) intersections and if $\cm,\cn\in\rl$ then their upper
bound $\cR$ is constructed as follows: if $\rp(\cm),\rp(\cn)$ are
the sets of minimal projections of $\cm,\cn$ then we define
$\rp(\cR)$ as the set consisting of the non-zero projections of the
form $PQ$ with $P\in\rp(\cm),Q\in\rp(\cn)$ and take $\cR$ equal to
the linear span of $\rp(\cR)$. The total algebra $\co$ is the norm
closure of the union of the algebras in $\rl$, because each
$A\in\co$ is normal, its spectral measure $E_A$ has values in $\co$,
and so is a norm limit of finite sums of the form
$B=\sum_kz_kE_A(\Delta_k)$ with $z_k\in\mbc$ and
$\Delta_k\subset\mbc$ Borel sets. Note also that the standard
construction of such sums will produce operators with
$\|B\|\leq\|A\|$.

From Proposition \ref{pr:elpr} we see that $\{\rf(\cm)\mid
\cm\in\rl\}$ is a filtered increasing family of $C^*$-subalgebras of
$\rf(\co)$.  The definition \eqref{eq:wo}, Lemma \ref{lm:eles}, and
the remark made above concerning the norm of $B$ imply that
$\rf(\co)$ is the norm closure of the union of these subalgebras.
In other terms, $\rf(\co)$ is the inductive limit of the net
$\{\rf(\cm)\}_{\cm\in\rl}$. Lemma \ref{lm:ofin*} gives us for each
$\cm\in\rl$ a canonical morphism $\cp_\cm$ and from the Remark
\ref{re:cmor} it follows that $\cp_\co(T)\equiv\cp_\cm(T)$ is
independent of $\cm$ if $T\in\cup_\cm\rf(\cm)$. It remains to extend
$\cp_\co$ to all $\rf(\co)$ by continuity and to check condition (i)
of Proposition \ref{pr:carp} by an obvious density and continuity
argument.  \qed

\begin{remark}\label{re:skano}{\rm
This is a natural extension of Remark \ref{re:skan}. Let $\chi$ be a
state on a $C^*$-algebra $\co\subset B(\ch)$ and let $\{e\}$ be a
net of unit vectors in $\ch$ such that $e\rarrowup0$ and such that
the state associated to $e$ on $\co$ converges weakly to $\chi$ 
(G.\ Skandalis has shown me that each state $\chi$ on a
$C^*$-algebra $\co$ with $\co\cap K(\ch)=\{0\}$ can be
expressed in this way). Then
$$
\slim_{e\rarrowup0}a(e)\left[W(u)\Gamma(T)\ind_n\right]a^*(e)=
\chi(T)W(u)\Gamma(T)\ind_{n-1} \hspace{2mm} \text{ for all }
u\in\ch \text{ and } T\in\co. 
$$
}\end{remark}

Denote $\text{I}_\co$ the identity morphism on $\co$ and for each
integer $k\geq1$ let us define
\begin{equation}\label{eq:iter}
\cp_{k}=\rm\text{I}_\co^{\otimes (k-1)}\otimes\cp:
\co^{\otimes (k-1)}\otimes\rf(\co)
\rarrow\co^{\otimes k}\otimes\rf(\co).
\end{equation}
This is a morphism with $\co^{\otimes (k-1)}\otimes\rk(\ch)$  as
kernel (tensor product with an abelian algebra preserves exact
sequences). Note that 
$\co^{\otimes (k-1)}\otimes\rk(\ch)\subset
B(\ch^{\otimes (k-1)}\otimes\Gamma(\ch))$ 
does not contain compact operators if $k\geq1$ and if we are in the
conditions of Theorem \ref{th:cmo}. The following extends Lemma
\ref{lm:minj}.

\begin{proposition}\label{pr:iter}
Under the conditions of Theorem \ref{th:cmo} the map
\begin{equation}\label{eq:itar}
\cp^k=\cp_{k}\circ\dots\circ\cp_1:\rf(\co)\rarrow
\co^{\otimes k}\otimes\rf(\co)
\end{equation}
is a morphism uniquely determined by the property:
$\cp^k(W(u)\Gamma(A))=A^{\otimes k}\otimes[W(u)\Gamma(A)]$ if
$u\in\ch$ and $A\in\co$, $\|A\|<1$. We have
$\lim_{k\rarrow\infty}\|\cp^k(T)\|=0$ for all $T\in\rf(\co)$.
\end{proposition}
\proof
It remains only to prove the last relation. Clearly it suffices to
consider only operators of the form $T=W(u)\Gamma(A)$. But then we
have $\|\cp^k(W(u)\Gamma(A))\|\leq\|A\|^k\|\Gamma(A)\|$.
\qed

We mention a description of the canonical morphism $\cp$ in the
spirit of Proposition \ref{pr:woo}. Below $\co$ is any $C^*$-algebra
on $\ch$.  At point (ii) we use the extension of the action of $\cp$
to unbounded operators affiliated to $\rf(\co)$ (see section
\ref{s:saa}): so (ii) is just (i) written at generator level (see
the proof of Proposition \ref{pr:eqfh}).

\begin{proposition}\label{pr:carp}
Assume that $\co$ contains a positive injective operator.  If
$\cp:\rf(\co)\rarrow\co\otimes\rf(\co)$ is a morphism then $\cp$ is
the canonical morphism if and only if it satisfies the following
equivalent conditions:
\begin{itemize}
\item[{\rm(i)}]
$\cp\left(\rme^{-H}\right)=\rme^{-h}\otimes\rme^{-H}$ if
$H=\dd\Gamma(h)+V$ is an elementary QFH;
\item[{\rm(ii)}]
$\cp(H)=h\otimes\ind_{\Gamma(\ch)}+\ind_\ch\otimes H$
if $H=\dd\Gamma(h)+V$ is an elementary QFH.
\end{itemize}
\end{proposition}
\proof Note that $\cp$ is uniquely determined by the condition (i)
because of Proposition \ref{pr:woo}. If $H=\rmd\Gamma(h)+V\equiv
H_0+V$ then \eqref{eq:ro*} holds in norm because $H_0$ is bounded
from below and $V$ is bounded. If $\cp$ is the canonical morphism,
and since $\rme^{-tH_0}=\Gamma(\rme^{-th})$, we obtain (i) from:
\begin{equation}\label{eq:calc}
\cp\left[\left(\rme^{-V/n}\rme^{-H_0/n}\right)^n\right]=
\left[\cp\left(\rme^{-V/n}\rme^{-H_0/n}\right)\right]^n=
\left[\rme^{-h/n}\otimes
\left(\rme^{-V/n}\rme^{-H_0/n}\right)\right]^n.
\end{equation}
Reciprocally, assume that $\cp$ is a morphism and (i) holds.  Let
$H$ be as in (i) and set $\widetilde
H=h\otimes\ind_{\Gamma(\ch)}+\ind_\ch\otimes H$. The operators
$H,\widetilde H$ are bounded from below and
$\cp(\rme^{-H})=\rme^{-\widetilde H}$. Since $\cp$ is a morphism and
the function $x\mapsto\rme^{-x}$ algebraically generates
$\Co([a,\infty[)$ if $a\in\mbr$, we get
$\cp(\theta(H)=\theta(\widetilde H)$ for all $\theta\in\Co(\mbr)$.
In particular, if $z$ is a complex number with sufficiently large
negative real part we can take $\theta(x)=(z-x)^{-1}$ and get
$\cp[(z-H)^{-1}]=(z-\widetilde H)^{-1}$. Denote $R_z=(z-H_0)^{-1}$
and $\widetilde R_z=(z-\widetilde H_0)^{-1}$, where $\widetilde
H_0=h\otimes\ind_{\Gamma(\ch)}+\ind_\ch\otimes H_0$.  Then we make a
norm convergent series expansion to get:
$$
\cp\textstyle\sum_kR_z[VR_z]^k=\textstyle\sum_k
\widetilde R_z[(\ind_\ch\otimes V)\widetilde R_z]^k.
$$ 
We replace $V$ by $sV$ and take derivatives at $s=0$ to obtain
$\cp[R_zVR_z]=\widetilde R_{z}(\ind_\ch\otimes V)\widetilde R_{z}$.
On the other hand, by taking $V=0$ in this argument  we get
$\cp(\theta(H_0))=\theta(\widetilde H_0)$ for all
$\theta\in\Co(\mbr)$. Thus
$$
\cp\left[\theta(H_0)R_zVR_z\right)]=
\theta(\widetilde H_0)
\widetilde R_{z}(\ind_\ch\otimes V)\widetilde R_{z}.
$$
By arguments already used in the proof of Proposition \ref{pr:woo}
we get first 
$$
\cp\left[\eta(H_0)VR_z\right]=\eta(\widetilde H_0)
(\ind_\ch\otimes V)\widetilde R_{z}
$$ for $\eta\in\Co(\mbr)$ and then we see that this relation remains
true for $\eta=1$. Thus we have
$\cp\left[VR_z\right]=(\ind_\ch\otimes V)\widetilde R_{z}$ for all
complex numbers $z$ with sufficiently large negative real part. By
standard arguments we then get
$\cp[V\theta(H_0)]=(\ind_\ch\otimes V)\theta(\widetilde H_0)$ for
all $\theta\in\Co(\mbr)$, in particular 
$\cp\left[V\rme^{-H_0}\right]=
(\ind_\ch\otimes V)\rme^{-\widetilde H_0}$. But this is the same as
$$
\cp\left[V\Gamma(\rme^{-h})\right]=
(\ind_\ch\otimes V)(\rme^{-h}\otimes\Gamma(\rme^{-h})=
\rme^{-h}\otimes\left[V\Gamma(\rme^{-h})\right].
$$ Thus $\cp[V\Gamma(A)]=A\otimes[V\Gamma(A)]$ if $A=\rme^{-h}$.  By
first choosing $h$ conveniently and then by using the same argument
as in the last part of the proof of Proposition \ref{pr:woo} we see
that the preceding relation holds for all $A\in\co$ with $\|A\|<1$
and $A\geq0$. As in Example \ref{ex:cmor} this implies 
$$
n\cp[V(A_1\vee\dots\vee A_n)]=\textstyle\sum_k A_k\otimes
[V(A_1\vee\dots\vee A_{k-1}\vee A_{k+1}\vee\dots A_n)]
$$
first for $A_k\geq0$ and then for all $A_k\in\co$. Thus 
$\cp[VA^{\vee n}]=A\otimes[V A^{\vee(n-1)}]$ for all $A\in\co$ from
which we clearly get 
$\cp[V\Gamma(A)]=A\otimes[V\Gamma(A)]$ if $A\in\co$ and
$\|A\|<1$. That this holds also for $V=W(u)$ follows easily as in
the proof of Proposition \ref{pr:wo}. So $\cp$ is the canonical
morphism. 
\qed

We give one more characterization of $\cp$ which is sometimes useful
(e.g.\ it implies Theorem \ref{th:max}). The proof involves the
same ideas as that of Proposition \ref{pr:wot} so we do not give
details.

\begin{lemma}\label{lm:cmd}
$\rf(\co)$ coincides with the $C^*$-algebra generated by the
operators of the form $\phi(u)^n\Gamma(A)$ with $u\in\ch,n\in\mbn$
and $A\in\co$ with $\|A\|<1$. A morphism 
$\cp:\co\rarrow\co\otimes\rf(\co)$ is the canonical morphism
if and only if it satisfies 
$\cp\left(\phi(u)^n\Gamma(A)\right)=
A\otimes\left[\phi(u)^n\Gamma(A)\right]$ for all such $u,n,A$.
\end{lemma}

\section{The fermionic case}\label{s:ffock}
\protect\setcounter{equation}{0}

\noindent{\bf 1.} The fermionic version of the theory seems to me
most pleasant esthetically speaking and certainly much easier.  As
before $\ch$ is a complex Hilbert space with scalar product
$\lag\cdot|\cdot\rag$. A \emph{representation of the CAR over
$\ch$}, or a \emph{Clifford system over $\ch$}, is a couple
$(\rh,\phi)$ consisting of a Hilbert space $\rh$ and an
$\mbr$-linear map $\phi:\ch\rarrow B(\rh)$ which satisfies
\begin{equation}\label{eq:cs}
\phi(u)^*=\phi(u) \hspace{2mm} \text{ and }
\phi(u)^2=\|u\|^2  \hspace{2mm} 
\textrm{ for all } u\in\ch.
\end{equation}
We set $[A,B]_+=AB+BA$. Then the second condition above is
equivalent to:
\begin{equation}\label{eq:cs*}
[\phi(u),\phi(v)]_+=2\Re\lag u|v \rag \hspace{2mm}
\textrm{ for all } u,v\in\ch.
\end{equation}
Note that the map $\phi:\ch\rarrow B(\rh)$ is an isometry, which
makes the theory much simpler. We define the \emph{annihilation} and
\emph{creation} operators associated to the one particle state $u$
by the relations \eqref{eq:an}, so that $\phi(u)=a(u)+a^*(u)$.  Then
$a^*:\ch\rarrow B(\rh)$ is a linear continuous map, $a:\ch\rarrow
B(\rh)$ is antilinear and continuous, and $a^*(u)$ is just the
adjoint of the operator $a(u)$. We have
\begin{equation}\label{eq:acf}
[a(u),a^*(v)]_+=\lag u|v \rag, \hspace{2mm}
 [a(u),a(v)]_+=0, \hspace{2mm}
[a^*(u),a^*(v)]_+=0.
\end{equation}
A \emph{number operator for the Clifford system $(\rh,\phi)$} is
a self-adjoint operator $N$ on $\rh$ satisfying
\begin{equation}\label{eq:fno}
\rme^{\rmi tN}\phi(u)\rme^{-\rmi tN}=\phi(\rme^{\rmi t}u)
\hspace{2mm} \textrm{ for all } t\in\mbr \textrm{ and } u\in\ch. 
\end{equation}
As in the bosonic case we have:
\begin{equation}\label{eq:fnf}
[N,\rmi\phi(u)]=\phi(\rmi u), \hspace{2mm}
(N+1)a(u)=a(u)N, \hspace{2mm} (N-1)a^*(u)=a^*(u)N.
\end{equation}
A \emph{vacuum state} for the Clifford system $(\rh,\phi)$
is a vector $\Omega\in\rh$ with $\|\Omega\|=1$ such that the map
$u\mapsto\phi(u)\Omega$ is linear and this condition is
equivalent to $a(u)\Omega=0$ for all $u$.

\noindent{\bf 2.}
We define the \emph{Clifford algebra over $\ch$} by
\begin{equation}\label{eq:cah}
\rf(\ch)=C^*(\phi(u)\mid u\in\ch).
\end{equation}
We refer to \cite{PR} for a presentation of the theory of Clifford
algebras suited to our context. In their terminology, $\rf(\ch)$ is
the Clifford algebra generated by the \emph{real} vector space $\ch$
equipped with the scalar product $\Re\lag u|v \rag$. In particular,
if the (complex) dimension of $\ch$ is $n$ then $\rf(\ch)$ is of
dimension $2^{2n}$. The $C^*$-algebras $\rf(\ch)$ associated to two
Clifford systems over $\ch$ are canonically isomorphic in a natural
sense, which explains why $(\rh,\phi)$ is not included in the
notation. The algebra $\rf(\ch)$ has a rich and interesting
structure: it is central and simple, it has a unique tracial state,
and it is $\mbz_2$-graded ($\mbz_2=\mbz/2\mbz$), i.e.\ there is a
unique automorphism $\gamma$ of $\rf(\ch)$ such that
$\gamma(\phi(u))=-\phi(u)$ for all $u\in\ch$. Clearly
$\gamma^2=1$ and if we set $\rf_\pm(\ch)=\{T\in\rf(\ch)\mid
\gamma(T)=\pm T\}$ then we get a linear direct sum
decomposition $\rf(\ch)=\rf_+(\ch)+\rf_-(\ch)$.

If $\ck$ is a closed vector subspace of $\ch$ we
identify $\rf(\ck)$ with the $C^*$-subalgebra of $\rf(\ch)$
generated by the operators $\phi(u)$ with $u\in\ck$.   If
$E\subset F$ are finite dimensional subspaces of $\ch$ then
$\rf(E)\subset\rf(F)$ are finite dimensional $*$-subalgebras of
$\rf(\ch)$ and
\begin{equation}\label{eq:fu}
\rf(\ch)=\overline{\cup_E\rf(E)}
\end{equation}
where $E$ runs over the set of finite dimensional subspaces of
$\ch$. In particular, $\rf(\ch)$ is nuclear.

\noindent{\bf 3.}  One defines the Fock representation exactly as in
the bosonic case; the uniqueness modulo canonical isomorphisms is
obvious. The construction of the ``particle Fock realization'' is
parallel to that in the Bose case, one just has to replace
``symmetric'' and the symbol $\vee$ by ``antisymmetric'' and
$\wedge$ (the details can be found in \cite{PR}). So
$\ch^{\wedge}_{\text{alg}}$ is the antisymmetric (or exterior)
algebra\symbolfootnote[2]{\ The definition is similar to that in the
symmetric case, cf.\ the footnote on page \pageref{p:sa}, just
replace the commutativity condition $\xi(u)\xi(v)=\xi(v)\xi(u)$ by
$\xi(u)\xi(v)=-\xi(v)\xi(u)$.} over the vector space $\ch$, we use
the notation $uv$ for the product of two elements $u,v$ of
$\ch^{\wedge}_{\text{alg}}$ (or $u\wedge v$ if ambiguities occur in
concrete situations), and the unit element is denoted either $1$ or
$\Omega$. Then $\ch^{\wedge n}_{\text{alg}}$ is the linear subspace
spanned by the products $u_1\dots u_n$ with $u_i\in\ch$ and
$\ch^{\wedge}_{\text{alg}}$ is equal to the linear direct sum
$\sum_{n\in\mbn}\ch^{\wedge n}_{\text{alg}}$.  We shall equip
$\ch^{\wedge}_{\text{alg}}$ with the unique scalar product such that
$\ch^{\wedge n}_{\text{alg}}\perp \ch^{\wedge m}_{\text{alg}}$ for
$n\neq m$ and:
\begin{equation}\label{eq:fsc}
\lag u_1\dots u_n| v_1\dots v_n \rag
=\textstyle\sum_{\sigma\in\gs(n)}\varepsilon_\sigma
\lag u_1 | v_{\sigma(1)} \rag\dots \lag u_n | v_{\sigma(n)} \rag
\end{equation}
where $\varepsilon_\sigma$ is the signature of the permutation
$\sigma$. The estimate \eqref{eq:prod} remains 
valid in the present situation.

We define the \emph{Fock space $\Gamma(\ch)\equiv\ch^\wedge$ over
$\ch$} as the completion of $\ch^{\wedge}_{\text{alg}}$ for the
scalar product defined by \eqref{eq:fsc}.  Then $\ch^{\wedge n}$ is
the closure of $\ch^{\wedge n}_{\text{alg}}$ in $\Gamma(\ch)$, we
have $\Gamma(\ch)=\bigoplus_{n}\ch^{\wedge n}$ (Hilbert space direct
sum), and the spaces $\Gamma_n(\ch)$ and $\Gamma_{\text{fin}}(\ch)$
are defined as in the symmetric case. Similarly for the number
operator $N$ and the projections $\ind_n,\ind^n,\omega$. Note that
$\gfin(\ch)$ is a unital algebra but not abelian: it is a
$\mbz$-graded \emph{anticommutative} algebra, i.e.\ we have
$uv=(-1)^{nm}vu$ if $u\in\ch^{\wedge n}$ and $v\in\ch^{\wedge m}$.

The creation-annihilation operators $a^{(*)}(u)$ and the field
operator $\phi(u)$ are defined exactly as in the bosonic
case. Important differences are the boundedness of these operators:
$\|a^{(*)}(u)\|=\|u\|$, and the fact that $a(u)$ is an
\emph{antiderivation}:
\begin{equation}\label{eq:ader}
a(u)(vw)=(a(u)v)w+(-1)^nv(a(u)w) \hspace{2mm} \text{ if }
v\in\ch^{\wedge n},w\in\Gamma_{\text{fin}}(\ch).
\end{equation}

If $A_1,\dots,A_n\in B(\ch)$ then there is a unique operator
$A_1\wedge\dots\wedge A_n\in B(\ch^{\wedge n})$ such that
\begin{equation}\label{eq:Au}
(A_1\wedge\dots\wedge A_n)(u_1\dots u_n)=(n!)^{-1}\textstyle
\sum_{\sigma\in\gs(n)} \varepsilon_\sigma
(A_1u_{\sigma(1)})\dots(A_nu_{\sigma(n)})
\end{equation}
for all $u_1,\dots,u_n\in\ch$.  We extend it to $\Gamma(\ch)$ by
identifying 
$A_1\wedge\dots\wedge A_n\equiv A_1\wedge\dots\wedge A_n\ind^n$.
If $A_1=\dots=A_n\equiv A$ we denote $A^{\wedge n}$ this  operator.
Note that  $A^{\wedge n}$ is uniquely defined by the relation
$A^{\wedge n}(u_1\dots u_n)=(Au_1)\dots(Au_n)$ for all
$u_1,\dots,u_n\in\ch$. Observe that $A_1\wedge\dots\wedge A_n$ is a
symmetric function of $A_1,\dots,A_n$ hence one may use the
polarization formula in this case too.

As in the bosonic case, for each $A\in B(\ch)$ there is a unique
unital endomorphism $\Gamma(A)$ of the algebra
$\Gamma_{\text{fin}}(\ch)$ such that $\Gamma(A)u=Au$ for all
$u\in\ch$ and such that the restriction of $\Gamma(A)$ to each
$\Gamma_n(\ch)$ be continuous. In fact $\Gamma(A)=\oplus_{n\geq0}
A^{\wedge n}$. Clearly $\Gamma(AB)=\Gamma(A)\Gamma(B)$,
$\Gamma(1)=1$, $\Gamma(0)=\omega$, and $z^N=\Gamma(z)$ for
$z\in\mbc$. The relations \eqref{eq:ga}-\eqref{eq:ga*} remain valid.
The operator $\Gamma(A)$ is bounded on $\Gamma(\ch)$ if
$\|A\|\leq1$.  Finally, there is a unique derivation $\rmd\Gamma(A)$
of the algebra $\Gamma_{\text{fin}}(\ch)$ such that
$\rmd\Gamma(A)u=Au$ if $u\in\ch$. Hence
$
\rmd\Gamma(A)(u_1\dots u_n)=\textstyle\sum_k u_1\dots(Au_k)\dots u_n 
$ 
if $n\geq1$ and $\rmd\Gamma(A)\Omega=0$. We denote also by
$\rmd\Gamma(A)$ the closure of this operator.  If $A$ is not bounded
but generates a contractive $C_0$-semigroup on $\ch$ then
$\rmd\Gamma(A)$ is defined by
$\Gamma(\rme^{tA})=\rme^{t\rmd\Gamma(A)}$.

If $\ck\subset\ch$ is a closed subspace we identify
$\ck^\wedge_{\text{alg}}$ with the subalgebra of
$\ch^\wedge_{\text{alg}}$ generated by $\ck$ and then by taking the
closure in $\Gamma(\ch)$ we get an isometric embedding
$\Gamma(\ck)\subset\Gamma(\ch)$. The scalar product \eqref{eq:fsc}
has been chosen such that
$$
\lag uv \mid u'v'\rag=\lag u \mid u'\rag\lag v \mid v'\rag=
\lag u\otimes v \mid u'\otimes v'\rag  \hspace{2mm}
\text{ for all } 
u\in\gfin(\ck),v\in\gfin(\ck^\perp) 
$$ hence the linear map
$\gfin(\ck)\otimes_{\text{alg}}\gfin(\ck^\perp)\rarrow\gfin(\ch)$
associated to the bilinear map $(u,v)\mapsto uv$ extends to a linear
bijective isometry
$\Gamma(\ck)\otimes\Gamma(\ck^\perp)\rarrow\Gamma(\ch)$. This gives
us a canonical Hilbert space identification
$\Gamma(\ch)=\Gamma(\ck)\otimes\Gamma(\ck^\perp)$.  Note that the
product on $\gfin(\ck)\otimes_{\text{alg}}\gfin(\ck^\perp)$ induced
by the embedding in $\gfin(\ch)$ is the anticommutative tensor
algebra product, see \cite{Bo}. Note that
$\Omega_\ch=\Omega_\ck\otimes\Omega'_{\ck}$ and everything we said
starting with \eqref{eq:abc} until the end of Section \ref{s:fock}
remains valid.

It is also trivial to check that, as in bosonic case, for each
$u\in\ck$ we have $a^{(*)}_\ch(u)=a^{(*)}_\ck(u)\otimes1$ and
$\phi_\ch(u)=\phi_\ck(u)\otimes1$ relatively to the
factorization $\Gamma(\ch)=\Gamma(\ck)\otimes\Gamma(\ck^\perp)$.
On the other hand, if $u\in\ck^\perp$ it is easy to check that
$a^{(*)}_\ch(u)=(-1)^{N_\ck}\otimes a^{(*)}_{\ck^\perp}(u)$. Thus
for  $u\in\ck$ and $v\in\ck^\perp$ we have:
\begin{equation}\label{eq:fkh}
\phi_\ch(u+v)=\phi_\ck(u)\otimes1+
(-1)^{N_\ck}\otimes \phi_{\ck^\perp}(u)
\end{equation}

\noindent{\bf 4.}  The theory developed in Sections
\ref{s:wo}-\ref{s:cmo} has a complete analog in the present
setting. Many things become in fact simpler and look more natural
due to the boundedness of the field operators. So in what follows we
state the results and make some comments concerning the proofs.

If $\co$ is a $C^*$-algebra on $\ch$ then $\Gamma(\co)$ is defined
as in \eqref{eq:kef} and Proposition \ref{pr:kef} (with $\vee$
replaced by $\wedge$) remains true because $A_1\wedge\dots\wedge
A_n$ is a symmetric function of $A_1,\dots,A_n$. Then we define:
\begin{equation}\label{eq:fco*}
\rf(\co)=C^*(S\Gamma(A)\mid S\in\rf(\ch),A\in\co,\|A\|<1)
\end{equation}
and we set $\ra(\ch)=\rf(\mbc\ind_\ch)$. \emph{If $\co$ is
non-degenerate then we have} 
\begin{equation}\label{eq:fco}
\rf(\co)=\llb\rf(\ch)\cdot\Gamma(\co)\rrb
\end{equation}
The proof is a much simplified version of that of Proposition
\ref{pr:wot}. We now consider Proposition \ref{pr:elpr}.

\noindent{\bf Proof of the fermionic version of Proposition
\ref{pr:elpr}:} $\rf(\{0\})$ is the $C^*$-algebra generated by the
operators $\phi(u_1)\dots\phi(u_n)\omega$ (where the product may be
empty) and the linear span of these operators coincides with the
linear span of $a^*(u_1)\dots a^*(u_n)\omega=|u_1\dots
u_n\rag\lag\Omega|$, from which (2) of Proposition \ref{pr:elpr} in
the Fermi case follows easily. Now we prove (3) of Proposition
\ref{pr:elpr}. Basically this follows from
\begin{eqnarray*}
\phi(u)\Gamma(A) &=& \left(\phi(u_1)\otimes\ind+
(-1)^{N_1}\otimes \phi(u_2)\right)
\Gamma(A_1)\otimes\Gamma(A_2)\\ &=&
[\phi(u_1)\Gamma(A_1)]\otimes\Gamma(A_2)+
\Gamma(-A_1)\otimes[\phi(u_2)\Gamma(A_2)].
\end{eqnarray*}
but the complete argument is complicated by the fact that we have to
consider arbitrary polynomials in the fields. Consider a product
$\phi(w_1)\dots\phi(w_n)\Gamma(A)$ and decompose $w_k=u_k+v_k$,
$A=B\oplus C$ with $u_1,\dots,u_n\in\ch_1$, $v_1,\dots,v_n\in\ch_2$,
and $B\in\co_1,C\in\co_2$ with norms $<1$. Due to \eqref{eq:fkh}
and since $(-1)^{N_{\ch_1}}=\Gamma(-1_{\ch_1})$ we have  
$\phi(w_k)=\phi(u_k)\otimes\ind+ \Gamma(-1)\otimes \phi(v_k)$
with some simplifications in the notations. If we develop the
product  $\phi(w_1)\dots\phi(w_n)$ and if we take into account the
relation $\Gamma(-1)\phi(u_k)=\phi(-u_k)\Gamma(-1)$
we clearly get a sum of terms of the form (ordered  products)
$$
\left[\textstyle\prod_{j\in\alpha}\phi(\widetilde u_j)\right]
\otimes\left[\textstyle\prod_{k\in\beta}\phi(v_k)\right]\cdot
\Gamma(\pm1)\otimes\ind
$$ 
where $\alpha$ is a subset of $\{1,\dots, n\}$, $\beta$ is the
complementary subset, and  $\widetilde u_j$ is either $u_j$ or
$-u_j$. Since 
$\Gamma(\pm1)\otimes\ind\cdot\Gamma(A)=\Gamma(\pm
B)\otimes\Gamma(C)$ we see that
$\phi(w_1)\dots\phi(w_n)\Gamma(A)\in\rf(\co_1)\otimes\rf(\co_2)$ and
the proof is finished by an obvious argument.
\qed

We mention one more fact, which is also true in the bosonic case but
with a more complicated proof.

\begin{proposition}\label{pr:frem}
If $\co$ is non-degenerate then $\rf(\co)$ is the $C^*$-algebra
generated by the operators of the form $\Gamma(A)$ or
$\phi(u)\Gamma(A)$ with $u\in\ch$ and $A\in\co$, $A\geq0$,
$\|A\|<1$. 
\end{proposition}
\proof We give the proof under the supplementary assumption that
$\co$ contains a positive injective operator (this is the only
situation relevant in field theory; in general one has to use an
approximate unit as in the proof of Proposition \ref{pr:wot}).  Let
$\rc$ be the $C^*$-algebra generated by the operators of the form
$\Gamma(A)$ or $\phi(u)\Gamma(A)$ with $u\in\ch$ and $A\in\co$,
$A\geq0$, $\|A\|<1$. Due to \eqref{eq:late} it is sufficient to show
that any product $\phi(u_1)\dots\phi(u_n)\Gamma(A)$ with $A$ as
above belongs to $\rc$. We show this in the case of two field
factors $\phi(u)\phi(v)\Gamma(A)$, the general case is similar. We
have $A=(\sqrt A)^2$ and $\sqrt A\in\co$, is positive, and has norm
strictly less than $1$. By writing $\phi(u)\phi(v)\Gamma(A)=
\phi(u)[\phi(v)\Gamma(\sqrt A)]\Gamma(\sqrt A)$ we see that it
suffices to show the following: for each $v\in\ch$ and $B\in\co$
with $B\geq0,\|B\|<1$, the operator $\phi(v)\Gamma(B)$ belongs to
the norm closure $\rl$ of the linear span of the operators of the
form $\Gamma(A)\phi(u)$ with $u,A$ as before. We have
$\phi(v)\Gamma(B)=a(v)\Gamma(B)+a^*(v)\Gamma(B)$ and so it suffices
to have $a^{(*)}(v)\Gamma(B)\in\rl$. In the case of $a(u)\Gamma(B)$
this is obvious by \eqref{eq:ga}. Now let $S\in\co$ be positive and
injective and let $\varepsilon>0$ real. Then \eqref{eq:ga} implies
$a^*((B+\varepsilon S)w)\Gamma(B+\varepsilon S)=
\Gamma(B+\varepsilon S)a^*(w)\in\rl$ for all $w\in\ch$. The operator
$B+\varepsilon S$ is positive and injective hence it has dense
range. The map $u\mapsto a^*(u)\in B(\Gamma(\ch))$ is norm
continuous, hence we get $a^*(v)\Gamma(B+\varepsilon S)\in\rl$ for
all $v\in\ch$. From Lemma \ref{lm:eles} we easily get
$\Gamma(B+\varepsilon S)\rarrow\Gamma(B)$ in norm as
$\varepsilon\rarrow0$, hence $a^*(v)\Gamma(B)\in\rl$.
\qed

One may define elementary QFH as in Definition \ref{df:qfh} by
asking $V\in\rf(\ch)$ or $V\in\rf(E)$ for some finite dimensional
subspace $E$ of $\ch$. And then Proposition \ref{pr:woo} remains
true (only a minor modification of the end of the proof is
required). 
We may now state the fermionic version of our main result.

\begin{theorem}\label{th:fcmo}
If $\co$ is an abelian $C^*$-algebra on $\ch$ and its strong closure
does not contain finite rank operators, then there is a unique
morphism $\cp:\rf(\co)\rarrow\co\otimes\rf(\co)$ such that
\begin{equation}\label{eq:cmof}
\cp[S\Gamma(A)]=A\otimes[S\Gamma(A)] \hspace{2mm} \text{ if }
S\in\rf(\ch) \text{ and } A\in\co, \|A\|<1.
\end{equation}
We have $\ker\cp=\rk(\ch)$, which gives us a canonical embedding
\begin{equation}\label{eq:fcmo}
\rf(\co)/\rk(\ch)\hookrightarrow \co\otimes\rf(\co).
\end{equation} 
\end{theorem}
\emph{If $\co$ is non-degenerate then one may require
\eqref{eq:cmof} to hold only for $S=\phi(u)^k$} (the powers
$\phi(u)^k$ with $k\in\mbn$ are multiples of $\phi(u)$ or of the
identity). The second characterization of $\cp$ presented in
Proposition \ref{pr:carp} remains valid. The canonical endomorphism
$\cp$ of $\ra(\ch)$ satisfies $\cp(S\theta(N))=S\theta(N+1)$ for all
$S\in\rf(\ch)$ and $\theta\in\Co(\mbn)$.

The strategy of the proof of Theorem \ref{th:fcmo} is identical to
that from the symmetric case. We first treat the case of $\ra(\ch)$
as in Section \ref{s:wa} with the help of the algebras
$$
\ra_E(\ch)=\llb\rf(E)\cdot\Co(N)\rrb
=\rk(E)\otimes\Co(N_E') \text{ relatively to }
\Gamma(\ch)=\Gamma(E)\otimes\Gamma(E^\perp).
$$ Here $E$ is finite dimensional and $\rf(E)\equiv\rf(E)\otimes
1_{E^\perp}$ the $\rf(E)$ from the right hand side being the algebra
of all operators on the finite dimensional space $\Gamma(E)$. In
particular we now have $N_E\in\rf(E)$, in fact $N_E=\sum_{k=0}^n
a^*(e_k)a(e_k)$ if $e_1,\dots,e_n$ is an orthonormal basis of $E$.
For a general algebra $\co$ we proceed as in Section \ref{s:cmo}.

We now prove that $\ra(\ch)$ has a natural $\mbz_2$-grading and we
state the fermionic version of Remark \ref{re:skan}.

\begin{proposition}\label{pr:skan}
There is a unique automorphism $\gamma$ of $\ra(\ch)$ such that
$\gamma(S\theta(N))=\gamma(S)\theta(N)$ for all $S\in\rf(\ch)$ and
$\theta\in\Co(\mbn)$. We have $\gamma^2=1$ and for each
$T\in\ra(\ch)$:
\begin{equation}\label{eq:skaf}
\cp(T)=\slim_{e\rarrowup0}a(e)\gamma(T)a^*(e)
\end{equation}
\end{proposition}
\proof From the fermionic version of \eqref{eq:ind} it follows that
it suffices to define $\gamma$ on $\ra_E(\ch)$ for each finite
dimensional $E$. Since, as explained above, we then have
$\ra_E(\ch)=\rk(E)\otimes\Co(N_E')$, the existence is rather
obvious. However, the following explicit construction, cf.\
\cite[Theorem 1.1.10]{PR}, gives more information. Observe first
that if $e\in\ch$ and $\|e\|=1$ then $\phi(e)\phi(\rmi
e)=\rmi[a(e),a^*(e)]$, hence $\phi(e)\phi(\rmi
e)=\phi(ze)\phi(\rmi ze)$ for all complex $z$ with
$|z|=1$. Let $e_1,\dots,e_n$ be an orthonormal basis of $E$ and
$w=\phi(e_1)\phi(\rmi e_1)\dots\phi(e_n)\phi(\rmi e_n)$.
It is clear that $w$ is a unitary element of $\rf(E)$ with $w^*=w$
if $n$ is even and $w^*=-w$ if $n$ is odd. The relation
$wSw^*=\gamma(S)$ for $S\in\rf(E)$ is easy to check (or see Theorem
1.1.10 in \cite{PR}). By using the expression given above for $N_E$
we get $wN_Ew^*=N_E$ and it is clear that $wN'_Ew^*=N'_E$. Thus
$wNw^*=N$ and we may define $\gamma(T)=wTw^*$ for all $T\in\rf(E)$.

We have $a(e)u_0\dots u_n=\sum_k(-1)^ku_0\dots\lag e|u_k\rag\dots
u_n$ hence $\slim_{e\rarrowup0}a(e)=0$. From the anticommutation
relation $a(e)a^*(e)+a^*(e)a(e)=1$ we get
$\slim_{e\rarrowup0}a(e)a^*(e)=1$. Thus $\cp$ \emph{defined} by
\eqref{eq:skaf} is an endomorphism of $\ra(\ch)$. Note that
$$
\|a(e)\phi(u)+\phi(u)a(e)\|=|\lag e|u\rag|\rarrow0
\text{ if } e\rarrowup0. 
$$
Finally, by using \eqref{eq:fnf} it follows easily
that $\cp$ is the canonical endomorphism of $\ra(\ch)$.  \qed

It is clear that everything we said in Section \ref{s:cmo} starting
with Proposition \ref{pr:iter} remains true or has an analog in the
fermionic case.

\section{ Self-adjoint operators affiliated to $\rf(\co)$}
\label{s:saa}
\protect\setcounter{equation}{0}

\noindent{\bf 1.}  
It will be convenient to use the notion of observable affiliated to
a $C^*$-algebra as introduced in \cite{BG2} and further studied in
\cite{ABG,DG}. In this paper a self-adjoint operator is supposed to
be densely defined but not densely defined operators appear by
taking (norm) resolvent limits or images through $C^*$-algebra
morphisms. An observable is a Hilbert space independent formulation
of the notion of ``not necessarily densely defined self-adjoint
operator''.

An \emph{observable affiliated to a $C^*$-algebra $\rc$} is a
morphism $H:\Co(\mbr)\rarrow\rc$. We set $H(\theta)=\theta(H)$
although $H$ cannot be realized as a self-adjoint operator in
general.  Observables have the advantage that one can consider their
images through morphisms: if $\cp:\rc\rarrow\rd$ is a morphism, then
$\cp(H)$ is the observable affiliated to $\rd$ defined by
$\theta(\cp(H))=\cp(\theta(H))$ (this  operation makes no sense
at the Hilbert space level). The \emph{spectrum} of $H$ is the set
$\sigma(H)$ of real points $\lambda$ such that $\theta(H)\neq0$ if
$\theta(\lambda)\neq0$. A sequence $\{H_n\}$ of observables
affiliated to $\rc$ is \emph{convergent} if $\lim_n\theta(H_n)$
exists (in norm) for each $\theta\in\Co(\mbr)$.  Then
$\theta(H)=\lim_n\theta(H_n)$ is an observable affiliated to $\rc$
and we write $H=\lim_n H_n$.

Let $\rc$ be a $C^*$-algebra of operators on a Hilbert space $\rh$.
We say that a self-adjoint operator $H$ on $\rh$ is 
{\em affiliated}\,\symbolfootnote[2]{\ This should not be confused
        with the terminology of Woronowicz, see \cite{DG}.}
to $\rc$ if $(H-z)^{-1}\in\rc$ for some
$z\in\mbc\setminus\sigma(H)$.  This is equivalent to
$\theta(H)\in\rc$ for all $\theta\in\Co(\mbr)$ and this gives us a
morphism $\theta\mapsto\theta(H)$, hence $H$ defines an observable
affiliated to $\rc$ and this observable determines the self-adjoint
operator $H$ uniquely. So the set of self-adjoint operators
affiliated to $\rc$ is a subset of the set of observables affiliated
to $\rc$.  But there are observables affiliated to $\rc$ which do
not correspond to self-adjoint operators on $\rond H$ (and these
could be physically interesting).  See \cite[page 364]{ABG} and
\cite{BGS} for details on this question.

It is clear that the spectrum of $H$ as self-adjoint operator on
$\rh$ and as observable affiliated to $\rc$ are identical. If
$\{H_n\}$ is a sequence of self-adjoint operators affiliated to
$\rc$ then the sequence of observables $H_n$ converges if and only
if the sequence of operators $H_n$ converges in norm resolvent
sense.

If one insists in working with self-adjoint operators the following
notion is useful.  We say that an observable or a self-adjoint
operator $H$ is \emph{strictly affiliated to $\rc$} if the linear
space generated by the products $\theta(H)T$ with
$\theta\in\Co(\mbr)$ and $T\in\rc$ is dense in $\rc$.  If there is a
self-adjoint operator on $\rh$ affiliated to $\rc$ then $\rc$ is
non-degenerate on $\rh$.

We refer to \cite[Appendix]{DG} for a proof of the following fact:
\emph{if $H$ is a self-adjoint operator strictly affiliated to $\rc$
and if $\cp$ is a non-degenerate representation of $\rc$ on a
Hilbert space $\rk$, then there is a unique self-adjoint operator
$\cp(H)$ on $\rk$ such that $\cp(\phi(H))=\phi(\cp(H))$ for
all $\phi\in\Co(\mbr)$. Moreover, $\cp(H)$ is strictly affiliated
to the $C^*$-algebra $\cp(\rc)$}.

From now on we assume that $\rc\subset B(\rh)$ is non-degenerate on
$\rh$. Then the \emph{multiplier algebra}\symbolfootnote[2]{\ This is
isomorphic with the abstractly defined multiplier algebra, cf.\ 
\cite{La}, but we shall not use this fact.} of $\rc$ is defined by:
\begin{equation}\label{eq:ma}
\rM=\{B\in B(\rh)\mid BC\in\rc \text{ and } CB\in\rc 
\text{ if } C\in\rc\}.
\end{equation}
Each non-degenerate representation $\cp$ of $\rc$ on a Hilbert space
$\rk$ extends in a unique way to a representation (also denoted
$\cp$) of $\rM$ on $\rk$ such that $\cp(B)\cp(C)=\cp(BC)$ for all
$B\in\rM$ and $C\in\rc$.

\begin{lemma}\label{lm:maf}
Assume that $H_0$ is a self-adjoint operator (strictly) affiliated to
$\rc$ and that $V=V^*$ belongs to the multiplier algebra of $\rc$.
Then $H=H_0+V$ is (strictly) affiliated to $\rc$.  If $\cp$ is a
non-degenerate representation of $\rc$ then
$\cp(H)=\cp(H_0)+\cp(V)$.
\end{lemma}

This is an easy consequence of $R(z)=\sum
R_0(z)\left(VR_0(z)\right)^k$ for large $z$, where $R(z)=(z-H)^{-1}$
and $R_0(z)=(z-H_0)^{-1}$. See \cite{DG} for the proof of the strict
affiliation.

We quote below several affiliation criteria which are convenient for
quantum field models.

\begin{theorem}\label{th:ro}
Let $H_0$ and $V$ be bounded from below self-adjoint operators on
$\rh$ such that the operator $H=H_0+V$ with domain $D(H_0)\cap D(V)$
is self-adjoint (in particular, the intersection has to be dense in
$\rh$).  If $\rme^{-tH_0}\rme^{-2tV}\rme^{-tH_0}\in\rc$ for all
$t>0$ then $H$ is affiliated to $\rc$.
\end{theorem}
This follows from a result of Rogava \cite{Ro} (see \cite{IT}
for more recent results) which says that
\begin{equation}\label{eq:ro}
\rme^{-2tH}=\lim_{n\rarrow\infty}
\left[\rme^{-tH_0/n}\rme^{-2tV/n}\rme^{-tH_0/n}\right]^n
=\lim_{n\rarrow\infty}
\left[\left(\rme^{-tV/n}\rme^{-tH_0/n}\right)^*
\left(\rme^{-tV/n}\rme^{-tH_0/n}\right)\right]^n
\end{equation}
holds in norm for all $t>0$. Under the same conditions we also have
norm convergence in:
\begin{equation}\label{eq:ro*}
\rme^{-tH}=\lim_{n\rarrow\infty}
\left[\rme^{-tV/n}\rme^{-tH_0/n}\right]^n.
\end{equation}
Other affiliation criteria can be found in \cite{DG}, for example:

\begin{theorem}\label{th:dag}
Let $H_0\geq0$ be a self-adjoint operator affiliated to $\rc$ and
let $V$ be a symmetric form such that $-aH_0-b\leq V \leq bH_0+b$
for some real numbers $0<a<1$ and $b>0$. Assume that
$U\equiv(H_0+1)^{-1/2}V(H_0+1)^{-1/2}$ belongs to the multiplier
algebra $\rM$. Then $H=H_0+V$ defined in form sense is a
self-adjoint operator affiliated to $\rc$. If $H_0$ is strictly
affiliated to $\rc$ then $U\in\rM$ if and only
$\theta(H_0)V(H_0+1)^{-1/2}\in\rc$ for all 
\mbox{\rm $\theta\in\Cc(\mbr)$} and then $H$ is strictly affiliated
to $\rc$. 
\end{theorem}

Now let us fix a probability measure space $Q$ and consider the
associated scale of $L^p$ spaces. Let $H_0$ be a positive
self-adjoint operator on $L^2$ which generates a
\emph{hypercontractive} semigroup in the following sense: for each
$t>0$ the operator $\rme^{-tH_0}$ is a contraction in each $L^p$ and
there are $p>2$ and $t>0$ such that $\rme^{-tH_0}L^2\subset L^p$. We
shall say that a real function $V$ on $Q$ is \emph{admissible} if
$V$ and $\rme^{-V}$ belong to $L^p$ for all $p<\infty$ (observe that
if $V$ is bounded from below the second condition is automatically
satisfied).  Under these conditions on $H_0$ and $V$ it can be shown
that $H_0+V$ is essentially self-adjoint on $D(H_0)\cap D(V)$ and
its closure $H$ is bounded from below, see \cite[Theorem
X.58]{RS}. Then \cite[Theorem X.60]{RS}:

\begin{theorem}\label{th:shk}
Assume that $H$ is as above, let $\{V_n\}$ be a sequence of
admissible functions, and let $H_n$ be the closure of the operator
$H_0+V_n$. Assume that there is $p>2$ such that
$\|V_n-V\|_{L^p}\rarrow0$ and
$\sup_n\|\rme^{-V_n}\|_{L^p}<\infty$. Then $\lim H_n=H$ in norm
resolvent sense.
\end{theorem}

\noindent{\bf 2.}  We consider now the case of interest in this
paper. Let $\ch$ be a complex Hilbert space and $\co$ an abelian
\emph{non-degenerate} $C^*$-algebra on $\ch$ such that $\co''\cap
K(\ch)=\{0\}$. We take $\rh=\Gamma(\ch)$, which is either the
bosonic or the fermionic Fock space, and $\rc=\rf(\co)$. Then
according to Theorems \ref{th:cmo} and \ref{th:fcmo} we have a
canonical morphism $\cp:\rf(\co)\rarrow\co\otimes\rf(\co)$ whose
kernel is $\rk(\ch)\equiv\ K(\Gamma(\ch))$. The algebra
$\co\otimes\rf(\co)$ is naturally realized on the Hilbert space
$\ch\otimes\Gamma(\ch)$ and thus we get an embedding
\begin{equation}\label{eq:can}
\rf(\co)/\rk(\ch)\hookrightarrow \co\otimes\rf(\co)\subset
B(\ch\otimes\Gamma(\ch)).
\end{equation}
Thus we may think of $\cp$ as a representation of $\rf(\co)$ on
$\ch\otimes\Gamma(\ch)$ with range $\rf(\co)/\rk(\ch)$
included (strictly in general) in $\co\otimes\rf(\co)$.
\begin{lemma}\label{lm:ndg}
$\rf(\co)$ is non-degenerate on $\Gamma(\ch)$ and the representation
$\cp$ of $\rf(\co)$ on $\co\otimes\Gamma(\ch)$ is non-degenerate.
If $h\geq m>0$ is a self-adjoint operator on $\ch$ strictly
affiliated to $\co$ then $H_0=\dd\Gamma(h)$ is strictly affiliated
to $\Gamma(\co)$ and to $\rf(\co)$.
\end{lemma}
\proof The action of the algebra $\rf(\co)$ on $\Gamma(\ch)$ is
non-degenerate because $\rk(\ch)\subset\rf(\co)$. The action of
$\cp(\rf(\co))$ on $\ch\otimes\Gamma(\ch)$ is also non-degenerate
because this algebra contains the operators of the form
$S\otimes\Gamma(S)$ with $S\in\co$ and $\|S\|<1$ and if we take a
sequence $\{S_n\}$ of such operators with $S_n\rarrow\ind_\ch$
strongly then $S_n\otimes\Gamma(S_n)$ converges strongly to the
identity operator on $\ch\otimes\rh$.

If $h$ is strictly affiliated to $\co$ then the linear span of the
operators $\theta(h)T$ with $\theta\in\Co(\mbr)$ and $T\in\co$ is
dense in $\co$. If $h$ is also bounded from below this clearly
implies $\|\rme^{-\veps h}T-T\|\rarrow0$ as $\veps\rarrow0$ (and
reciprocally). If $h\geq m>0$ then from Lemma \ref{lm:eles} we
clearly get $\|\rme^{-\veps H_0}\Gamma(A)-\Gamma(A)\|\rarrow0$ as
$\veps\rarrow0$ if $A\in\co,\|A\|<1$, and from this we deduce that
$H_0$ is strictly affiliated to $\Gamma(\co)$. Finally, we make a
general remark: \emph{if $H$ is an observable strictly affiliated to 
$\Gamma(\co)$ then it is strictly affiliated to $\rf(\co)$}. Indeed,
we have $\Gamma(\co)\subset\rf(\co)$ and the natural (left or right)
action of $\Gamma(\co)$ on $\rf(\co)$ is non-degenerate, cf.\
Proposition \ref{pr:wo}.
\qed

Thus, if $H$ is a self-adjoint operator on $\Gamma(\ch)$ strictly
affiliated to $\rf(\co)$ then $\cp(H)$ is a self-adjoint operator on
$\ch\otimes\Gamma(\ch)$ strictly affiliated to the quotient algebra
$\rf(\co)/\rk(\ch)$. If $H$ is only affiliated to $\rf(\co)$ then
$\cp(H)$ is only an observable affiliated to $\rf(\co)/\rk(\ch)$ and
in general can not be realized as a self-adjoint operator on
$\ch\otimes\Gamma(\ch)$. In any case, as the simplest application in
spectral theory of Theorems \ref{th:cmo} and \ref{th:fcmo}, we have
the following description of the essential spectrum of $H$.

\begin{theorem}\label{th:ess}
  We have \mbox{\rm$\se(H)=\sigma(\cp(H))$} if $H\in\rf(\co)$ or $H$
  is affiliated to $\rf(\co)$.
\end{theorem}

This result can be made more explicit in the following terms. Since
$\co$ is an abelian $C^*$-algebra its spectrum $\rx$ is a locally
compact topological space and we have a canonical identification
\begin{equation}\label{eq:ox}
\co\otimes\rf(\co)\cong \Co(\rx;\rf(\co)),
\end{equation}
where $\Co(\rx;\rf(\co))$ is the $C^*$-algebra of norm continuous
functions $F:\rx\rarrow\rf(\co)$ which tend to zero at
infinity. Assume for simplicity that $\wtilde H\equiv\cp(H)$ is a
self-adjoint operator on $\ch\otimes\Gamma(\ch)$ (which holds if $H$
is strictly affiliated to $\rf(\co)$), then $\wtilde H$ is
identified with a continuous family $\{\wtilde H(x)\}_{x\in\rx}$ of
self-adjoint operators affiliated to $\rf(\co)$ and we have
\begin{equation}\label{eq:ox*}
\se(H)=\textstyle\bigcup_{x\in\rx}\sigma(\wtilde H(x)).
\end{equation}
See \cite[8.2.4]{ABG} for details and for the proof that the
union is closed ($\widetilde H$ could be only an observable).

\noindent{\bf 3.} The simplest operators affiliated to $\rf(\co)$
are the elementary QFH, and their images through $\cp$ are described
in Proposition \ref{pr:carp}. We give other examples below and in
later sections.  Since we think of $\rf(\co)$ as the $C^*$-algebra
of energy observables of a quantum field, any observable affiliated
to it should be interpreted as the Hamiltonian of some quantum field
model with one particle kinetic energy affiliated to $\co$. Thus
Theorem \ref{th:ess} and the formula \eqref{eq:ox*} should cover a
large class of models.  However, the Hamiltonians of the usual models
are of the same nature as the elementary QFH (only much more
singular). We isolate this class of operators in the next
definition.

\begin{definition}\label{df:dqfh}
A self-adjoint operator $H$ on $\Gamma(\ch)$ is a \emph{standard
quantum field Hamiltonian (SQFH)} if $H$ is bounded from below and
affiliated to $\rf(\co)$ and if there is a self-adjoint operator
$h\geq0$ on $\ch$ affiliated to $\co$ such that
$\cp(H)=h\otimes\ind_{\Gamma(\ch)}+\ind_\ch\otimes H$. Under these
conditions we shall also say that \emph{$H$ is of type $\co$} and
that $h$ is the \emph{one particle kinetic energy} and $m=\inf h$
the \emph{one particle mass} associated to $H$.
\end{definition}

If we apply Theorem \ref{th:ess} to SQFH Hamiltonians we get:
\begin{theorem}\label{th:esz}
If $H$ is a SQFH with one particle kinetic energy $h$ and one
particle mass $m$ then:
\begin{equation}\label{eq:ess}
\mbox{\rm$\se(H)$}=\sigma(h)+\sigma(H)=
\{\lambda+\mu\mid\lambda\in\sigma(h), \mu\in\sigma(H)\}.
\end{equation}
In particular, if $m>0$ then $\inf H$ is an eigenvalue of finite
multiplicity of $H$ isolated from the rest of the spectrum. 
If $\sigma(h)=[m,\infty[$ then \mbox{\rm$\se(H)=[m+\inf H,\infty[$}.
\end{theorem}

The class of SQFH is quite large and many singular physically
interesting Hamiltonians are affiliated to it. We shall give such
examples in the next sections and we devote the rest of this section
to some preliminary results in this direction.

\begin{lemma}\label{lm:mra}
The multiplier algebra of $\rf(\co)$ contains \mbox{\rm
$\rw_{\text{max}}(\ch)$} in the bosonic case and $\rf(\ch)$ in the
fermionic case. If $V$ belongs to one of these classes we have
$\cp(V)=\ind_\ch\otimes V$.
\end{lemma}
\proof In the bosonic case it suffices to consider $V=W(f)$ with $f$
a bounded Borel regular measure on $\ch$ and to show that 
for $T=\Gamma(A)S$ with $S\in\rw(\ch)$ and $A\in\co,\|A\|<1$
we have $VT\in\rf(\co)$ and $\cp(VT)=(\ind_\ch\otimes V)\cp(T)$.
We have $VT=\int W(u)\Gamma(A)S df(u)$ the integral being convergent
in norm by Lemma \ref{lm:eles}, and $W(u)\Gamma(A)S\in\rf(\co)$,
hence $VT\in\rf(\co)$ and 
\begin{eqnarray*}
\cp(VT) &=& \int \cp(W(u)\Gamma(A)S) df(u)
=\int A\otimes(W(u)\Gamma(A)S) df(u)\\ &=&
A\otimes(V\Gamma(A)S)=(\ind_\ch\otimes
V)(A\otimes(\Gamma(A)S)=(\ind_\ch\otimes V)\cp(T).
\end{eqnarray*}
The proof in the fermionic case is similar and easier.
\qed

\begin{proposition}\label{pr:eqfh}
Let $h$ be a self-adjoint operator on $\ch$ affiliated to $\co$ and
such that $\inf h>0$. Let $V=V^*$ be an element of the multiplier
algebra of $\rf(\co)$. Then $H=\dd\Gamma(h)+V$ is affiliated to
$\rf(\co)$
and we have
$\cp(H)=h\otimes\ind_{\Gamma(\ch)}+
\ind_\ch\otimes\dd\Gamma(h)+\cp(V)$.  
In particular, if \mbox{\rm $V\in\rw_{\text{max}}(\ch)$} in the
bosonic case and $V\in\rf(\ch)$ in the fermionic case, then we have
$\cp(H)=h\otimes\ind_{\Gamma(\ch)}+ \ind_\ch\otimes H$, so $H$ is a
SQFH. 
\end{proposition}
\proof The operator $H_0=\rmd\Gamma(h)$ has the property
$\rme^{-tH_0}=\Gamma(\rme^{-th})$ for $t>0$ and $\rme^{-th}\in\co$
and has norm $<1$, so that 
$$
\cp\left(\rme^{-tH_0}\right)=\rme^{-th}\otimes\Gamma(\rme^{-th})=
\rme^{-th}\otimes\rme^{-tH_0}.
$$ Thus $\cp(H_0)=h\otimes\ind_{\Gamma(\ch)}+\ind_\ch\otimes H_0$
and then we use Lemmas \ref{lm:maf} and \ref{lm:mra}.
\qed

\begin{proposition}\label{pr:sqfh}
Let $V$ be a bounded from below self-adjoint operator on
$\Gamma(\ch)$ affiliated to $\rw_{\text{max}}(\ch)$ in the Bose case
and to $\rf(\ch)$ in the Fermi case. Let $h$ be a self-adjoint
operator on $\ch$ affiliated to $\co$ with $h\geq m>0$ and let us
set $H_0=\dd\Gamma(h)$. If $H=H_0+V$ is self-adjoint on $D(H_0)\cap
D(V)$ then $H$ is a SQFH of type $\co$ with $h$ as one particle
kinetic energy.
\end{proposition}
\proof That $H$ is affiliated to $\rf(\co)$ is a consequence of
Theorem \ref{th:ro}. Then $\wtilde H=\cp(H)$ is an observable
affiliated to $\rf(\co)$ but we do not yet know if it can be
realized as a self-adjoint operator on $\ch\otimes\Gamma(\ch)$. In
any case, the semigroup $\{\rme^{-t\wtilde H}\}_{t>0}$ is well
defined (it could be zero on a nontrivial subspace) and
\eqref{eq:ro*} implies:
\begin{eqnarray*}
\rme^{-t\wtilde H} &=& \cp\left(\rme^{-t H}\right)=
\lim_{n\rarrow\infty}
\left[\cp\left(\rme^{-tV/n}\rme^{-tH_0/n}\right)\right]^n 
\\ &=&
\lim_{n\rarrow\infty}
\left[
\cp\left(\rme^{-tV/n}\Gamma\left(\rme^{-th/n}\right)\right)\right]^n
\\ &=&
\lim_n
\left[\rme^{-th/n}\otimes\left(\rme^{-tV/n}
\Gamma\left(\rme^{-th/n}\right)\right)\right]^n
\\ &=&
\lim_n
\rme^{-th}\otimes
\left[\rme^{-tV/n}\rme^{-tH_0/n}\right]^n
=\rme^{-th}\otimes\rme^{-tH}.
\end{eqnarray*} 
Since this holds for all $t>0$ we get
$\wtilde H=h\otimes\ind_{\Gamma(\ch)}+\ind_\ch\otimes H$.
\qed 

The fact that the class of SQFH contains singular physically
interesting Hamiltonians is mainly due to its stability under norm
resolvent convergence.

\begin{proposition}\label{pr:nrl}
  Assume that $\{H_n\}$ is a sequence of SQFH of type $\co$ with the
  same one particle kinetic energy $h$ and such that $H_n\rarrow H$
  in norm resolvent sense, where $H$ is a self-adjoint operator on
  $\Gamma(\ch)$. Then $H$ is SQFH of type $\co$ with one particle
  kinetic energy $h$.
\end{proposition}
\proof
Due to norm resolvent convergence the operators $H_n$ are uniformly
bounded from below and $\rme^{-tH_n}\rarrow\rme^{-tH}$ in norm for
each $t>0$. Thus $\rme^{-tH}\in\rf(\co)$ hence $H$ is affiliated to
$\rf(\co)$ and we have
$$
\cp\left(\rme^{-tH}\right)=\lim_n\cp\left(\rme^{-tH_n}\right)=
\rme^{-th}\otimes\rme^{-tH_n}=\rme^{-th}\otimes\rme^{-tH}
$$
for all $t>0$. This is equivalent to
$\cp(H)=h\otimes\ind_{\Gamma(\ch)}+\ind_\ch\otimes H$.
\qed

\section{Mourre estimate for operators affiliated to $\rf(\co)$}
\label{s:me}
\protect\setcounter{equation}{0}

\noindent{\bf 1.}  We begin with some basic facts concerning the
Mourre estimate as presented in \cite[Ch.\ 7]{ABG}. Improvements of
the theory including an extension to conjugate operators $A$ which
are only maximal symmetric can be found in \cite{GGM} (this is
especially useful for the treatment of zero mass fields).

Fix a self-adjoint operator $A$ (the conjugate operator) on a
Hilbert space $\rh$. An  operator $S\in B(\rh)$ is of class $C^1(A)$
if the map $t\mapsto\rme^{-\rmi tA}S\rme^{\rmi tA}$ is strongly
$C^1$.  If this map is of class $C^1$ in norm, we say that $S$ is of
class $C^1_{\mathrm u}(A)$. It is easy to see that $S$ is of class
$C^1(A)$ if and only if the commutator $[A,S]$, which is well
defined as sesquilinear form on $D(A)$, extends to a bounded
operator $[A,S]^\circ$ on $\rh$.

Now let $H$ be a second self-adjoint operator on $\rh$ (the
Hamiltonian). We say that $H$ is of class $C^1(A)$ or
$C^1_{\mathrm{u}}(A)$ if $(H-z)^{-1}$ has the corresponding property
(here $z$ is any number not in the spectrum of $H$). It is possible
to characterize the $C^1(A)$ property in terms of the commutator
$[A,H]$, we recall here only what is strictly necessary (see
\cite{GGM}).  If $H$ is of class $C^1(A)$ then $D(H)\cap D(A)$ is a
core for $H$ and the commutator $[A,H]$, defined as sesquilinear
form on $D(H)\cap D(A)$, extends to a continuous sesquilinear form
$[A,H]^\circ$ on $D(H)$ equipped with the graph topology
\cite[Proposition 2.19]{GGM}. Moreover, we have:
\begin{equation}\label{eq:inv}
[A,(H-z)^{-1}]^\circ=-(H-z)^{-1}[A,H]^\circ(H-z)^{-1}.
\end{equation}
From now on we keep the notation $[A,H]$ for the extension
$[A,H]^\circ$.

We define $\wtilde\rho_H^A:\mbr\rarrow(-\infty,\infty]$ as
follows: $\wtilde{\rho}_H^A(\lambda)$ is the upper bound of the
numbers $a$ for which there are a real function $\theta\in
\Cc(\mbr)$ with $\theta (\lambda)\neq0$ and a compact operator $K$
such that
$$
\theta(H)[H, \rmi A]\theta(H)  \geq a \theta(H)^2 + K
$$ In other terms, $\wtilde{\rho}_H^A(\lambda)$ is the best constant
in the Mourre estimate. Then let $\rho_H^A(\lambda)$ be the upper
bound of the numbers $a$ such that the preceding inequality holds
for some $\theta$ and $K=0$.  So we get a second function
$\rho_H^A:\mbr\rarrow(-\infty,\infty]$ such that
$\rho_H^A\leq\wtilde{\rho}_H^A$. We have $\rho^A_H(\lambda)<\infty$
if and only if $\lambda\in\sigma(H)$ and
$\wtilde\rho^A_H(\lambda)<\infty$ if and only if $\lambda\in\se(H)$,
see Lemma 7.2.1 and Proposition 7.2.6 in \cite{ABG}. If
$\lambda\nin\tau_A(H)$ we say that \emph{$A$ is conjugate to $H$ at
$\lambda$}.

The two functions defined above are lower semi-continuous. Thus the
set $\tau_A(H)$ where $\wtilde{\rho}_H^A(\lambda)\leq0$ is closed and
will be called the set of $A$-{\em thresholds} of $H$.  The closed
set $\kappa_A(H)$ of $A$-{\em critical points} of $H$ is given by
the condition $\rho_H^A(\lambda)\leq0$.

Clearly $\tau_A(H)\subset\kappa_A(H)$. In order to understand how
much differ these sets we introduce the following notion.  Say that
$\lambda\in\mbr$ is an {\em M-eigenvalue} of $H$ if it is an
eigenvalue and $\wtilde {\rho}_{H}^A (\lambda )>0$.  By the virial
theorem, these eigenvalues are of finite multiplicity and are not
accumulation points of eigenvalues. Thus the set $\mu_A(H)$ of all
M-eigenvalues of $H$ is discrete.  The next result \cite[Theorem
7.2.13]{ABG} says that the functions $\rho_H^A$ and $\wtilde\rho_H^A$
differ only on the small set $\mu^A(H)$. Let $\sigma_{\text{p}}(H)$
be the set of eigenvalues of $H$.

\begin{proposition}\label{pr:imp}
We have $\rho _{H}^A (\lambda )=0$ if $\lambda$ is a M-eigenvalue of
$H$ and otherwise $\rho _H^A (\lambda )=\wtilde {\rho}_{H}^A
(\lambda)$.  Moreover, $\rho _{H}^A (\lambda )>0$ if and only if
$\wtilde {\rho}_{H} (\lambda )>0$ and $\lambda \notin
\sigma_{\text{p}}(H)$.  In particular ($\sqcup$ means disjoint
union):
\begin{equation}\label{eq:thre}
\kappa_A(H)=\tau_A(H)\cup\sigma_{\text{p}}(H)=
\tau_A(H)\sqcup\mu_A(H).
\end{equation}
\end{proposition}

We shall also need the following result, which is a particular case
of \cite[Theorem 8.3.6]{ABG} (see also \cite[Theorem 3.4]{BG1*} for
a simpler proof in an important particular case).

\begin{proposition}\label{pr:tens}
Let $\rh=\rh_1\otimes \rh_2$ and let $H_i,A_i$ be self-adjoint
operators on $\rh_i$ such that $H_i$ is bounded from below and of
class $C^1_{\mathrm u}(A_i)$. Consider the self-adjoint
operators $H=H_1\otimes1+1\otimes H_2$ and $A=A_1\otimes1+1\otimes
A_2$ on $\rh$. Then $H$ is of class $C^1_{\mathrm u}(A)$ and
\begin{equation}\label{eq:tens}
\rho_H^A(\lambda)=\inf_{\lambda=\lambda_1+\lambda_2}
\left[\rho_{H_1}^{A_1}(\lambda_1)+
\rho_{H_2}^{A_2}(\lambda_2)\right].  
\end{equation}
\end{proposition}

\noindent{\bf 2.}  We shall explain now how one may compute the
function $\wtilde\rho^A_H$ using $C^*$-algebra methods. This
technique has been introduced in \cite{BG2} in the context of the
$N$-body problem and further developed in \cite[Ch.\ 8]{ABG}.  The
main point of this approach is that it avoids the use of auxiliary
objects like partitions of unity. The presentation below is adapted
to our needs, that from \cite{BG2,ABG} is more general since it does
not require the quotient algebra to be represented on a Hilbert
space.

Let $\rc$ be a $C^*$-algebra such that $K(\rh)\subset\rc\subset
B(\rh)$. Then the quotient $C^*$-algebra $\wtilde\rc=\rc/K(\rh)$ is
well defined. If $H$ is a self-adjoint operator on $\rh$ affiliated
to $\rc$ then one can consider its image $\wtilde H=\cp(H)$ through
the canonical morphism $\cp:\rc\rarrow \wtilde{\rc}$. Then $\wtilde
H$ is an observable affiliated to $\wtilde{\rc}$ and the essential
spectrum of $H$ is equal to the spectrum of $\wtilde H$. We shall
assume that a faithful non-degenerate realization of $\wtilde{\rc}$
on some Hilbert space $\wtilde\rh$ is given and that the observable
$\wtilde H$ is realized as a self-adjoint operator (which we denote
also by $\wtilde H$) on $\wtilde{\rh}$.

Let $A$ be a self-adjoint operator on $\rh$ with $\rme^{-\rmi
tA}\rc\rme^{\rmi tA}=\rc$ for each real $t$ and such that the map
$t\mapsto\rme^{-\rmi tA}S\rme^{\rmi tA}$ be norm continuous for each
$S\in\rond C$. Since $\rme^{-\rmi tA}K(\rh)\rme^{\rmi tA}=K(\rh)$,
there is a norm continuous one-parameter group of automorphisms
$\alpha_t$ of $\wtilde{\rc}$ such that
$\cp\left(\rme^{-\rmi tA}S\rme^{\rmi tA}\right)=\alpha_t(\wtilde S)$ 
for all $t$ and $S\in \rc$.  Finally, assume that the group
$\alpha_t$ is unitarily implemented in the representation on
$\wtilde{\rh}$ (this is not needed in the more abstract theory
presented in \cite{BG2,ABG}). More precisely, our hypotheses are:
$$ \label{p:C}
\leqno{\mbox{\bf(CA)}}\hspace{1mm}
\left\{
\begin{array}{ll}
A\text{ is a self-adjoint operator on }\rh \text{ with }
\rme^{-\rmi tA}\rc\rme^{\rmi tA}=\rc \text{ for all } t;
\\[1mm]
\text{ the map } t\mapsto\rme^{-\rmi tA}S\rme^{\rmi tA}
\text{ is norm continuous for each } S\in\rond C;
\\[1mm]
\wtilde A \text{ is self-adjoint on } \wtilde{\rh} \text{ and }
\cp\left(\rme^{-\rmi tA}S\rme^{\rmi tA}\right)=
\rme^{-\rmi t\wtilde A}\,\cp(S)\,\rme^{\rmi t\wtilde A}
\text{ for all } t \text{ and } S\in \rc.
\end{array}
\right.
$$
The next proposition follows immediately from the
preceding definitions and comments.

\begin{proposition}\label{pr:mest}
Assume that $H$ is a self-adjoint operator on $\rh$ affiliated to
$\rc$ and of class $C^1_{\mathrm u}(A)$. If $\wtilde H$ is a
self-adjoint operator on $\wtilde \rh$ then $\wtilde H$ is of class
$C^1_{\mathrm u}(\wtilde A)$ and $\wtilde{\rho}_H^A=\rho_{\wtilde
H}^{\wtilde A}$.
\end{proposition}

\noindent{\bf 3.}  We shall apply the preceding general theory in
the situation of interest for us in this paper. Let $\ch$ be a
complex Hilbert space and $\co$ an abelian non-degenerate
$C^*$-algebra of operators on $\ch$ such that $\co''\cap
K(\ch)=\{0\}$. Let $\rh=\Gamma(\ch)$ be the symmetric or
antisymmetric Fock space over $\ch$ and $\rc=\rf(\co)$.  We shall
consider only conjugate operators of the form: \label{p:A}
$$
\leqno{\mbox{\bf(OA)}}\hspace{1mm}
\left\{
\begin{array}{ll}
A=\dd\Gamma(\g{a}) \text{ where }\g{a} \text{ is a self-adjoint
operator on } \ch \text{ such that }
\rme^{-\rmi t\g{a}}\co\rme^{\rmi t\g{a}}=\co 
\\[1mm] 
\text{ and such that the map } 
t\mapsto\rme^{-\rmi t\g{a}}S\rme^{\rmi t\g{a}} 
\text{ is norm continuous for all } S\in\co.
\end{array}
\right.
$$

\begin{lemma}\label{lm:cor}
We have $\rme^{-\rmi tA}\rf(\co)\rme^{\rmi tA}=\rf(\co)$ for all
real $t$ and  the map $t\mapsto\rme^{-\rmi tA}T\rme^{\rmi tA}$ is
norm continuous for all $T\in\rf(\co)$.
\end{lemma}
\proof
Note that $\rme^{\rmi tA}=\Gamma(\rme^{\rmi t\g{a}})$.
In the bosonic case it suffices to take $T=W(u)\Gamma(S)$ with
$u\in\ch$ and $S\in\co$ with $\|S\|<1$. Then, due to \eqref{eq:gf},
we have:
\begin{equation}\label{eq:act}
\rme^{-\rmi tA}T\rme^{\rmi tA}=W(\rme^{-\rmi t\g{a}}u)
\Gamma(\rme^{-\rmi t\g{a}}S\rme^{\rmi t\g{a}})
\end{equation}
and we get norm continuity by Lemma \ref{lm:eles}. In the
fermionic case we may assume $T=\phi^k(u)\Gamma(S)$ with $k=0,1$
and the argument is even simpler.  \qed

\begin{lemma}\label{lm:c1u}
Let $H$ be a self-adjoint operator affiliated to $\rf(\co)$. Then
$H$ is of class $C^1_{\mathrm u}(A)$ if and only if $H$ is of class
$C^1(A)$ and the operator $[A,(H-z)^{-1}]$ given by \eqref{eq:inv}
belongs to $\rf(\co)$.
\end{lemma}
\proof If $S=(H-z)^{-1}$ then $S(t)\equiv\rme^{-\rmi tA}S\rme^{\rmi
tA}$ belongs to $\rf(\co)$ for all real $t$. If $H$ is of class
$C^1_{\mathrm u}(A)$ then $[S,\rmi A]$ is the norm derivative at
$t=0$ of the map $t\mapsto S(t)$ hence belongs to $\rf(\co)$. On the
other hand, if $H$ is of class $C^1(A)$ then $[S(t),\rmi A]$ is the
strong derivative of the map $t\mapsto S(t)$ hence we have
$S(t)-S=\int_0^t \rme^{-\rmi \tau A}[S,\rmi A]\rme^{\rmi \tau A}$ in
the strong topology. If $[S,\rmi A]\in\rf(\co)$ then by Lemma
\ref{lm:cor} the integrand here is norm continuous, hence the
integral exists in norm, so $t\mapsto S(t)$ is norm $C^1$.
\qed

From Theorems \ref{th:cmo} and \ref{th:fcmo} and from relations like
\eqref{eq:act} (bosonic case) we get canonical identifications:
\begin{equation}\label{eq:real}
\wtilde\rc\equiv\cp(\rf(\co))\subset\co\otimes\rf(\co),\hspace{2mm}
\wtilde\rh=\ch\otimes\rh\equiv\ch\otimes\Gamma(\ch), \hspace{2mm}
\wtilde A=\g{a}\otimes\ind+\ind\otimes A.
\end{equation}

Our main result on the Mourre estimate for SQFH follows.

\begin{theorem}\label{th:mou}
Let $H$ be a SQFH of type $\co$ with one particle kinetic energy $h$
and one particle mass $m=\inf h>0$. Assume that condition (OA) from
page \pageref{p:A} is fulfilled, that $H$ is of class $C^1_{\mathrm
u}(A)$, and that $h$ is of class $C^1_{\mathrm u}(\g{a})$ and such
that $\rho^{\g{a}}_h\geq0$.  Then
$\kappa_{\g{a}}(h)=\tau_{\g{a}}(h)$, we have $\rho^A_H\geq0$ and:
\begin{equation}\label{eq:ths}
\tau_A(H)=
\big[\ccup_{n=1}^\infty \tau^n_{\g{a}}(h)\big]+\sigma_{\text{p}}(H),
\end{equation}
where $\tau^n_{\g{a}}(h)=\tau_{\g{a}}(h)+\dots+\tau_{\g{a}}(h)$ 
($n$ terms). Alternatively, if we set $H_0=\dd\Gamma(h)$ then:
\begin{equation}\label{eq:ths*}
\tau_A(H_0)=
\ccup_{n=1}^\infty \tau^n_{\g{a}}(h)
\hspace{2mm} \text{ and } \hspace{2mm}
\tau_A(H)=\tau_A(H_0)+\sigma_{\text{p}}(H). 
\end{equation}
\end{theorem}
\proof The operator $h$ cannot have eigenvalues of finite
multiplicity because the corresponding spectral projection would be
in $\co''$ which does not contain finite dimensional
projections. Hence from Proposition \ref{pr:imp} we get
$\wtilde\rho^{\g{a}}_h=\rho^{\g{a}}_h$, in particular
$\kappa_{\g{a}}(h)=\tau_{\g{a}}(h)$. Since $H$ is a SQFH we have
$\wtilde H=h\otimes\ind_{\Gamma(\ch)}+\ind_\ch\otimes H$. By taking
into account \eqref{eq:real} we deduce from Propositions
\ref{pr:mest} and \ref{pr:tens} that:
\begin{equation}\label{eq:qten}
\wtilde\rho_H^A(\lambda)=\inf_{\lambda=\lambda_1+\lambda_2}
\left[\rho_{h}^{\g{a}}(\lambda_1)+
\rho_{H}^{A}(\lambda_2)\right]
=\textstyle\inf_\mu
\left[\rho_{h}^{\g{a}}(\lambda-\mu)+ \rho_{H}^A(\mu)\right].  
\end{equation}
In this proof we simplify notations and set
$\wtilde\rho=\wtilde\rho_H^A$, $\rho=\rho_{H}^{A}$, and
$\rho_{h}=\rho_{h}^{\g{a}}$.  Also, without loss of generality, we
shall assume that $\inf H=0$.  Then $\se(H)\subset[m,\infty[$ due to
Theorem \ref{th:esz}.  Thus the functions $\rho$ on the interval
$\lambda<0$ and $\wtilde\rho$ and $\rho_h$ on $\lambda<m$ are equal
to infinity, in particular
\begin{equation}\label{eq:rten}
\wtilde\rho(\lambda)=\textstyle\inf_{0\leq\mu\leq\lambda-m}
\left[\rho_{h}(\lambda-\mu)+ \rho(\mu)\right] 
\end{equation}
with the convention that the infimum over an empty set is equal to
infinity. Observe that if $\lambda<m$ then $\lambda$ is either in
the resolvent set of $H$, and then $\rho(\lambda)=\infty$, or
$\lambda$ is in the discrete spectrum of $H$, hence is an
M-eigenvalue of $H$, so $\rho(\lambda)=0$ by Proposition
\ref{pr:imp}. Thus $\rho(\lambda)\geq0$ if $\lambda<m$.  Assume now
that we have shown that $\rho(\lambda)\geq0$ if $\lambda< km$ for
an integer $k\geq1$. If $\lambda<km+m$ then in \eqref{eq:rten}
only $\mu<km$ will appear and so $\rho(\mu)\geq0$. Since 
$\rho_{h}\geq0$ by hypothesis, we get $\rho(\lambda)\geq0$ if
$\lambda< (k+1)m$. By induction we finally obtain
$\rho(\lambda)\geq0$ for all $\lambda$.

We thus have $0\leq\rho\leq\wtilde\rho$ and $\rho_h\geq0$.  Hence
$\tau(H)\equiv\tau_A(H)$ is the set of $\lambda$ such that
$\widetilde\rho(\lambda)=0$ and $\kappa(H)\equiv\kappa_A(H)$ is the
set of $\lambda$ such that $\rho(\lambda)=0$. Moreover,
$\tau(h)\equiv\tau_{\g{a}}(h)=\kappa_{\g{a}}(h)$ is the set of
$\lambda$ such that $\rho_h(\lambda)=0$. Then the first equality in
\eqref{eq:qten} clearly gives: $\rho(\lambda)=0$ if and only if one
can write $\lambda=\lambda_1+\lambda_2$ with $\rho_h(\lambda_1)=0$
and $\rho(\lambda_2)=0$ (these functions are lower
semi-continuous). Finally, from \eqref{eq:thre} we obtain:
\begin{equation}\label{eq:sten}
\tau(H)=\tau(h)+\kappa(H)=\tau(h)+
\left[\tau(H)\cup\sigma_{\text{p}}(H)\right]=
\left[\tau(h)+\sigma_{\text{p}}(H)\right]\ccup
\left[\tau(h)+\tau(H)\right].
\end{equation}
This equation for the set $\tau(H)$ has as unique solution
$\ccup_{n=1}^\infty\big[\tau^n_{\g{a}}(h)+\sigma_{\text{p}}(H)\big]$
obtained by iteration. This gives \eqref{eq:ths}, for
\eqref{eq:ths*} note that $0$ is the only eigenvalue of $H_0$.  \qed

\begin{remark}\label{re:phy}{\rm
The relation \eqref{eq:ths} describing the set $\tau_A(H)$ of
$A$-thresholds of $H$ has a simple physical interpretation. It says
that an energy $\lambda$ is an $A$-threshold if and only if one can
write it as a sum $\lambda=\lambda_1+\dots+\lambda_n+\mu$ where the
$\lambda_k$ are $\g{a}$-threshold energies of the free particle and
$\mu$ is the energy of a bound state of the field. This means that
at energy $\lambda$ one can pull out $n$ free particles from the
field, each one having an $\g{a}$-threshold energy, such that the
field remains in a bound state.
}\end{remark}

\begin{remark}\label{re:count}{\rm
Outside the threshold set $\tau_A(H)$ one expects $H$ to have nice
spectral properties. A rather weak condition which implies the
absolute continuity of the spectrum of $H$ outside $\tau_A(H)$ (and
many other properties) is that $H$ be of class $\cc^{1,1}(A)$, which
means that the map $t\mapsto\rme^{-\rmi tA}(H+\rmi)^{-1}\rme^{\rmi
tA}$ is of Besov class $B_\infty^{1,1}$ in norm (this is slightly
more restrictive than the $C^1_{\text{u}}(A)$ class; the
boundedness of the double commutator $[A,[A,(H-z)^{-1}]]$ implies
it).  In particular, in order to exclude the existence of the
singularly continuous spectrum, it is important to be sure that
$\tau_A(H)$ is a small set. Note that $\tau_A(H)$ is always closed
and that it is countable if $\tau_{\g{a}}(h)$ is countable and $\ch$
separable.  In fact, in the most important physical cases we have
$\tau_{\g{a}}(h)=\{m\}$ and then
$\tau_A(H)=m\mbn^*+\sigma_{\text{p}}(H)$.  }\end{remark}

As an example, we consider the important particular case when $\ch$
is a Sobolev space over an Euclidean space $X=\mbr^s$, e.g.\
$\ch=L^2(X)$. The $P(\varphi)_2$ model as treated in \cite{DG2} is
covered by this example. Then we take $\co=\Co(X^*)$ (space of
continuous functions of the momentum operator $P$ which tend to zero
at infinity). A self-adjoint operator $h$ on $\ch$ with $\inf h=m>0$
is strictly affiliated to $\Co(X^*)$ if and only if $h=h(P)$ where
$h:X\rarrow\mbr$ is a continuous function such that
$|h(p)|\rarrow\infty$ when $|p|\rarrow\infty$.

We shall assume that $h:X\rarrow\mbr$ is a function of class $C^1$
in the usual sense. Let $\tau(h)$ be the set of critical values of
the function $h$ in the usual sense, i.e.\ the numbers of the form
$h(p)$ with $\nabla h(p)=0$.  In this context it is natural to
consider one particle conjugate operators of the form
$\g{a}=F(P)Q+QF(P)$ with $F$ a vector field of class
$\Cc^\infty(X)$. The corresponding operators $A=\rmd\Gamma(\g{a})$
will be called \emph{of class VF} (vector fields). The following is
a consequence of Theorem \ref{th:mou}.

\begin{corollary}\label{co:mou}
  In the preceding framework, let $H$ be a SQFH with one particle
  kinetic energy $h$. Then \mbox{\rm$\se(H)=[m+\inf H,\infty[$}.
  Assume that $H$ is of class $C^1_{\text{u}}(A)$ if $A$ is of class
  VF and let
\begin{equation}\label{eq:thse}
\tau(H)=
\big[\ccup_{n=1}^\infty \tau^n(h)\big]+\sigma_{\text{p}}(H),
\end{equation}
where $\tau^n(h)=\tau(h)+\dots+\tau(h)$ ($n$ terms).  Then $H$
admits a conjugate operator of class VF at each point not in
$\tau(H)$. If $H$ is of class $\cc^{1,1}(A)$ (e.g.\ if
$[A,[A,(H-z)^{-1}]]$ is bounded) for each operator $A$ of class VF
then $H$ has no singular continuous spectrum outside $\tau(H)$.
\end{corollary}

\begin{remark}\label{re:disp}{\rm
It is possible to prove the Mourre estimate for more general
Hamiltonians $H$ affiliated to $\rf(\co)$ if the operator $A$
satisfies the condition (OA).  We use again Proposition
\ref{pr:mest} by taking into account the identifications made in
\eqref{eq:real}. But now one step in the preceding arguments is
missing because in general $\widetilde H$ is no more representable
in the form $h\otimes\ind_{\Gamma(\ch)}+\ind_\ch\otimes H'$ with
operators $h$ and $H'$ affiliated to $\co$ and $\rf(\co)$
respectively, so we cannot use the Proposition \ref{pr:tens}.
However, by using the techniques from \cite[Sections 5 and 6]{DG*}
one can sometimes overcome this difficulty. For example, if
$\widetilde H=h\otimes M+\ind_\ch\otimes H'$ with $M\geq c>0$ then
one can proceed as in \cite[Section 6]{DG*} (in fact, the situation
here is much simpler). The main point is that Proposition
\ref{pr:mest} shows that we only have to estimate from below the
commutator $[\widetilde H,\rmi \widetilde A]$ which has the
following special structure:
\begin{equation}\label{eq:gme}
[\widetilde H,\rmi \widetilde A]=
[\widetilde H,\rmi \g{a}\otimes\ind_{\Gamma(\ch)}]+
[\widetilde H,\ind_\ch\otimes\rmi A].
\end{equation}
As already mentioned in the comments after Theorem \ref{th:ess}, if
$H$ is strictly affiliated to $\rf(\co)$ the quotient $\widetilde H$
is identified to a continuous family $\{\widetilde H(x)\}_{x\in\rx}$
of self-adjoint operators $\widetilde H(x)$ on $\Gamma(\ch)$
strictly affiliated to $\rf(\co)$. Since $\g{a}$ ``acts'' only on
the variable $x$ (by condition (OA)) and due to Lemma \ref{lm:cor},
each term on the right hand side of \eqref{eq:gme} formally belongs
to $\rf(\co)$ and one may impose conditions which ensure strict
positivity of the sum. All this can be done rigorously by working
with the resolvent of $H$ instead of $H$, as in \cite[Section
5]{DG*},  and in fact the situation here is simpler than in the case
of an $N$-body dispersive Hamiltonian. 
}\end{remark}

\section[Lagrangian subspaces of $\ch$ and QFH 
($P(\varphi)_2$ model)]
{QFH associated to Lagrangian subspaces of $\ch$}
\label{s:lgr}
\protect\setcounter{equation}{0}

Our purpose in this section is to show that Hamiltonians like that
of the $P(\varphi)_2$ model are covered by our formalism.  We shall
consider only the bosonic situation. We first recall another
classical procedure for constructing realizations of the Fock
representation of the CCR, the so-called \emph{field
realizations}. The idea is to use maximal abelian subalgebras of the
Weyl algebra $\rw(\ch)$ defined on page \pageref{p:w}. Note that
$\rw(\ch)$ depends (modulo canonical isomorphisms) only on the
symplectic structure of $\ch$ defined by the symplectic form
$\sigma(u,v)=\Im\lag u|v \rag$.  We recall that a real linear
subspace of $\ch$ is called \emph{isotropic} if $\sigma(u,v)=0$ for
all $u,v\in\ce$ and that a maximal isotropic subspace is called
\emph{Lagrangian}. A straightforward argument gives:

\begin{lemma}\label{lm:iso}
For any isotropic subspace $\ce$ we have $\ce\cap\rmi\ce=\{0\}$ and
$\|u+\rmi v\|^2=\|u\|^2+\|v\|^2$ for all $u,v\in\ce$; and $\ce$ is
Lagrangian if and only if $\ch=\ce+\rmi\ce$ and then $\ce$ is closed.
If $c$ is a conjugation (antilinear isometry such that $c^2=1$) then
$\ch_c=\{u\in\ch\mid cu=u\}$ is a Lagrangian subspace of $\ch$ and
reciprocally, each Lagrangian subspace of $\ch$ is of this form for
a uniquely determined $c$. 
\end{lemma}

%
%
%

For each real linear subspace $\ce\subset\ch$ let $\rw(\ce)$ be the
closed linear subspace of $\rw(\ch)$ generated by the operators
$W(u)$ with $u\in\ce$. This is obviously a $C^*$-subalgebra of
$\rw(\ch)$. 

\begin{lemma}\label{lm:iso1}
Let $\ce$ be a real linear subspace of $\ch$. Then $\rw(\ce)$ is
abelian if and only if $\ce$ is isotropic and $\rw(\ce)$ is maximal 
abelian in $\rw(\ch)$ if and only if $\ce$ is Lagrangian.
\end{lemma}

\proof Assume that $\rw(\ce)$ is abelian and let $u,v\in\ce$. From
\eqref{eq:ws} we get $\rme^{\rmi\Im\lag u|tv\rag}=1$ for all
$t\in\mbr$ hence $\Im\lag u|v\rag=0$, so $\ce$ is isotropic. If
$\ce$ is Lagrangian then $\rw(\ce)$ is maximal abelian in $\rw(\ch)$
because $\rw(\ce)''$ is maximal abelian on the Fock space
$\Gamma(\ch)$. Finally, assume that $\ce$ is not Lagrangian, so that
$\ck=\ce+\rmi\ce\neq\ch$. If $u\in\ch\setminus\ck$ then, as shown in
the proof of Proposition 5.2.9 from \cite{BR}, one has
$W(u)\nin\rw(\ck)$ so $W(u)\nin\rw(\ce)$. If $\ck$ is not dense in
$\ch$ we may choose $u\perp\ck$ and get $W(u)$ in the commutant of
$\rw(\ce)$ but not in $\rw(\ce)$. If $\ck$ is dense in $\ch$ then
$\ce$ cannot be closed and we choose $u$ in the closure of $\ce$ but
not in $\ce$. Since the closure of $\ce$ is isotropic we see that
$[W(u),W(v)]=0$ for all $v\in\ce$. But since the sum
$\ck=\ce+\rmi\ce$ is direct $W(u)\nin\rw(\ce)$.  \qed

In the rest of this section we fix a Lagrangian subspace $\ce$ of
$\ch$. It is not difficult to show that the Von Neumann algebra
$\rw(\ce)''$ generated by $\rw(\ce)$ on $\Gamma(\ch)$ is maximal
abelian and that $\Omega$ is a cyclic and separating vector for it.
Then $\lag T\rag=\lag\Omega |T\Omega \rag$ defines a faithful state
on $\rw(\ce)''$ and we denote $L^p(\ce)$ the $L^p$ spaces associated
to the couple $(\rw(\ce)'',\lag\cdot\rag)$. These spaces are
intrinsically defined by abstract integration theory \cite{Ne} and
can be realized as usual $L^p$ spaces over a probability measure
space $Q$ which we shall not specify\symbolfootnote[2]{\ We
emphasize that if $\ch$ is infinite dimensional one can never take
$Q=\ce$ in any natural sense, so the notation $L^p(\ce)$ could be
misleading. Of course, one may take $Q$ equal to the spectrum of the
$C^*$-algebra $\rw(\ce)$, but this is not a really convenient
choice. On the other hand, the theory of Gaussian cylindrical
measures on $\ce$ offers many useful realizations.} because this is
of no interest here (we refer to \cite{DG2,Si} for details on these
questions). However, we mention that at the abstract level we have
canonical identifications $L^\infty(\ce)=\rw(\ce)''$ and if $1\leq
p<\infty$ then $L^p(\ce)$ is the completion of $L^\infty(\ce)$ for
the norm $\|T\|_p=\lag|T|^p\rag^{1/p}$.  Moreover, from $\lag
W(v)^*W(u)\rag=\lag W(v)\Omega|W(u)\Omega\rag$ it follows that the
map $W(u)\mapsto W(u)\Omega$ extends to a unitary map
$L^2(\ce)\rarrow\Gamma(\ch)$ which will be used from now on to
identify these two Hilbert spaces. Thus we have
\begin{equation}\label{eq:qed}
\rw(\ce)''\equiv L^\infty(\ce)\subset L^p(\ce)\subset L^2(\ce)
\equiv\Gamma(\ch)\subset L^q(\ce)\subset L^1(\ce) \hspace{2mm}
\text{ if } 1<q<2<p<\infty.
\end{equation}

We get a realization on $L^2(\ce)$ of the Fock representation by
transport from $\Gamma(\ch)$ with the help of the identification map
defined above.  This \emph{$\ce$-realization} is a ``field
realization'' in the sense that the field operators $\phi(u)$ are
realized as operators of multiplication by (equivalence classes of)
real measurable functions defined on a probability space $Q$. Note
that the ``momentum operators'' defined by
$$
\pi(u)=\phi(\rmi u)=\rmi(a^*(u)-a(u)) \hspace{2mm} \text{ for }
u\in\ce
$$
can be realized as differential operators for certain choices of
$Q$. One has the commutation relations 
$$
[\phi(u),\phi(v)]=[\pi(u),\pi(v)]=0 \hspace{2mm}\text{ and }
\hspace{2mm} [\phi(u),\pi(v)]=2\rmi\lag u|v\rag \hspace{2mm}
\text{ if } u,v\in\ce.
$$

\begin{example}\label{ex:ken}{\rm
This is the most elementary situation which is of physical
interest. Let $h$ be a self-adjoint operator on $\ch$ which leaves
$\ce$ invariant (i.e.\ is real with respect to the conjugation
associated to $\ce$) and has pure point spectrum. Then there is an
orthonormal basis $\{e_k\}_{k\in K}$ of the real Hilbert space $\ce$
and a function $h:K\rarrow\mbr$ such that 
$h=\sum_kh(k)|e_k\rag\lag e_k|$ as operator on $\ch$. Let us set
$a_k=a(e_k)$, $\phi_k=\phi(e_k/\sqrt2)$, and
$\pi_k=\pi(e_k/\sqrt2)$. Then  $H_0=\rmd\Gamma(h)$ has the following
familiar expression: 
$$
H_0=\textstyle\sum_kh(k)\rmd\Gamma(|e_k\rag\lag e_k|)=
\textstyle\sum_kh(k)a^*_ka_k=
\frac{1}{2}\textstyle\sum_kh(k)(\pi_k^2+\phi_k^2-1)
$$
where $\phi_k,\pi_k$ are self-adjoint operators satisfying 
the commutation relations $[\phi_j,\phi_k]=[\pi_j,\pi_k]=0$ and
$[\phi_j,\pi_k]=\rmi\delta_{jk}$. This is the kinetic energy
operator of the (discretized) field and the total Hamiltonian is
obtained by adding a ``generalized polynomial'' $V$ in the field
operators $\phi_k$.
}\end{example}

We want to show that much more general Hamiltonians constructed by
procedures similar to that of Example \ref{ex:ken} are SQFH in our
sense. Let $\co$ be an abelian non-degenerate $C^*$-algebra on $\ch$
such that $\co''\cap K(\ch)=\{0\}$. In the statement of the next
result we use the terminology of abstract integration theory; we
refer to \cite{Ne} for a short review of the main facts.

\begin{theorem}\label{th:bit}
  Let $H_0=\dd\Gamma(h)$ where $h$ is a self-adjoint operator on
  $\ch$ affiliated to $\co$ and satisfying $m\equiv\inf h>0$ and
  $h^{-1}\ce\subset\ce$.  Let $V$ be a self-adjoint operator on
  $\Gamma(\ch)$ which is bounded from below, affiliated to
  $\rw(\ce)''$, and has the property $V\in L^p(\ce)$ for all
  $p<\infty$. Then $H_0+V$ is essentially self-adjoint on
  $D(H_0)\cap D(V)$ and its closure $H$ is a SQFH of type $\co$ with
  one particle kinetic energy $h$.
\end{theorem}
\proof We shall use Theorem \ref{th:shk} with $H_0=\dd\Gamma(h)$.
The conditions imposed on $h$ imply that $H_0$ generates a
hypercontractive semigroup due to Nelson's theorem \cite[Theorem
1.17]{Si}. Then $V$, viewed as function on $Q$, is admissible by
hypothesis, so $H$ is essentially self-adjoint on $D(H_0)\cap D(V)$.
Now assume that $V\in L^\infty=\rw(\ce)''$. Kaplansky's density
theorem \cite[Theorem 4.3.3]{Mu} implies that the closed ball of
radius $\|V\|$ in $\rw(\ce)$ is strongly dense in the closed ball of
radius $\|V\|$ in $\rw(\ce)''$. Since the function
$\ind\equiv\Omega$ belongs to $L^2$ it follows that there is a
\emph{sequence} $\{V_n\}$ of self-adjoint operators $V_n$ in
$\rw(\co)$ with $\|V_n\|\leq\|V\|$ such that
$\|V_n-V\|_{L^2}\rarrow0$. But we have
$\|V_n-V\|_{L^\infty}\leq2\|V\|$ hence we get by interpolation
$\|V_n-V\|_{L^p}\rarrow0$ for all $p<\infty$. Let $H_n=H_0+V_n$,
then Theorem \ref{th:shk} implies that $H_n\rarrow H$ in norm
resolvent sense. From Proposition \ref{pr:eqfh} it follows that each
$H_n$ is a SQFH hence $H$ is a SQFH of type $\co$ with one particle
kinetic energy $h$ by Proposition \ref{pr:nrl}. In the general case,
we consider the operators $V_n=\inf(V,n)\in L^\infty$ which
obviously have the properties required in Theorem \ref{th:shk}. Thus
$H_n\rarrow H$ in norm resolvent sense and we use again Proposition
\ref{pr:nrl}.  \qed

The preceding theorem covers $P(\varphi)_2$ models with a spatial
and an ultraviolet cutoff in any dimension. In space-time dimension
$2$ it is possible to remove the ultraviolet cutoff staying in the
Fock space. The fact that the corresponding Hamiltonian is a SQFH in
the sense of Definition \ref{df:dqfh} follows from:

\begin{theorem}\label{th:byte}
  Let $H_0$ be as in Theorem \ref{th:bit} and let $V$ be a
  self-adjoint operator on $\Gamma(\ch)$ affiliated to $\rw(\ce)''$
  with the property $V\in L^p(\ce)$ for all $p<\infty$. Assume that
  there is a sequence of operators $V_n$ with the same properties as
  $V$ and that there is some $q>2$ such that: (i) each $V_n$ is
  bounded from below; (ii) $\sup_n\|\rme^{-V_n}\|_{L^q}<\infty$;
  (iii) $\|V_n-V\|_{L^q}\rarrow0$.  Then $H_0+V$ is essentially
  self-adjoint on $D(H_0)\cap D(V)$ and its closure $H$ is a SQFH of
  type $\co$ with one particle kinetic energy $h$.
\end{theorem}
This follows immediately from Theorems \ref{th:bit} and \ref{th:shk}
and Proposition \ref{pr:nrl}.  Christian G\'erard sent
me\symbolfootnote[2]{\ By fax, on March 15, 2001 ({\it sic}).}
a short proof of the fact that the conditions of this theorem are
satisfied in the two dimensional $P(\varphi)_2$ model with a spatial
cutoff with $V_n$ defined with the help of ultraviolet cutoffs.

\section{Coupling of systems and Pauli-Fierz model}
\label{s:cs}
\protect\setcounter{equation}{0}

\noindent{\bf 1.}  Our treatment of the coupling between several
fields and other external systems is based on the following
elementary fact (which follows by induction from \cite[Theorem
2.3]{GI1}). By \emph{ideal} we mean a closed bilateral ideal.

\begin{proposition}\label{pr:int}
Assume that $\rc_1,\dots,\rc_n$ are nuclear
$C^*$-algebras equipped with ideals $\rj_1,\dots,\rj_n$. Let
$\cp_k:\rc_k\rarrow\widetilde{\rc_k}\equiv\rc_k/\rj_k$ be the
canonical surjection and let $\cp_k'=
\ind_{\rc_1}\otimes\dots\otimes\cp_k\otimes\dots\otimes\ind_{\rc_n}$
be the tensor product of this morphism with the identity maps, so
that 
$$
\cp_k':\rc_1\otimes\dots\otimes\rc_n\rarrow
\rc_1\otimes\dots\otimes\widetilde{\rc_k}\otimes\dots\otimes\rc_n
$$ 
is a morphism. Then the kernel of the morphism
$$
\cp\equiv\textstyle\bigoplus_{k=1}^n\cp'_k:
\rc_1\otimes\dots\otimes\rc_n\rarrow
\bigoplus_{k=1}^n
\rc_1\otimes\dots\otimes\widetilde {\rc_k}\otimes\dots\otimes\rc_n
$$
is equal to $\rj_1\otimes\dots\otimes\rj_n$.
\end{proposition}

\begin{corollary}\label{co:int}
Assume that each $\rc_k$ is realized on a Hilbert space $\rh_k$ and
$\rj_k=K(\rh_k)$. Let $H$ be a self-adjoint operator on
$\rh=\rh_1\otimes\dots\otimes\rh_n$ affiliated to
$\rc=\rc_1\otimes\dots\otimes\rc_n$ and let us denote $\widetilde
H_k=\cp'_k(H)$, which is an observable affiliated to
$\rc_1\otimes\dots\otimes\widetilde{\rc_k}\otimes\dots\otimes\rc_n$.
Then:
\begin{equation}\label{eq:cess}
\mbox{\rm$\se(H)$}=\ccup_k\sigma(\widetilde H_k).
\end{equation}
\end{corollary}
For this it suffices to note that
$K(\rh)=K(\rh_1)\otimes\dots\otimes K(\rh_n)$.

For simplicity we take $n=2$, we assume that we are in the framework
of Corollary \ref{co:int}, and that the quotient $\widetilde{\rc_k}$
is realized on a Hilbert space $\widetilde\rh_k$. Then
$\cp=\cp'_1\oplus\cp'_2$ gives an embedding of the quotient algebra
$\widetilde{\rc}=\rc/K(\rh)$ as follows:
\begin{equation}\label{eq:csm1}
\widetilde{\rc}\subset
\left(\widetilde{\rc_1}\otimes\rc_2\right) \oplus 
\left(\rc_1\otimes\widetilde{\rc_2}\right).
\end{equation} 
The $C^*$-algebra from right hand side is realized on the
Hilbert space 
\begin{equation}\label{eq:csm2}
\wtilde\rh=\left(\wtilde\rh_1\otimes\rh_2\right) \oplus
\left(\rh_1\otimes\wtilde\rh_2\right).
\end{equation}
Thus if $H$ is a self-adjoint operator on $\rh$ affiliated to $\rc$
then its image 
$\cp(H)=\widetilde H_1\oplus\widetilde H_2 \equiv \wtilde H$,
an observable affiliated to $\wtilde\rc$, is expected to be realized
as a self-adjoint operator on $\wtilde\rh$ (this is always the case
if we accept not densely defined self-adjoint operators).

We shall explain now how to prove the Mourre estimate in such
situations. We assume that the data 
$\rc_k,\cp_k,\rh_k,A_k,\widetilde \rh_k,\widetilde A_k$ 
satisfy condition (CA) page \pageref{p:C}.
If $A=A_1\otimes\ind_{\rh_2}+\ind_{\rh_1}\otimes A_2$ on $\rh$ then
$\rme^{\rmi tA}=\rme^{\rmi tA_1}\otimes\rme^{\rmi tA_2}$, hence
$\rme^{-\rmi tA}\rc\rme^{\rmi tA}=\rc$ and the map
$t\mapsto\rme^{-\rmi tA}T\rme^{\rmi tA}=\rc$ is norm continuous for
all $T\in\rc$. Let us set
\begin{equation}\label{eq:csm3}
A^\circ_1=\wtilde A_1\otimes\ind_{\rh_2}+
\ind_{\widetilde\rh_1}\otimes A_2, 
\hspace{2mm}
A^\circ_2=A_1\otimes\ind_{\widetilde\rh_2}+
\ind_{\rh_1}\otimes \widetilde A_2, 
\hspace{2mm}
\wtilde A=A^\circ_1 \oplus A^\circ_2.
\end{equation}
Then $\wtilde A$ is a self-adjoint operator on $\wtilde\rh$ such
that $\cp\left(\rme^{-\rmi tA}T\rme^{\rmi tA}\right)= \rme^{-\rmi t
\wtilde A}\cp(T)\rme^{\rmi t\wtilde A}$ for all $T\in\rc$.  So if
$H$ is of class $C^1_{\rm{u}}(A)$ then $\widetilde H$ is of class
$C^1_{\rm{u}}(\wtilde A\,)$, each $\wtilde H_k$ is of class
$C^1_{\rm{u}}(A^\circ_k)$. Let us set
$\rho_k=\rho^{A^\circ_k}_{\wtilde H_k}$. Then, by using Proposition
\ref{pr:mest} and \cite[Proposition 8.3.5]{ABG} we obtain:
\begin{equation}\label{eq:mmm}
\wtilde\rho^A_H=\rho^{\wtilde A}_{\wtilde H}= \min(\rho_1,\rho_2).
\end{equation} 
Thus we are reduced to finding estimates from below for the
functions $\rho_k$ which can be done by using its relation with the
corresponding function $\wtilde\rho_k$ as explained in the first
part of Section \ref{s:me}. For this we need to know more about the
operators $\widetilde H_k$ and we shall consider this question below
only in the much more elementary case of the Pauli-Fierz
Hamiltonians. Couplings with $N$-body systems as in
\cite{BFS,BFSS,Sk} should be covered by the preceding formalism (we
did not check the details).

\noindent{\bf 2.}  An often studied situation is that of a field
coupled with a small confined system. Confinement means that the
Hamiltonian of the small system has purely discrete spectrum, hence
we take as $C^*$-algebra of energy observables of the small system
the algebra of compact operators. Since taking tensor products with
a nuclear algebra preserves short exact sequences, we have slightly
more than in the general case.

\begin{proposition}\label{pr:cint}
Let $\rc$ be a $C^*$-algebra of operators on a Hilbert space $\rh$
such that $K(\rh)\subset\rc$ and let us denote
$\wwtilde{\rc}=\rc/K(\rh)$. Let $\rl$ be a second Hilbert space
and $H$ a self-adjoint operator on $\rh\otimes\rl$ affiliated to
$\rc\otimes K(\rl)$. Let $\widetilde H=\cp(H)$ where
$\cp\equiv\cp\otimes\Id:\rc\otimes K(\rl)\rarrow\widetilde\rc\otimes
K(\rl)$ is the canonical morphism. Then
\mbox{\rm$\se(H)=\sigma(\widetilde H)$}.
\end{proposition}

We apply this to a bosonic or fermionic field coupled with a
confined system. The next result is an immediate consequence of
Theorems \ref{th:cmo} and \ref{th:fcmo} and of the Proposition
\ref{pr:cint}.

\begin{theorem}\label{th:cint}
Let $\ch$ be a Hilbert space and $\co\subset B(\ch)$ a
non-degenerate abelian $C^*$-algebra such that 
$\co''\cap K(\ch)=\{0\}$. Let $\rl$ be a second Hilbert space and
$\rh=\Gamma(\ch)\otimes\rl$. Then there is a unique morphism
$\cp:\rf(\co)\otimes K(\rl)\rarrow\co\otimes\rf(\co)\otimes K(\rl)$
such that $\cp[(F\Gamma(A))\otimes L]=A\otimes(F\Gamma(A))\otimes L$
for all $F\in\rf(\ch)$, $A\in\co$ with $\|A\|<1$, and $L\in K(\rl)$.
One has $\ker\cp=K(\rh)$. If $H$ is a self-adjoint operator on $\rh$
affiliated to $\rf(\co)\otimes K(\rl)$ then
\mbox{\rm$\se(H)=\sigma(\cp(H))$}.
\end{theorem}

\begin{remark}\label{re:pf}{\rm
We shall adopt, in the framework of Theorem \ref{th:cint}, exactly
the same definition of \emph{standard QFH} as in Definition
\ref{df:dqfh}, we just replace the algebra $\rf(\co)$ with
$\rf(\co,\rl)$. Then clearly \emph{Theorem \ref{th:esz} remains true
without any change}. The conjugate operators which are well adapted
to the present situation are of the form $A\otimes\ind_\rl$ where
$A$ is as in assumption (OA) page \pageref{p:A}. We keep the notation
$A$ for them and note that \emph{Theorem \ref{th:mou} and Corollary
\ref{co:mou} remain valid without any change}.  
}\end{remark}

Our purpose now is to show that the Hamiltonians of the massive
Pauli-Fierz models are covered by Theorem \ref{th:cint}. We shall
consider the abstract version of this model introduced in \cite{DG1}
and further studied in \cite{Ge2,DJ,GGM*,BD}. We treat only the case
of a boson field, the fermionic case is easier (just replace $\vee$
by $\wedge$ and note that many assertions become obvious). The
following is a standard fact.

\begin{lemma}\label{lm:tp}
For each $p,q\in\mbn$ there is a unique linear continuous map
$\cs_{p,q}:\ch^{\vee p}\otimes\ch^{\vee q}
\rarrow\ch^{\vee(p+q)}$ such that $\cs_{p,q}(u\otimes v)=uv$ for
all $u\in\ch^{\vee p}$ and $v\in\ch^{\vee q}$. One has
$\|\cs_{p,q}\|=\binom{p+q}{p}^{1/2}$.
\end{lemma}

We consider the framework of Theorem \ref{th:cint} (bosonic case)
and take $\rf(\co,\rl)=\rf(\co)\otimes K(\rl)$ as algebra of energy
observable of our system. We recall \cite{DG1} that for each
operator $u\in B(\rl,\ch\otimes\rl)$ the creation operator $a^*(u)$
acting in $\rh$ is defined as the closure of the algebraic direct
sum of the operators
\begin{equation}\label{eq:oc}
a^*_n(u):\ch^{\vee n}\otimes\rl\rarrow\ch^{\vee(n+1)}\otimes\rl
\hspace{2mm}\text{ defined by }
a^*_n(u)=(\cs_{n,1}\otimes\ind_\rl)\circ
(\ind_{\ch^{\vee n}}\otimes u).
\end{equation}
The difference in coefficients with respect to \cite[(3.1)]{GGM*} is
due to our choice of scalar product in the Fock space. Since no
ambiguity may occur we shall identify $N=N\otimes\ind_\rl$. Then
clearly we have:
\begin{equation}\label{eq:oc1}
\|a^*(u)(N+1)^{-1/2}\|=\|u\|_{B(\rl,\ch\otimes\rl)}
\end{equation} 
Let $a(u)$ be the adjoint of the operator $a^*(u)$ and let
$\phi(u)=a(u)+a^*(u)$. The domains of these operators contain
$\rh_{\text{fin}}$, the algebraic direct sum of the spaces
$\ch^{\vee n}\otimes\rl$, and it is easy to see that $\phi(u)$ is
essentially self-adjoint on this domain; we use the same notation
for its closure. It is clear that the commutation relations
\eqref{eq:nf} remain valid. Below and later on we shall identify
$\Gamma(A)=\Gamma(A)\otimes\ind_\rl$ except in the situations when
the clarity of the text requires more precision.

\begin{lemma}\label{lm:oco}
If $u\in K(\rl,\ch\otimes\rl)$ and $A\in\co,\|A\|<1$, then
$a^{(*)}(u)\Gamma(A)\in\rf(\co,\rl)$ and
\begin{equation}\label{eq:oco}
\cp[a^{(*)}(u)\Gamma(A)]=A\otimes[a^{(*)}(u)\Gamma(A)]
\hspace{2mm}\text{ on } \ch\otimes\rh.
\end{equation}
\end{lemma}
\proof
From \eqref{eq:oc1} we get 
$$
\|a^{(*)}(u)\Gamma(A)\|\leq\|a^{(*)}(u)(N+1)^{-1/2}\|
\|(N+1)^{1/2}\Gamma(A)\|\leq C\|u\|_{B(\rl,\ch\otimes\rl)}
$$
hence the map $u\mapsto a^{(*)}(u)\Gamma(A)$ is norm continuous on
$B(\rl,\ch\otimes\rl)$. Thus it suffices to prove the assertions of
the lemma for $u$ of the form $u=f\otimes K$ with $f\in\ch$ and $K$
a compact  operator on $\rl$. More precisely, 
$u\in B(\rl,\ch\otimes\rl)$ is the defined by: $u(e)=f\otimes K(e)$.
Then it is easy to check that $a^{(*)}(u)=a^{(*)}(f)\otimes K$
hence 
$a^{(*)}(u)\Gamma(A)=[a^{(*)}(f)\Gamma(A)]\otimes K
\in\rf(\co)\otimes K(\rl)$.
\qed

\begin{lemma}\label{lm:oc} For each $u\in B(\rl,\ch\otimes\rl)$ the
  following relations are satisfied.\\
  {\rm(i)} Let $S,T\in B(\rl)$ and $A\in B(\ch)$ with $\|A\|<1$.
  Then
\begin{equation}\label{eq:oc2}
(\Gamma(A)\otimes S)a^*(u)(\ind_{\Gamma(\ch)}\otimes T)=
a^*((A\otimes S)uT)(\Gamma(A)\otimes\ind_\rl).
\end{equation}
{\rm(ii)} Let $h, L$ be self-adjoint operators on $\ch$ and $\rl$
respectively such that $h\geq m>0$ and $L\geq0$ and let
$H_0=\dd\Gamma(h)\otimes\ind_\rl+\ind_{\Gamma(\ch)}\otimes L$.
Then for all \mbox{\rm $f\in\rh_{\text{fin}}$} and all numbers $r>0$
we have: 
\begin{equation}\label{eq:oc3}
|\lag f|\phi(u)f\rag|\leq C(u,r)\lag f| (H_0+r)f\rag
\end{equation}
where $C(u,r)=\|(h^{-1/2}\otimes\ind_\rl)u(L+r)^{-1/2}\|^2$ and the
right hand side is allowed to be $+\infty$.
\end{lemma}
The proof of (i) is a mechanical application of the definitions;
note that both sides of \eqref{eq:oc2} are bounded operators.  The
second assertion is a particular case of \cite[Proposition
4.1]{GGM*}, but see also \cite[Proposition 4.1]{DJ} and
\cite[Theorem 2.1]{BD}.

The second part of the Lemma \ref{lm:oc} allows us to define
$\phi(u)$ as a continuous sesquilinear form on $D(H_0^{1/2})$ for an
arbitrary continuous linear map\symbolfootnote[2]{\ 
The theory of Pauli-Fierz Hamiltonians for such ``form factors'' has
first been developed in \cite{BD}, but we shall not follow their
method. However, the reader might prefer the direct arguments
and the more detailed presentation from \cite{BD}.}
$u:\rl_1\rarrow \ch_1^*\otimes\rl$.  Here
$\rl_1=D(L^{1/2})$ and $\ch_1=D(h^{1/2})$ are equipped with the
graph topologies, $\ch_1^*$ is the space adjoint to $\ch_1$, and we
embed as usual $\ch_1\subset\ch\subset \ch_1^*$. Then
$B(\rl,\ch\otimes\rl)\subset B(\rl_1,\ch_1^*\otimes\rl)$ densely in
the strong operator topology and if $B(R)$ is the closed ball of
radius $R$ in $B(\rl_1,\ch_1^*\otimes\rl)$ then $B_0(R)=B(R)\cap
B(\rl,\ch\otimes\rl)$ is strongly dense\,\symbolfootnote[3]{\
Indeed, it suffices to approximate $T$ with $[(\ind+\varepsilon
h)^{-1}\otimes\ind_\rl] T(\ind+\varepsilon L)^{-1}$} in $B(R)$.

Let, for example, $\rd$ be the symmetric algebra over $\ch_1$
algebraically tensorized with $\rl_1$. This is a core for
$H_0^{1/2}$ consisting of linear combinations of decomposable
vectors. Fix $f\in\rd$ and consider the map $u\mapsto\lag
f|\phi(u)f\rag$ defined for the moment only on
$B(\rl,\ch\otimes\rl)$. It is clear from the definition
\eqref{eq:oc} that this map is continuous for the strong operator
topology induced by $B(\rl_1,\ch_1^*\otimes\rl)$.  Thus, by the
preceding considerations, \eqref{eq:oc3} remains valid for $u\in
B(\rl_1,\ch_1^*\otimes\rl)$ with the same constant $C(u,r)$.

One can define $\phi(u)$ in a second way (which below gives the same
$H$). The graph norm on $\ch_1$ defined by $h^{1/2}$ is such that
the embedding $\ch_1\subset\ch$ is contractive. Then we get
injective contractive linear maps 
$\ch_1\hookrightarrow\ch\hookrightarrow\ch^*_1$
hence contractive dense embeddings
$\Gamma(\ch_1)\subset\Gamma(\ch)\subset\Gamma(\ch^*_1)$. On the
other hand, we have a natural identification
$\Gamma(\ch_1)^*=\Gamma(\ch^*_1)$.  If $u:\rl_1\to\ch_1^*\otimes\rl$
then \eqref{eq:oc} clearly gives a continuous map
$a^*_n(u):\ch^{\vee
n}\otimes\rl_1\to(\ch_1^*)^{\vee(n+1)}\otimes\rl$ hence we obtain as
usual a linear map $a^*(u):\Gamma_{\text{fin}}(\ch)\otimes\rl_1\to
\Gamma_{\text{fin}}(\ch_1^*)\otimes\rl$. Then we define $\phi(u)$ as
a quadratic form on $\Gamma_{\text{fin}}(\ch_1)\otimes\rl_1$ (which
is a core for $H_0$) by taking 
$\lag f|\phi(u)f\rag=2\Re\lag f|a^*(u)f\rag$.

We summarize below our assumptions concerning massive Pauli-Fierz
models: \label{p:pf}

$$
\leqno{\mbox{\bf(PF)}}\hspace{1mm}
\left\{
\begin{array}{ll}
\ch \text{ and } \rl \text{ are Hilbert spaces, }
\Gamma(\ch) \text{ is the symmetric Fock space, }
\rh=\Gamma(\ch)\otimes\rl;
\\[3mm]
\co\subset B(\ch) \text{ is a non-degenerate abelian $C^*$-algebra
such that } \co''\cap K(\ch)=\{0\};
\\[3mm]
h\geq m>0 \text{ is a self-adjoint operator on } \ch 
\text{ strictly affiliated to } \co;
\\[3mm] 
L\geq 0\text{ is a self-adjoint operator on } \rl
\text{ with purely discrete spectrum};
\\[3mm]
v\in B(D(L^{1/2}),D(h^{1/2})^*\otimes\rl) \text{ is such that }
\lim_{r\rarrow\infty}C(v,r)<1;
\\[3mm]
(h+L)^{-\alpha}v(L+1)^{-1/2}
\hspace{1mm}\text{ and }\hspace{1mm}
(h+L)^{-1/2}v(L+1)^{-\alpha}
\text{ are compact operators if } \alpha>1/2.
\end{array}
\right.
$$
Here and later we use the abbreviation 
$h+L=h\otimes\ind_\rl+\ind_\ch\otimes L$. 

\begin{theorem}\label{th:pf}
Assume that conditions (PF) are fulfilled. Then
$H_0=\dd\Gamma(h)\otimes\ind_\rl+\ind_{\Gamma(\ch)}\otimes L$ is a
positive self-adjoint operator on $\rh$ strictly affiliated to
$\rf(\co,\rl)$ and $\phi(v)$ is a symmetric quadratic form on
$D(H_0^{1/2})$ such that $\pm\phi(v)\leq aH_0+b$ for some $0<a<1$,
$b>0$. The form sum $H=H_0+\phi(v)$ is a self-adjoint operator on
$\rh$ strictly affiliated to $\rf(\co,\rl)$ and $H$ is a standard
QFH with $h$ as one particle kinetic energy (see Remark
\ref{re:pf}). In particular
$\mbox{\rm$\se(H)$}=\sigma(h)+\sigma(H)$. Finally, assume that $A$
is as in condition (A) page \pageref{p:A} and let us identify
$A\otimes\ind_\rl=A$. If $H$ is of class $C^1_{\rm{u}}(A)$ and $h$
is of class $C^1_{\rm{u}}(\g{a})$ with $\rho^{\g{a}}_h\ge0$, then
the conclusions of Theorem \ref{th:mou} are valid. 
\end{theorem}
\proof We assume, without loss of generality, that $L\geq1$.  We
have
$\rme^{-tH_0}=\Gamma(\rme^{-th})\otimes\rme^{-tL}\in\rf(\co,\rl)$
for all $t>0$ and strict affiliation follows by noting that
$\|\rme^{-tH_0}T\otimes K-T\otimes K\|\rarrow0$ if $t\rarrow0$ for
all $T\in\rf(\co)$ and $K\in K(\rl)$, see the proof of Lemma
\ref{lm:ndg}. The assertion concerning the existence of $H$ as
self-adjoint operator is clear by the preceding discussion (see also
\cite{BD}). We shall now prove the strict affiliation of $H$ to
$\rf(\co,\rl)$ and we do this by checking the conditions of Theorem
\ref{th:dag}, more precisely we shall prove that
$\theta(H_0)\phi(v)H_0^{-1/2}\in\rf(\co,\rl)$ if
$\theta\in\Co(\mbr)$. We shall prove by two different methods that
$\rme^{-H_0}a^*(v)H_0^{-1/2}\in\rf(\co,\rl)$ and
$H_0^{-1/2}a^*(v)\rme^{-H_0}\in\rf(\co,\rl)$, which clearly
suffices.

We first show that $LH_0^{-1}$ belongs to the multiplier algebra of
$\rf(\co,\rl)$, where $L\equiv \ind_{\Gamma(\ch)}\otimes L$.
It suffices to prove that 
$(LH_0^{-1})(S\otimes T)\in\rf(\co)\otimes K(\rl)$ 
for dense sets of operators $S$ and $T$ in $\rf(\co)$ and $K(\rl)$
respectively. Note that the linear span of the operators $T=L^{-1}K$
with $K$ compact on $\rl$ is dense in $K(\rl)$ because it contains
the rank one operators of the form $|f \rag\lag g|$ with $f$ in the
range of $L^{-1}$, which is dense in $\rl$. Since 
$(LH_0^{-1})(S\otimes T)=H_0^{-1}(S\otimes K)$ for such $T$, it
suffices to prove that 
$\rme^{-H_0}(S\otimes K)\in\rf(\co)\otimes K(\rl)$, because then 
this will remain valid if $\rme^{-H_0}$ is replaced by any
$\theta(H_0)$ with $\theta\in\Co(\mbr)$. But
$\rme^{-H_0}(S\otimes K)=(\Gamma(\rme^{-h})S)\otimes(\rme^{-L}K)$
clearly belongs to $\rf(\co)\otimes K(\rl)$.

Now by using \eqref{eq:oc2} we get:
\begin{eqnarray*}
\rme^{-H_0}a^*(v)H_0^{-1/2} &=& (\Gamma(\rme^{-h})\otimes\rme^{-L})
a^*(v)(\ind_{\Gamma(\ch)}\otimes L^{-1/2}) \cdot (LH_0^{-1})^{1/2} \\
&=&
a^*(\rme^{-h-L}vL^{-1/2})\Gamma(\rme^{-h}) \cdot (LH_0^{-1})^{1/2}
\end{eqnarray*}
where $LH_0^{-1}$ is interpreted as above. Since
$\rme^{-h-L}vL^{-1/2}$ is compact we can use Lemma \ref{lm:oco} and
then it suffices to note that $(LH_0^{-1})^{1/2}$ is also a
multiplier for the algebra $\rf(\co,\rl)$.

Next we consider the case of $H_0^{-1/2}a^*(v)\rme^{-H_0}$. In order
to simplify the writing we shall sometimes identify
$\ind_n\equiv\ind_n\otimes\ind_\rl$ and similarly for $\ind^n$.
Since $H_0\ind_n^\perp\geq (n+1)m\ind_n^\perp$ we easily see that
$H_0^{-1/2}a^*(v)\rme^{-H_0}$ is the norm limit as $n\rarrow\infty$
of $H_0^{-1/2}a^*(v)\rme^{-H_0}\ind_n$. But $\ind_n$ is a finite sum
of projections $\ind^k$, so it suffices to show that $T\equiv
H_0^{-1/2}a^*(v)\rme^{-H_0}\ind^n$ belongs to $\rf(\co,\rl)$ for
each $n$. From \eqref{eq:oc} we get:
\begin{eqnarray*}
T &=& H_0^{-1/2}(\cs_{n,1}\otimes\ind_\rl)(\ind^n\otimes v)
\left[\Gamma(\rme^{-h})\ind^n\right]\otimes\rme^{-L}
\\ &=&
H_0^{-1/2}\left(\cs_{n,1}\otimes\ind_\rl\right)
\left(\ind^n\otimes M\right)
\left(\ind^n\otimes\left[M^{-1}vL^{-\alpha}\right]\right)
\Big(\Gamma(\rme^{-h})\otimes L^{\alpha}\rme^{-L}\Big).
\end{eqnarray*}
where $M=h^{1/2}+L^{1/2}$ is an operator acting in $\ch\otimes\rl$
such that $(h+L)^{1/2}\le M\le \sqrt2(h+L)^{1/2}$.  Thus, by
hypothesis, $v_0=M^{-1}vL^{-\alpha}$ is a compact operator
  $\rl\to\ch\otimes\rl$. 
In the rest of
this proof we realize $\ch^{\vee k}$ as the subspace of
$\ch^{\otimes k}$ consisting of symmetric tensors (the norm being
modified by a factor $\sqrt{k!}$, but this does not matter here),
and then we have $H_0^{-1/2}\left(\cs_{n,1}\otimes\ind_\rl\right)=
\left(\cs_{n,1}\otimes\ind_\rl\right) H_0^{-1/2}$ in a natural sense
and we have:
$$
T= (\cs_{n,1}\otimes\ind_\rl) H_0^{-1/2}(\ind^n\otimes M)
(\ind^n\otimes v_0)
\Big(\Gamma(\rme^{-h})\otimes L^{\alpha}\rme^{-L}\Big).
$$ 
The operator $ H_0^{-1/2}(\ind^n\otimes M)$, acting in
$\ch^{\otimes(n+1)}\otimes\rl$, is bounded and $v_0$ is norm limit
of linear combinations of operators of the form $u_0\otimes K_0$
where $u_0\in\ch$ and $K_0\in K(\rl)$ (see the proof of Lemma
\ref{lm:oco}).  Thus it suffices to prove that $T\in\rf(\co,\rl)$
under the assumption $v_0=u_0\otimes K_0$ and clearly we may also
assume $u_0\in D(h^{1/2})$ and $K=L^{1/2}K_0$ compact. If we set
$u=h^{1/2}u_0$ then we obtain:
\begin{eqnarray*}
T &=& H_0^{-1/2}(\cs_{n,1}\otimes\ind_\rl)(\ind^n
\otimes[u\otimes K_0+u_0\otimes K])
\Big(\Gamma(\rme^{-h})\otimes L^{\alpha}\rme^{-L}\Big)\\
&=& H_0^{-1/2}a^*(u\otimes K_0+u_0\otimes K)
\cdot \Gamma(\rme^{-h})\otimes \ind_\rl \cdot 
\ind_{\Gamma(\ch)}\otimes \left(L^{\alpha}\rme^{-L}\right).
\end{eqnarray*}
From Lemma \ref{lm:oco}, and since
$\ind_{\Gamma(\ch)}\otimes \left(L^{\alpha}\rme^{-L}\right)$ is
multiplier for $\rf(\co,\rl)$, we get $T\in\rf(\co,\rl)$.

To prove that $H$ is a SQFH it remains to show that
$\cp(H)=h\otimes\ind_\rh+\ind_\ch\otimes H$ (then the formula for
the essential spectrum is a consequence, cf.\ Remark
\ref{re:pf}). Let $\lambda\geq0$ real and let us set
$\Lambda=(H_0+\lambda)^{-1/2}$ (recall that in this proof we assume
$H_0\ge1$) and $U=\Lambda\phi(v)\Lambda$. By Theorem \ref{th:dag}
and by what we proved above, $U$ belongs to the multiplier algebra
$\rM$ of $\rf(\co,\rl)$. Indeed, this argument gives directly
$\Lambda\in\rM$ if $\lambda=0$ and for the general case it suffices
to write
$U=(H_0^{1/2}\Lambda)(H_0^{-1/2}\phi(v)H_0^{-1/2})(H_0^{1/2}\Lambda)$
and to note that $H_0^{1/2}\Lambda\in\rM$ because $H_0$ is strictly
affiliated to $\rf(\co,\rl)$.
We have $\rme^{-H_0}=\Gamma(\rme^{-h})\otimes\rme^{-L}$ hence from
Theorem \ref{th:cint} we get 
$\cp\left(\rme^{-H_0}\right)=\rme^{-h}\otimes\rme^{-H_0}$ hence
$$
\widetilde  H_0\equiv\cp(H_0)=h\otimes\ind_\rh+\ind_\ch\otimes H_0,
\hspace{2mm}
\widetilde \Lambda\equiv\cp(\Lambda)=
\left(\widetilde H_0+\lambda\right)^{-1/2}.
$$
We shall prove below that
\begin{equation}\label{eq:U}
\wtilde U\equiv\cp(U)=
\wtilde\Lambda(\ind_\ch\otimes\phi(v))\wtilde\Lambda\equiv
\wtilde\Lambda\wtilde\phi(v)\wtilde\Lambda
\end{equation}
where $\cp$ is canonically extended to $\rM$ as mentioned before
Lemma \ref{lm:maf}. Assuming that this has been done, choose
$\lambda$ such that $\|U\|<1$ (this is possible because
$\pm\phi(v)\leq aH_0+b$ with $a<1$). Then clearly we have a norm
convergent expansion
$$
(H+\lambda)^{-1}=\Lambda(1+U)^{-1}\Lambda=
\textstyle\sum(-1)^n\Lambda U^n\Lambda
$$
which implies
$$
\cp\left((H+\lambda)^{-1}\right)=
\textstyle\sum(-1)^n\cp(\Lambda)\cp(U)^n\cp(\Lambda)=
\textstyle\sum(-1)^n\wtilde\Lambda\wtilde U^n \wtilde\Lambda=
(\wtilde H+\lambda)^{-1}
$$
where $\wtilde H=\wtilde H_0+\wtilde\phi(v)$ and this finishes the
proof of the relation $\cp(H)=h\otimes\ind_\rh+\ind_\ch\otimes H$.
Note that $\pm\wtilde\phi(v)\leq a\wtilde H_0+b$ with the same $a,b$
as above.

It remains to prove \eqref{eq:U}. Since
$a^*(v)=(\phi(v)+\rmi\phi(\rmi v))/2$ we have 
$\Lambda a^*(v)\Lambda\in\rM$ and its adjoint is 
$\Lambda a(v)\Lambda$. Thus \eqref{eq:U} is a consequence of
\begin{equation}\label{eq:Ua}
\cp(\Lambda a^*(v) \Lambda)=
\wtilde\Lambda(\ind_\ch\otimes a^*(v))\wtilde\Lambda, 
\end{equation}
which is what we show now. From \eqref{eq:oc2} we have
$$
\rme^{-H_0}a^*(v)\left(\ind_{\Gamma(\ch)}\otimes L^{-1/2}\right)=
a^*\left(\rme^{-h-L}vL^{-1/2}\right)
\left(\Gamma(\rme^{-h})\otimes\ind_\rl\right).
$$
The operator $\ind_{\Gamma(\ch)}\otimes L^{-1/2}$ belongs
to $\rM$ and it is easy to check that
$$
\cp\left(\ind_{\Gamma(\ch)}\otimes L^{-1/2}\right)=
\ind_\ch\otimes\ind_{\Gamma(\ch)}\otimes L^{-1/2}.
$$
From now on we simplify notations and no more write the tensor
product symbols when they are obvious from the context. Then:
$$
\rme^{-\wtilde H_0}\cp\left(\Lambda a^*(v)\Lambda\right)L^{-1/2}=
\cp\left(\rme^{-H_0}\Lambda a^*(v)\Lambda L^{-1/2} \right) = 
\cp\left(\Lambda a^*\left(\rme^{-h-L}vL^{-1/2}\right)
\Gamma(\rme^{-h}) \Lambda\right).
$$
Due to \eqref{eq:oco} this is equal to:
$$
\cp(\Lambda) \cp\left(a^*\left(\rme^{-h-L}vL^{-1/2}\right)
\Gamma(\rme^{-h}) \right) \cp(\Lambda)=
\wtilde\Lambda \cdot \rme^{-h}\otimes\left[ 
a^*\left(\rme^{-h-L}vL^{-1/2}\right) \Gamma(\rme^{-h})\right]
\cdot \wtilde\Lambda
$$
which in turn is equal to
$$
\wtilde\Lambda \cdot \rme^{-h}\otimes\left[ \rme^{-H_0}
a^*(v)L^{-1/2}\right] \cdot \wtilde\Lambda=
\wtilde\Lambda \rme^{-\wtilde H_0} (\ind_\ch\otimes a^*(v))
 L^{-1/2}  \wtilde\Lambda.
$$
Thus we have proved:
$$
\rme^{-\wtilde H_0}\cp\left(\Lambda a^*(v)\Lambda\right)L^{-1/2}=
\wtilde\Lambda \rme^{-\wtilde H_0} (\ind_\ch\otimes a^*(v))
 L^{-1/2}  \wtilde\Lambda=
\rme^{-\wtilde H_0} \wtilde\Lambda (\ind_\ch\otimes a^*(v)) 
 \wtilde\Lambda  L^{-1/2}.
$$
Since the operators $\rme^{-\wtilde H_0}$ and $L^{-1/2}$ are
injective, we get \eqref{eq:Ua}.

The last assertion of the theorem concerns the Mourre estimate and
is clear by the Remark \ref{re:pf}.  
\qed

\begin{remark}\label{re:bd}{\rm
We note that the description of the essential spectrum given in
Theorem \ref{th:pf} is an improvement of the \emph{massive} case of
\cite[Theorem 2.3]{BD}, where it is assumed that
$h^{-1/2}v(L+1)^{-1/2}$ is compact, but not of \cite[Proposition
4.9]{GGM*}, which does not require $(L+1)^{-1}$ to be compact.
}\end{remark}

\section{Systems with a particle number cutoff}
\label{s:cut}
\protect\setcounter{equation}{0}

In this section we fix an abelian non-degenerate $C^*$-algebra $\co$
of operators on the infinite dimensional space $\ch$ with $\co''\cap
K(\ch)=\{0\}$ and let $\Gamma$ be the symmetric or antisymmetric
Fock space functor. We are interested in models where the number of
particles is at most $n$, a given positive integer. Then the Hilbert
space of the states of the system is $\Gamma_n(\ch)$ and the algebra
of energy observables must be a $C^*$-algebra of operators on this
space.  Let $\rk_n(\ch)=K(\Gamma_n(\ch))$ be the algebra of compact
operators on $\Gamma_n(\ch)$.

We define for each integer $n\geq0$ a $C^*$-subalgebra of
$\rf(\co)$ by the following rule: 
\begin{equation}\label{eq:ran}
\rf_n(\co)=\ind_n\rf(\co)\ind_n.
\end{equation}
Let $\rf_n(\co)=0$ for $n<0$. Thus $\rf_n(\co)$ lives in the
subspace $\Gamma_n(\ch)$ (i.e.\ it is non-degenerate on
$\Gamma_n(\ch)$ and its restriction to the orthogonal subspace is
zero) and: 
\begin{equation}\label{eq:ram}
\rf_0(\co)=\mbc\omega,\ \rf_n(\co)\subset\rf_{n+1}(\co)
\text{ and }\rf(\co)=\overline{\cup_n\rf_n(\co)}.
\end{equation}
Note that $\rk_n(\ch)=\ind_n\rk(\ch)\ind_n$ and this is an ideal of
$\rf_n(\co)$.  

In particular, the algebra
$\ra_n(\ch)=\rf_n(\mbc\ind_\ch)=\ind_n\ra(\ch)\ind_n$ is a
$C^*$-subalgebra of $\ra(\ch)$ which lives in the subspace
$\Gamma_n(\ch)$, has $\ind_n$ as unit element, and contains
$\rk_n(\ch)$ as an ideal. Moreover:
\begin{equation}\label{eq:rac}
\ra_0(\ch)=\mbc\omega,\hspace{2mm} \ra_n(\ch)\subset\ra_{n+1}(\ch),
\hspace{2mm} \ra(\ch)=\overline{\cup_n\ra_n(\ch)}.
\end{equation}
These algebras can be defined independently of the material from the
preceding sections. First, it is not difficult to prove that
$\ra_n(\ch)$ is the unital $C^*$-algebra generated by the operators
$\phi_n(u)=\ind_n\phi(u)\ind_n$. If $\Gamma_n(\co)$ is the
$C^*$-algebra generated by the operators $\Gamma_n(S)=\oplus_{k\leq
n} S^{\vee k}$ with $S\in\co$, then we have
$\rf_n(\co)=\llb\ra_n(\ch)\cdot\Gamma_n(\co)\rrb$.  With
$\cp_n=\cp|\rf_n(\co)$, we get from Theorems \ref{th:cmo} and
\ref{th:fcmo}:

\begin{proposition}\label{pr:fnp}
There is a unique morphism
$\cp_n:\rf_n(\co)\to\co\otimes\rf_{n-1}(\co)$ such that
\begin{equation}\label{eq:fnp}
\cp_n\left(\phi_n(u)^k\Gamma_n(S)\right)=
S\otimes\left(\phi_{n-1}(u)^k\Gamma_{n-1}(S)\right)
\end{equation}
 for all
$u\in\ch,k\geq0,S\in\co$.  
We have $\ker(\cp_n)=\rk_n(\ch)$, hence we get canonical embedding:
\begin{equation}\label{eq:qn}
\rf_n(\co)/\rk_n(\ch)\hookrightarrow\co\otimes\rf_{n-1}(\co).
\end{equation}
\end{proposition}

The case of the algebras $\ra_n(\ch)$ is particularly nice
(we use Remark \ref{re:skan}):

\begin{corollary}\label{co:anice}
There is a unique morphism $\cp_n:\ra_n(\ch)\rarrow\ra_{n-1}(\ch)$
such that $\cp_n[\phi_n(u)]=\phi_{n-1}(u)$ for all $u\in\ch$.  This
morphism is unital, surjective, it has $\rk_n(\ch)$ as kernel, and
is explicitly given by:
\begin{equation}\label{eq:anc2}
\cp_n(T)=\slim_{e\rarrowup0}a(e)Ta^*(e)
\hspace{3mm} \text{ for all } T\in\ra_n(\ch).
\end{equation}
Thus we get a sequence of canonical surjective morphisms
\begin{equation}\label{eq:nice}
0\leftarrow\ra_0(\ch)\leftarrow\ra_{1}(\ch)\dots
\leftarrow\ra_{n-1}(\ch)\leftarrow\ra_{n}(\ch)\leftarrow\cdots
\end{equation}
which induce canonical isomorphisms
$\ra_n(\ch)/\rk_n(\ch)\cong\ra_{n-1}(\ch)$. 
\end{corollary}

\begin{remark}\label{re:wr}{\rm
Theorem 1.2 from \cite{Ge} looks more general then Proposition
\ref{pr:fnp}, but I found a gap in my proof of that theorem, cf.\
the comment on page 162 in \cite{GI2}. In fact, I know how to deduce
Proposition \ref{pr:fnp} from \cite[Proposition 3.32]{GI2} (which is
elementary and easy to prove), but the argument is much more
involved than the methods used in the present paper (and the
assumptions that $\co$ is abelian and that there are no finite rank
operators in the Von Neumann algebra generated by $\co$ cannot be
avoided). 
}\end{remark}

We finish with some applications in spectral theory. An advantage in
having a particle number cutoff is that the strict positivity of the
one particle mass is no more necessary, in fact the one particle
kinetic energy $h$ can be an arbitrary bounded from below
self-adjoint operator affiliated to $\co$. On the other hand, the
notion of \emph{standard} QFH as introduced in Definition
\ref{df:dqfh} does not make sense now. Instead, in the present
context it is natural to consider the following class of
\emph{elementary} QFH with a particle number cutoff: these are the
operators of the form $H_n=\dd\Gamma_n(h)+V_n$ where $h$ is a
self-adjoint bounded from below operator affiliated to $\co$ and
$V_n\in\ra_n(\ch)$ is bounded and symmetric. It is clear that, as in
the preceding sections, one may consider much more general
interactions, but this is of no interest here. 

Such a $V_n$ being fixed, we define
$V_{k}=\cp^{n-k}(V_n)\in\ra_{k}(\ch)$ for $k\leq n$. Note that if
$V_n$ is a polynomial in the operators $\{\phi_n(u)\}_{u\in\ch}$
then $V_{k}$ is the same polynomial in which each $\phi_n(u)$ has
been replaced by $\phi_{k}(u)$. Or if $V_n=\ind_nV\ind_n$ for some
$V\in\rf(\ch)$, then $V_k=\ind_kV\ind_k$.

Let us set $H_k=\dd\Gamma_k(h)+V_k$, this is a self-adjoint operator
on $\Gamma_k(\ch)$.  Of course, $H_0=V_0=c\omega$ for some complex
number $c$. The techniques used before easily give that $H_k$ is
affiliated to $\rf_k(\co)$ and:
\begin{equation}\label{eq:fnqf}
\cp(H_k)=h\otimes\ind_{\Gamma_{k-1}(\ch)}+\ind_\ch\otimes H_{k-1}
\hspace{2mm} \text{ for } 1\leq k \leq n.
\end{equation}
In particular, we get an HVZ type description of the essential
spectrum of the operator $H_n$:
\begin{equation}\label{eq:fnes}
\se(H_n)=\sigma(h)+\sigma(H_{n-1}).
\end{equation}
Note how much simpler is this formula than in the $n$-body
situation. 

The treatment of the Mourre estimate is entirely similar to that
from Section \ref{s:me}, so we give only the result. We consider
only conjugate operators of the form $A_n=\dd\Gamma_n(\g{a})$ where
$\g{a}$ is as in condition (OA) page \pageref{p:A}. Exactly as in
the proof of Theorem \ref{th:mou} we now get:
\begin{equation}\label{eq:fnths}
\tau_{A_n}(H_n)=
\ccup_{k=1}^n\big[\tau^k_{\g{a}}(h)+\sigma_{\text{p}}(H_{n-k})\big]
\end{equation} 
where we make the convention $\sigma_{\text{p}}(H_0)=\{0\}$. Indeed,
if we abbreviate 
$\tau(h)=\tau_{\g{a}}(h)$ and $\tau(H_n)=\tau_{A_n}(H_n)$, then
\eqref{eq:fnths} follows by induction from the analogue in the
present context of \eqref{eq:sten}, namely:
\begin{equation}\label{eq:fnth}
\tau(H_n)=
\tau(h)+\left[\sigma_{\text{p}}(H_{n-1})\cup\tau(H_{n-1})\right]=
\left[\tau(h)+\sigma_{\text{p}}(H_{n-1})\right]\ccup
\left[\tau(h)+\tau(H_{n-1})\right].
\end{equation}

\addcontentsline{toc}{section}{
{References}}

\end{document}